\documentclass[%
reprint,
superscriptaddress,
 amsmath,amssymb,
aps,
]{revtex4-2}
\allowdisplaybreaks[4]
\usepackage[dvipsnames]{xcolor}
\usepackage{graphicx}
\usepackage{dcolumn}
\usepackage{bm}
\usepackage{subfig}
\usepackage{tabularx}
\usepackage{soul}
\soulregister{\ref}7
\soulregister{\cite}7
\soulregister{\eqref}7

\usepackage{lineno}
\usepackage{hyperref}

\hypersetup{
colorlinks=true, 
linkcolor=Blue, 
citecolor=Blue, 
filecolor=black, 
urlcolor=Blue 
}
\makeatletter
\pretocmd{\NAT@open}{\begingroup\color{\@citecolor}}{}{}
\apptocmd{\NAT@close}{\endgroup}{}{}

\begin{document}

\preprint{APS/123-QED}

\title{Relativistic model of binary extreme-mass-ratio inspiral systems and their gravitational radiation}
\author{Yucheng Yin}
\affiliation{Department of Astronomy, School of Physics, Peking University, 100871 Beijing, China}
\affiliation{Kavli Institute for Astronomy and Astrophysics at Peking University, 100871 Beijing, China}

\author{Josh Mathews}
\affiliation{Department of Physics, National University of Singapore, Singapore 117551}

\author{Alvin J. K. Chua}
\affiliation{Department of Physics, National University of Singapore, Singapore 117551}
\affiliation{Department of Mathematics, National University of Singapore, Singapore 119076}

\author{Xian Chen}%
\email{Corresponding author. \\ xian.chen@pku.edu.cn}
\affiliation{Department of Astronomy, School of Physics, Peking University, 100871 Beijing, China}
\affiliation{Kavli Institute for Astronomy and Astrophysics at Peking University, 100871 Beijing, China}
\date{\today}

\begin{abstract}
A binary extreme-mass-ratio inspiral (b-EMRI) is a hierarchical triple system 
consisting of a stellar-mass binary black hole (BBH) orbiting a central Kerr 
supermassive black hole (SMBH). Although predicted by 
several astrophysical models, b-EMRIs pose a challenge in
waveform modeling due to their complex three-body dynamics and 
strong relativistic effects. Here we take advantage of the hierarchical nature
of b-EMRI systems to transform the internal motion of the small binary into
global trajectories around the SMBH. This allows us to use black hole perturbation theory to calculate both the low-frequency gravitational waveform due to its EMRI nature and the high-frequency waveform generated by the inner motion of the BBH. 
When the inner binary's separation vanishes, our calculation recovers the standard relativistic adiabatic EMRI waveform.
Furthermore, by
including the
high-frequency perturbation, we find a correction to the waveform
as large as the adiabatic order when
the frequency matches the quasinormal
modes (QNMs) of the SMBH, therefore supporting an earlier proof-of-concept study claiming that
the small BBH can resonantly excite the QNMs of the SMBH. More importantly, we find that
b-EMRIs can evolve faster than regular EMRIs
due to this resonant dissipation through the high-frequency modes.
These characteristics distinguish b-EMRI waveform templates from regular EMRI templates for future space-based gravitational wave detectors.
\end{abstract}

\maketitle


\section{Introduction}

Extreme-mass-ratio inspirals (EMRIs) are an important target for future
milli-hertz gravitational wave (GW) detectors, such as the Laser Interferometer
Space Antenna (LISA) \cite{2018LRR....21....4A, Colpi:2024xhw}, or China's
Taiji \cite{10.1093/nsr/nwx116} and TianQin \cite{2016CQGra..33c5010L,Li:2024rnk}
projects. An EMRI normally consists of a stellar-mass black hole (BH) or neutron star (the `secondary') and a
SMBH (the `primary'). The mass ratio $\epsilon$ of the two bodies
falls in a typical range between $10^{-6}$ and $10^{-4}$.  By studying the GWs
radiated during the last $\mathcal{O}(1/\epsilon)$ orbits before plunge
\cite{PhysRevD.78.064028}, we can measure the spin and mass of the central
SMBH, and test the Kerr metric with unprecedented precision
\cite{PhysRevD.69.124022,PhysRevD.81.024030,2019BAAS...51c..42B,Zi:2021pdp,Babak:2017tow}. There have been significant efforts in developing both approximate and fully relativistic EMRI waveform models over the years, e.g.,~\cite{Barack:2003fp,2015CQGra..32w2002C, Fujita:2020zxe, 2021PhRvL.126e1102C, Katz:2021yft,2021PhRvD.103j4014H, 2022PhRvL.128w1101I,Wardell:2021fyy,Speri:2023jte,Babak:2006uv,Nasipak:2023kuf,Albertini:2023aol}, including many works on the finite-size effects of the compact secondary body
\cite{2010PhRvD..82h4013H, 2020PhRvD.102b4041P, 2022PhRvD.105l4040D, 2022PhRvD.105l4041D, Piovano:2021iwv, Skoupy:2021asz, Rahman:2021eay, 2022PhRvD.105h4031M, Skoupy:2022adh, Drummond:2023wqc,Skoupy:2023lih, Burke:2023lno, PhysRevD.110.084061,Albertini:2024rrs, Piovano:2024yks,2025arXiv250101413M}.
Recent astrophysical models predict that the small body in an EMRI, in fact,
could be a stellar-mass BBH. This type of BBH could be captured by
the SMBH during a close tidal interaction
\cite{2015arXiv150107856A,2018CmPhy...1...53C}, or delivered to the vicinity of
the SMBH by an accretion disk if the SMBH happens to be inside an active
galactic nucleus (AGN) \cite{2016ApJ...819L..17B,2021MNRAS.505.1324P,2020ApJ...898...25T,Pan:2021ksp}.
In such a triple system, the BBH is often referred to as the ``inner binary'' (IB),
and the binary formed by the IB and the SMBH is called the ``outer binary'' (OB).
Since the small BBH is now spiraling into the SMBH, the whole system is now known as the
``binary EMRI'' (or b-EMRI for short \cite{2018CmPhy...1...53C,2023LRR....26....2A}).

Although solid observational evidence for b-EMRIs is currently lacking, it has been
suggested that the event GW190521, discovered in the third observing run of the
Laser Interferometer GW Observatory (LIGO) and the Virgo detectors
\cite{PhysRevLett.125.101102,2021PhRvX..11b1053A}, could have derived from a BBH merger in the vicinity of a SMBH. In particular, this event involves one of the most massive BBHs
ever detected, whose final mass is approximately $150M_\odot$, much greater than
the typical mass of the BHs found in Galactic X-ray binaries
\cite{1998ApJ...499..367B,2001ApJ...554..548F}. The high mass could be
explained by a redshift of the GW signal if the merger occurred very close to a
SMBH \cite{2019MNRAS.485L.141C}.  Moreover, a potentially associated AGN flare
is reported by the Zwicky Transient Facility \cite{PhysRevLett.124.251102} and
further analysis of the flare \cite{2020ApJ...900L..13A} suggests that GW190521
could come from the accretion disk of that AGN.

The possibility of forming b-EMRIs in astrophysical environments has inspired
many studies on the waveform of such a special GW source. The methods adopted
by these earlier studies can be broadly divided into two categories. The first
approach starts from the waveform of an isolated stellar-mass BBH and adds to
it relativistic effects to account for the existence of a nearby SMBH \cite{Silva:2022blb}.
Previously considered effects include Doppler and gravitational redshifts
\cite{2017ApJ...834..200M,2019MNRAS.485L..29H,2019ApJ...878...75R,2019MNRAS.488.5665W,2023MNRAS.521.2919Z,10.1093/mnras/stae1093},
aberration due to the fast motion of the BBH
\cite{2019PhRvD.100f3012T,2023PhRvD.107j3044Y}, and gravitational lensing of
the GW signal by the SMBH either in the geometric optics limit
\cite{2021PhRvD.103f4047E,2023MNRAS.521.2919Z} or the wave optics one \cite{PhysRevD.110.044054}.
An important ingredient that is missing by these works is the gravitational
radiation of the OB, i.e., the EMRI signal.

The second approach starts with a standard EMRI and replaces the stellar-mass
secondary with a BBH. The original idea is to model the small binary as two
point particles orbiting each other with a centre-of-mass that is in turn orbiting the SMBH.  The waveform
is then computed using the Teukolsky formalism
\cite{PhysRevD.103.L081501,2021PhLB..82236654F}. More recently, Ref.~\cite{2024EPJC...84..478J} took a
slightly different approach, viewing the BBH in a b-EMRI as a mass distribution
with spin and other finite size effects and assessed the distinguishability of b-EMRI waveforms from standard EMRI systems. This perspective leverages the Mathisson-Papapetrou-Dixon-Harte formalism~\cite{Mathisson:1937zz, Mathisson:2010opl, Papapetrou:1951pa, Dixon:1970zz, Dixon:1974xoz,Harte:2011ku}, already used extensively to describe finite size effects of the secondary in EMRI models. If we were to `skeletonize' the extended stress-energy distribution of two orbiting point particles about a reference worldline via a multipole expansion, then it becomes more apparent how the approach relates to the MPD-Harte formalism. Further extension of this method is used to study ``dirty EMRIs'' \cite{Jiang:2024lwg}. Additionally, some efforts have been made to modify the ``numerical kludge'' (NK) waveform templates to mimic the GW signals of b-EMRIs \cite{2024arXiv240507113M}.

Two problems remain unsolved in the modeling of b-EMRI waveforms.  First,
Ref.~\cite{PhysRevD.103.L081501,2021PhLB..82236654F,2024arXiv240507113M} treat the trajectories of the small BHs as an algebraic
addition of the inner and outer orbits.  Although intuitive, this addition is
invalid for  motions in the relativistic regime. Both special relativistic
effects, such as the high velocity of the IB, and general relativistic effects,
like the curvature of spacetime, can cause the circular inner orbit observed
from the centre-of-mass (CM) of the IB to appear distorted when viewed from
infinity. A rigorous coordinate transformation to account for the above
relativistic effects is needed.

The second problem is partially 
realized in Ref.~\cite{PhysRevD.103.L081501} --- that the inner motion
of the small BHs could cause a perturbation which has almost the same frequency as one
of the quasi-normal modes (QNMs) of the SMBH \cite{2009CQGra..26p3001B}. In
this case, the IB can resonantly excite the SMBH as predicted by BH
perturbation theory \cite{PhysRevD.103.L081501}, but the effect requires precise treatment of the perturbing IB.
For example, in Ref.~\cite{2024EPJC...84..478J}, the resonant excitation of the
QNMs was not observed as the high-frequency modes of the perturber were neglected. In that work, the IB is closer to merger, corresponding to higher values of the inner frequency that may have been too high to excite the QNMs. That study would need to be extended to larger separations of the IB to discern QNM excitation.

Recent dynamics models for b-EMRIs have made significant progress in describing the
evolution of the IB and OB simultaneously
\cite{2022PhRvD.106j3040C,PhysRevD.107.084011,PhysRevD.108.123041,10.1063/1.3382338}.  The key
idea is to solve the evolution in a frame in which the CM of
the IB is in free fall. In this free-fall frame (FFF), by the equivalence principle, the
evolution of the BBH is dominated by its self-gravity. For our purposes, a
coordinate transformation from the FFF to the Boyer-Lindquist (BL) coordinates
\cite{1967JMP.....8..265B,2005CQGra..22.4729B,2008CQGra..25n5019K}, in
principle, should provide us with the trajectories of the two small BHs around
the SMBH. We can isolate the
high-frequency components of the trajectories due to the inner motion of the IB and evaluate their excitation of the QNMs. We use this scheme to produce a relativistic b-EMRI
waveform model.

Typical EMRI models employ a multi-scale (or two-timescale) expansion \cite{PhysRevD.78.064028, Pound:2021qin,LISAConsortiumWaveformWorkingGroup:2023arg,2025arXiv250101413M}. Within the multi-scale framework, the resulting waveforms are labelled by the accuracy of the waveform phase. This follows from the foundational scaling argument in Ref.~\cite{PhysRevD.78.064028}, that showed that the average waveform phase has the asymptotic small mass-ratio expansion
\begin{equation}
\phi=\epsilon^{-1}\phi^{(0)}+\epsilon^0\phi^{(1)}+\mathcal{O}(\epsilon).
\end{equation}
The first term is labelled as `adiabatic order' (0PA) and the subsequent subleading terms are `n$^{\text{th}}$ post-adiabatic order' (nPA). For simplicity, we neglect discussion of orbital resonances which include phase terms scaling with $\epsilon^{-1/2},\epsilon^{1/2}$ etc. For EMRI parameter estimation, it is anticipated that 1PA accurate models should be sufficient \cite{LISAConsortiumWaveformWorkingGroup:2023arg,Colpi:2024xhw}. In the model in this work, the b-EMRI parameters formally distinguish the waveform from a regular EMRI in a phase term that scales as $(d/M)^2\epsilon^{-1}$, where $d$ is the diameter of the IB's orbit. The Newtonian inner motion condition and the stability of the b-EMRI imply that $(d/M)^2\approx\epsilon$ (as discussed in Sec.~\ref{IVA} and Sec.~\ref{VB}), meaning that the distinguishing phase term effectively scales equivalently to a 1PA phase correction. However, we find that resonances between the IB's orbital frequency and the SMBH's QNMs can cause the b-EMRI phase correction to grow as large as a significant fraction of $\phi^{(0)}$.

This paper is organized as follows. In Sec.~\ref{II} we derive the trajectories
of the small black holes using a series of coordinate transformations, starting
from the FFF to the Local Inertia Frame (LIF) and finally to BL
coordinates. We use these trajectories to expand the Teukolsky source term into
a series expansion in the diameter of the IB. In addition, we present the schemes used
to compute the waveform snapshots and adiabatic waveforms. In Sec.~\ref{III},
we describe the steps we take to compute the numerical waveform.  In
Sec.~\ref{IV}, we show the results from our model, including the fluxes,
evolution of the OB, waveforms, and the properties of the QNMs in b-EMRI
systems. In Sec.~\ref{V}, we compare our results with those presented in
Ref.~\cite{PhysRevD.103.L081501,2024EPJC...84..478J,2024arXiv240507113M}. We
also explore the parameter space to find stable b-EMRI systems and examine the
validity of our model. Finally, in Sec.~\ref{VI} we summarize our main findings
and outline potential directions for future studies. Throughout this paper, we
set $G=c=1$. 

\section{\label{II}Theory}

\subsection{Definition of model parameters}

\begin{figure}[]
    \centering
    
    \subfloat[]{\label{fig:Scenario}\includegraphics[width=1.0\linewidth]{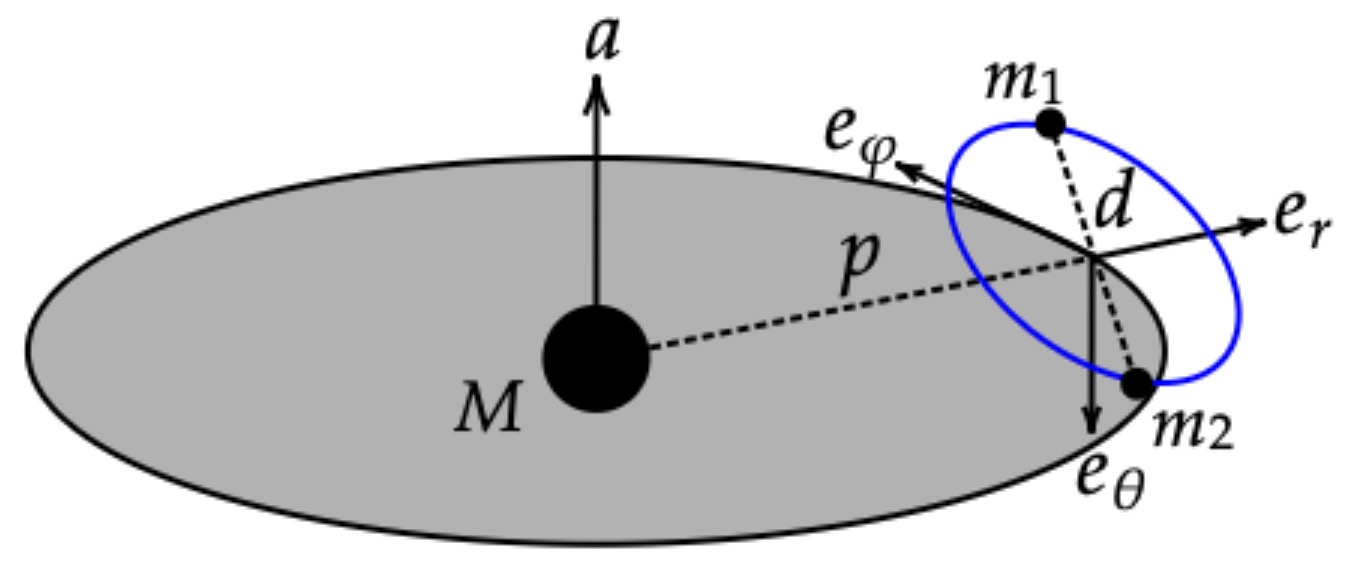}}

    \subfloat[]{\label{fig:EulerRotation}\includegraphics[width=0.8\linewidth]{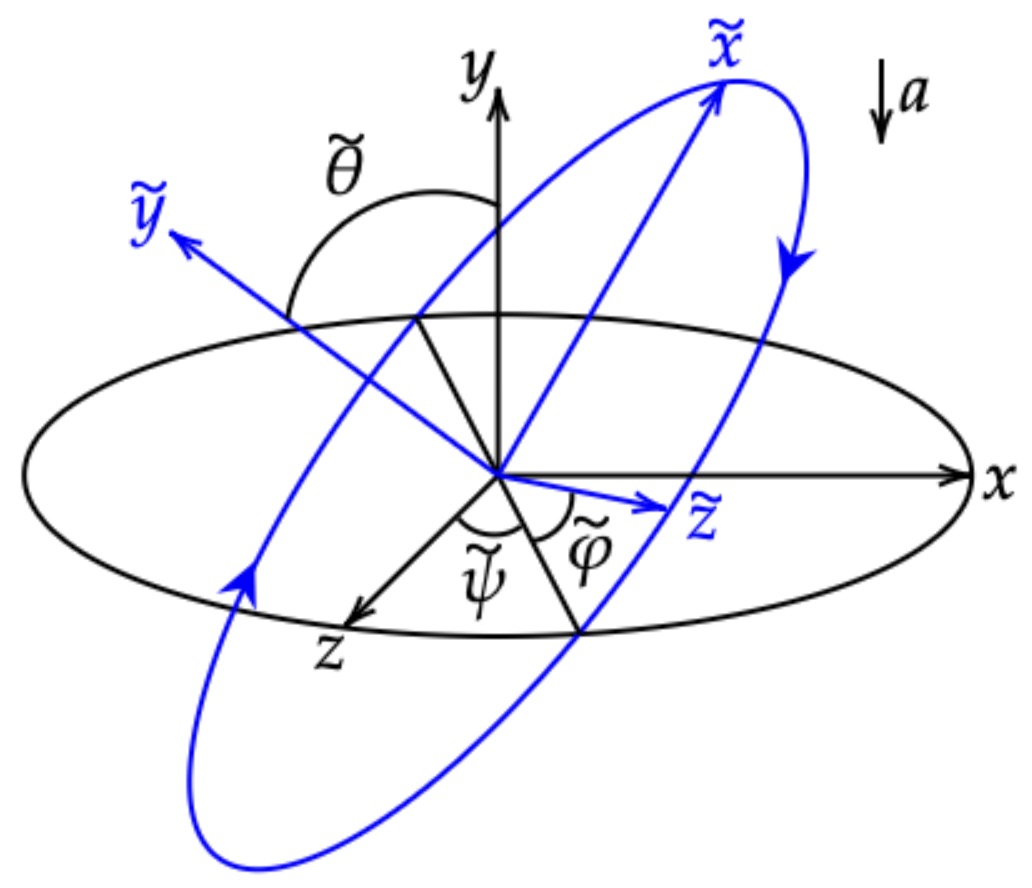}}

    \caption{A b-EMRI system and its parameters. Panel (a) illustrates the general scenario of a b-EMRI system. Panel (b) shows the Euler rotation between the inner orbital plane (in blue) and the outer orbital plane (in black). The blue arrows indicate the direction of motion of the small black holes. Note that the $y$-axis in panel (b) is geometrically aligned with the $e_\theta$-axis in the BL coordinates, which is opposite to the SMBH's spin direction for equatorial OB.}
    \label{fig:b-EMRI}
\end{figure}

Fig.~\ref{fig:Scenario} shows the physical picture of a b-EMRI system. The
central SMBH is a Kerr black hole with a mass of $M\sim 10^5-10^8M_\odot$ and a
spin parameter of $a$. We  denote $p$ as the semi-latus rectum of the OB, which
is equivalent to the radius of the OB if the orbit is circular.  As for the IB,
we denote the masses of the two stellar-mass BHs as $m_1$ and $m_2$. We use
$\mu$ to represent the total mass of the IB so that
$\mu\equiv m_1+m_2=M\epsilon$, and $m$ is reserved to denote a multipole
number, which is an integer. We assume $m_1=m_2$ and a circular orbit for the IB for simplicity
and denote its diameter as $d$.

In the following, we will assume a Newtonian orbit for the inner orbital motion
of the two stellar BHs, which requires that $d$ is relatively large compared to
$\mu$. However, if $d$ is too large, the IB will be unstable due to the Hill's mechanism. More delicate constraints on $d$ will be provides in Sec.~\ref{VB}. An appropriate choice is $d\approx\mbox{few thousand}\ \mu$ (see Sec.~\ref{IVA}). For such a large $d$, we need a fairly small $\epsilon$ (e.g., $\epsilon<10^{-4}$) to maintain the stability of the system. Other constraints on $d$, $\epsilon$, and $p$ will
also be discussed in detail in Sec.~\ref{VB}. 

As shown in Fig.~\ref{fig:EulerRotation}, there are three Euler angles
determining the orientation of the IB and we will see how they function in Eq.~\eqref{Eq.EulerRotation}. The $y$-axis of the FFF in Fig.~\ref{fig:EulerRotation} is naturally defined in the same direction as the $e_\theta$-axis of the BL coordinates in Fig.~\ref{fig:Scenario}, thus appearing downward when viewed in Fig.~\ref{fig:Scenario}. To facilitate our discussion of Euler rotation, we have reversed the system in Fig.~\ref{fig:EulerRotation} so that the spin direction of the SMBH is downward. However, the directions of the $x$- and $z$-axes do not coincide with those of $e_r$ and $e_\varphi$, but differ by a precession $\Omega$. To describe this precession, we define the LIF. The coordinate transformations between FFF, LIF, and BL will be thoroughly discussed in Sec.~\ref{IIC}. Form the perspective of Euler rotation, $\tilde{\psi}$ is the precession angle, $\tilde{\varphi}$ is the rotation angle and  $\tilde{\theta}$ is the nutation angle. Physically, $\tilde{\theta}$ is also the angle between the inner orbital angular momentum and the outer one. The $\tilde{\varphi}$ can be absorbed into the inner motion, resulting in a degeneracy with the initial phase of the inner circular motion. We therefore set the initial phase of IB to zero to simplify the calculations. Besides, $\tilde{\psi}$ degenerates with the azimuthal angle of the observer, $\varphi$, in the BL coordinates. We therefore set $\tilde{\psi}$ to zero for the same reason. In total there are eight intrinsic parameters in our b-EMRI model. They are listed in Table~\ref{tab1}.

\begin{table}[]
    \centering
    \begin{tabular}{c|l}
    \hline
         \textit{M}& mass of SMBH $\sim 10^5-10^8M_\odot$\\
         \textit{a}& spin of SMBH fixed at $0.9M$\\
         $\epsilon$& mass ratio $\sim10^{-6}-10^{-4}$\\
         \textit{p}& outer radius scaled by $M$\\
         \textit{d}& diameter of IB scaled by $\mu=M\epsilon$\\
         $\tilde{\psi}$& precession angle set to $0$\\
         $\tilde{\theta}$& inclination angle ranging from $0$ to $\pi$\\
         $\tilde{\varphi}$& rotation angle ranging from $0$ to $2\pi$\\
         \hline
    \end{tabular}
    \caption{Key parameters of b-EMRI in this paper, where $M$, $a$, $\epsilon$, $p$, $d$ are shown in Fig.~\ref{fig:Scenario}, and $\tilde{\psi}$, $\tilde{\theta}$, $\tilde{\varphi}$ are shown in Fig.~\ref{fig:EulerRotation}}
    \label{tab1}
\end{table}

\subsection{Orbital motion of the outer binary}

In this study, we assume that the orbit of the OB follows a circular-equatorial
orbit in Kerr spacetime, the metric of which is
\begin{equation}\label{Eq.KerrMetric}
    \begin{aligned}
        \mathrm{d}s^2=&-\left(1-\frac{2Mr}{\Sigma}\right)\mathrm{d}t^2-\frac{4Mar\sin^2\theta}{\Sigma}\mathrm{d}t\mathrm{d}\varphi+\frac{\Sigma}{\Delta}\mathrm{d}r^2\\
        &+\Sigma\mathrm{d}\theta^2+\sin^2\theta\left(r^2+a^2+\frac{2Ma^2r\sin^2\theta}{\Sigma}\right)\mathrm{d}\varphi^2,
    \end{aligned}
\end{equation}
where $\Delta = r^2 - 2Mr + a^2$ and $\Sigma = 1/(\rho\bar{\rho})$ with
$\rho=(r-ia\cos\theta)^{-1}$. 
Here overbar denotes complex conjugation. The complete form of the Kerr geodesics can be
found in Ref.~\cite{1973grav.book.....M,PhysRev.174.1559,Fujita_2009}, but here we focus on the
circular-equatorial case, where the Carter constant $\mathcal{Q}$ is set to
zero. We express the geodesic equations in terms of the four-velocity
$u^x=\mathrm{d}x/\mathrm{d}\tau \ (x=t,r,\theta,\varphi)$ as follows
\begin{subequations}\label{Eq.KerrGeodesics}
    \begin{align}
        &u^t=\frac{\mathcal{E}}{p^2}\Big[\frac{(p^2+a^2)^2}{\Delta}-a^2\Big]+\frac{a\mathcal{L}_z}{p^2}\Big(1-\frac{p^2+a^2}{\Delta}\Big),\\
    &u^r=0\qquad(\mbox{with}\quad r=p),\\
    &u^\theta=0\qquad(\mbox{with}\quad \theta=\pi/2),\\
    &u^{\varphi}=\frac{\mathcal{L}_z}{p^2}+\frac{a\mathcal{E}}{p^2}\left(\frac{p^2+a^2}{\Delta}-1\right)-\frac{a^2\mathcal{L}_z}{p^2\Delta},
    \end{align}
\end{subequations}
where $\mathcal{E}$ and $\mathcal{L}_z$ are dimensionless energy and angular
momentum of the OB. 

In the circular-equatorial case, the quantities $\mathcal{E}$ and $\mathcal{L}_z$ can be
written as
\begin{subequations}\label{Eq.E,Lz}
    \begin{align}
        \mathcal{E}&\equiv\frac{E}{\mu}=\frac{p^{3/2}-2 M p^{1/2}\pm a
   M^{1/2}}{p^{3/4} \left(p^{3/2}-3 M
   p^{1/2}\pm 2 a M^{1/2}\right)^{1/2}},\\
   \mathcal{L}_z&\equiv\frac{L_z}{\mu}=\frac{\pm M^{1/2}(p^2\mp2aM^{1/2}p^{1/2}+a^2)}{p^{3/4} \left(p^{3/2}-3 M
   p^{1/2}\pm 2 a M^{1/2}\right)^{1/2}},
    \end{align}
\end{subequations}
where the upper signs correspond to a prograde outer orbit and lower signs to a retrograde one. Throughout the paper, we assume the IB orbit is prograde. These expressions can be derived from 
the radial component of the Kerr geodesic
equations \cite{1973grav.book.....M}
\begin{equation}\label{Eq.r-Kerr}
    \begin{aligned}
        \Sigma^2\left(\frac{\mathrm{d} r}{\mathrm{d}\tau}\right)^2=&[\mathcal{E}(p^2+a^2)-a\mathcal{L}_z]^2\\
        &-\Delta[p^2+(\mathcal{L}_z-a\mathcal{E})^2]\equiv R,
    \end{aligned}
\end{equation}
where we have assumed $\mathcal{Q}=0$.  By setting $R=0$ and
$R'\equiv\partial_pR=0$ for a circular-equatorial orbit, we obtain
Eq.~\eqref{Eq.E,Lz}.

\subsection{Coordinate transformation for the inner orbit\label{IIC}}

The relative motion of the two stellar-mass BHs can be solved most easily in a
frame in which the binary's CM is freely falling in the curved spacetime around the BBH (the aforementioned FFF). In this
frame, the spacetime is sufficiently flat due to Einstein's Equivalence
Principle, and hence the two BH follows a near-Keplerian orbit. Together with
the previous assumption of a circular IB, we can calculate the inner
orbital frequency as $\omega_{\mathrm{IB}}=\sqrt{\mu/d^3}$ and write the coordinates of
the two BHs in their orbital plane as
\begin{subequations}\label{Eq.FFFCircular}
    \begin{align}
            &\tilde{x}=\frac{1}{2}d\cos(\omega_{\mathrm{IB}}\tau),\\
            &\tilde{y}=0,\\
            &\tilde{z}=\frac{1}{2}d\sin(\omega_{\mathrm{IB}}\tau),
    \end{align}
\end{subequations}
where $\tau$ is the proper time of the FFF, and we have assumed equal mass for the
inner binary so that the radius of the circular motion is $d/2$.
Furthermore, to account for an arbitrary orientation of the IB, we use   
\begin{equation} \label{Eq.EulerRotation} 
    \begin{aligned}
        \begin{bmatrix}
                x\\y\\z
        \end{bmatrix}=&\begin{bmatrix}
                \cos\tilde{\varphi}&0&\sin\tilde{\varphi}\\
                0&1&0\\
                -\sin\tilde{\varphi}&0&\cos\tilde{\varphi}
    \end{bmatrix}\begin{bmatrix}
                \cos\tilde{\theta}&-\sin\tilde{\theta}&0\\
                \sin\tilde{\theta}&\cos\tilde{\theta}&0\\
                0&0&1
    \end{bmatrix}\\
    &\times\begin{bmatrix}
                \cos\tilde{\psi}&0&\sin\tilde{\psi}\\
                0&1&0\\
                -\sin\tilde{\psi}&0&\cos\tilde{\psi}
    \end{bmatrix}\begin{bmatrix}
                \tilde{x}\\ \tilde{y}\\ \tilde{z}
    \end{bmatrix},
    \end{aligned}
\end{equation}
to rotate the orbital plane of the IB and get the coordinates in the FFF. Note that the rotation of $\tilde{\varphi}$ can be absorbed as an initial phase in Eqs.~\eqref{Eq.FFFCircular}, which has been set to zero in Eqs.~\eqref{Eq.FFFCircular}.

\begin{figure}[htb!]
    \centering
    \includegraphics[width=1.0\linewidth]{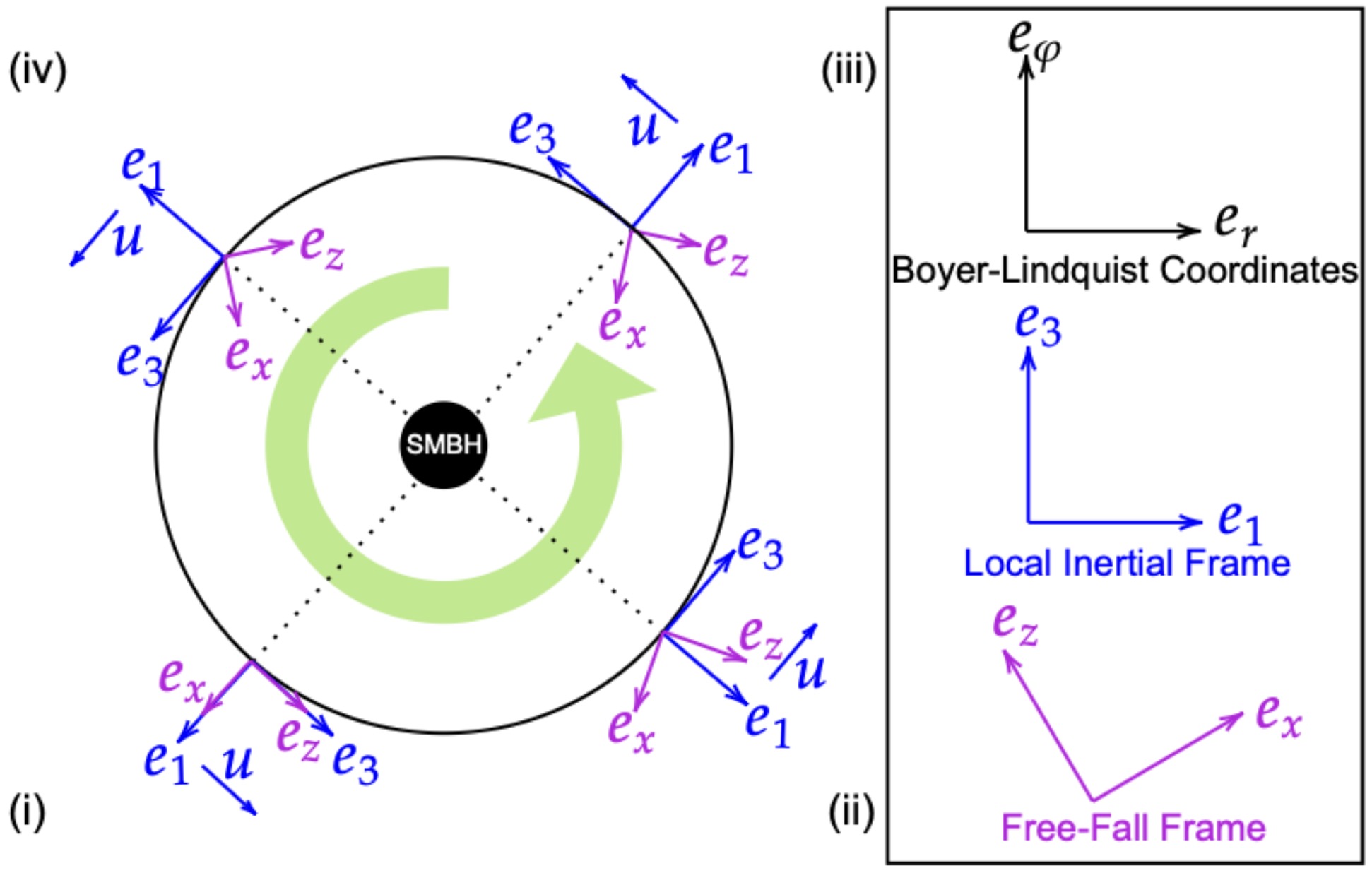}
    \caption{Different reference frames used in this paper and their relative orientation as they rotate around the SMBH. It is a top view of Fig. \ref{fig:Scenario} where $e_1,e_2,e_3$ are unit vectors in the LIF, aligned with $e_{r},e_{\theta},e_{\varphi}$ in the BL coordinates. The FFF and the LIF share the same origin, centered on the CM of the BBH, while the origin of the BL coordinates is centered on the Kerr SMBH which is not shown in the figure.}
    \label{fig:CoordinateTransformation}
\end{figure}

Since the Teukolsky formalism is derived in BL coordinates, we need to transform
the motion of the small BHs from the FFF to BL coordinates. This transformation is done
in two steps \cite{1990GReGr..22.1067R,2005CQGra..22.4729B}.
The relationships between different coordinates are 
illustrated in Fig.~\ref{fig:CoordinateTransformation}.

First, we transform the inner motion into the LIF. This frame is moving at the
same velocity as the FFF but its three spatial axes are aligned, respectively,
with the  $e_r$,  $e_\theta$, and  $e_\varphi$ axes of the BL
coordinates. It has been shown that the spatial axes of the FFF and LIF differ
by a precession, with an angular frequency of $\Omega=-\sqrt{M/p^3}$
\cite{1990GReGr..22.1067R}. It happens to be the Keplerian orbital frequency in
our circular-equatorial case. There is a minus sign because the $e_y$ axis points downward when aligned with the $e_\theta$ axis in the equatorial plane. Consequently, coordinates in these two frames
also differ by a rotational transformation
\begin{subequations}\label{Eq.FFFtoLIF}
    \begin{align}
        T&=\tau,\\
        X^1&=x\cos(\Omega\tau)-z\sin(\Omega\tau),\\
        X^2&=y,\\
        X^3&=x\sin(\Omega\tau)+z\cos(\Omega\tau).
    \end{align}
\end{subequations}
Therefore, the $e_x$ and $e_z$ axes in FFF will coincide with the $e_1$ and $e_3$ axes in LIF periodically, i.e., at the moments (i) in Fig.~\ref{fig:CoordinateTransformation}. By comparing Fig.~\ref{fig:CoordinateTransformation} (i) with Fig.~\ref{fig:Scenario}, we find that $\tilde{\psi}=0$ at these moments. We can choose these moments as reference points so that the initial precession information is contained in the azimuthal angle $\varphi$ between the BBH, the SMBH, and the observer. Therefore, the rotation with respect to $\tilde{\psi}$ in Eq.~\eqref{Eq.EulerRotation} becomes identical transformation.

Second, we transform the coordinates in the LIF into the BL coordinates.
Although the distance between each stellar-mass BH and the CM of the IB is
fixed in the LIF (due to our assumption of circular inner orbit), in the BL
coordinates the distance varies with time because of two reasons.  (i) The
coordinates in the LIF are measured by an observer at the CM of the IB, while
the BL coordinates are measured by an observer at infinity. There is a
coordinate transformation between the two.  (ii) Since the CM is moving, the
coordinate transformation also varies with time.  
It is important to mention that
such a variation of the distance from the CM has not been accounted for in 
the previous studies.
This varying distance 
can be generally expanded into an infinite series of terms
\cite{2008CQGra..25n5019K}, and it is sufficient to truncate 
the terms at the
second order in $X^{1,2,3}$ for our purposes. The resulting trajectory of one small BH has been derived in
Ref.~\cite{2005CQGra..22.4729B} and can be written as
\begin{subequations}\label{Eq.BLGeneralCoordinate}
    \begin{align}
        t=t_0+&\frac{(\mathcal{M}\mathcal{M}_\varphi+\gamma_{\varphi\varphi}^{1/2}\nu)X^3(\kappa_\varphi X^1+1)}{\gamma_{\varphi\varphi}^{1/2}\mathcal{M}/\gamma}\notag\\
        +&\frac{(\gamma_{\varphi\varphi}^{1/2}+\mathcal{M}\mathcal{M}_\varphi \nu)(T-\Omega X^1X^3)}{\gamma_{\varphi\varphi}^{1/2}\mathcal{M}/\gamma},\\
        r=r_0+&\frac{2X^1+\kappa_r (X^1)^2-\kappa_\theta(X^2)^2-\kappa_{\varphi}(X^3)^2}{2\gamma_{rr}^{1/2}},\\
        \theta=\theta_0+&\frac{\kappa_{\theta}X^1X^2+X^2}{\gamma_{\theta\theta}^{1/2}},\\
        \varphi=\varphi_0+&\frac{X^3(\kappa_{\varphi}X^1+1)+\nu(T-\Omega X^1X^3)}{\gamma_{\varphi\varphi}^{1/2}/\gamma},
    \end{align}
\end{subequations}
where $\mathcal{M}$, $\mathcal{M}_\varphi$, $\gamma_{rr}^{1/2}$, $\gamma_{\theta\theta}^{1/2}$, $\gamma_{\varphi\varphi}^{1/2}$, $\kappa_{r}$, $\kappa_{\theta}$, $\kappa_{\varphi}$, $\nu$, $\gamma$ are all functions of $M$, $a$, $p$.
Detailed expressions for these functions are provided in Appendix~\ref{appA}. 

From Eqs.~\eqref{Eq.EulerRotation}-\eqref{Eq.BLGeneralCoordinate}, the BL coordinate
$t$, $r$, $\theta$, $\varphi$ of each small BH 
can be expressed as a series expansion of $d$, which must be small to ensure the stability of the IB which we discuss in Sec.~\ref{VB}. The trajectory then takes the form of
\begin{subequations}\label{Eq.BLCompactCoordinate}
    \begin{align}
        &t=c_{00}+c_{01}d+c_{02}d^2+\mathcal{O}(d^3),\\
        &r=c_{10}+c_{11}d+c_{12}d^2+\mathcal{O}(d^3),\\
        &\theta=c_{20}+c_{21}d+c_{22}d^2+\mathcal{O}(d^3),\\
        &\varphi=c_{30}+c_{31}d+c_{32}d^2+\mathcal{O}(d^3),
    \end{align}
\end{subequations}
where the zeroth order terms $c_{00},\ c_{10},\ c_{20},$ and $c_{30}$ correspond to the outer motion with the geodesic relations:
\begin{subequations}\label{Eq.ZerothOrderTrajectory}
    \begin{align}
        &c_{00}=u^t\tau,\\
        &c_{10}=p,\\
        &c_{20}=\pi/2,\\
        &c_{30}=u^\varphi\tau.
    \end{align}
\end{subequations}
The first and second order coefficients $c_{01}$, $c_{02}$, $c_{11}$, $c_{12}$,
$c_{21}$, $c_{22}$, $c_{31}$, $c_{32}$ are derived in Appendix~\ref{appA}.
We note that $c_{01}$, $c_{02}$, $c_{11}$,
$c_{12}$, $c_{21}$, $c_{22}$, $c_{31}$, $c_{32}$ contain $\omega_{\mathrm{IB}} =
\sqrt{\mu / d^3}$, which is a function of $d$. It is important to stress that we perform all of the series expansions in $d$ at fixed values of $\omega_{\mathrm{IB}}$ for several reasons. First of all, that is the most straightforward way to ensure the expansions are well-behaved when treating the disparate time and length scales of the outer and inner binaries. Secondly, it avoids inaccurate expansions of the oscillatory functions of the inner binary's orbital phase.
Finally, by holding $\omega_{\mathrm{IB}}$ fixed, we avoid cumbersome expansions of $d$-dependence of the homogeneous radial Teukolsky solutions via $\omega_{\mathrm{IB}}$ in the discrete frequency spectrum, as described in the following subsection. Thus, we precalculate
$\omega_{\mathrm{IB}}$ and treat it as an independent parameter throughout. Its evolution is much slower than the outer binary's evolution and, therefore, may be neglected.

The expansion shown above reveals a much richer structure in the trajectory of
each small BH than a simple addition of the inner and outer motion as assumed
previously in Ref.~\cite{PhysRevD.103.L081501,2021PhLB..82236654F,2024arXiv240507113M}.  It allows
us to more accurately account for the deviation of each small BH from geodesic
motion and, in turn, compute the waveform resulting from the perturbed 
Kerr spacetime.

\subsection{Teukolsky formalism}

In this subsection, we review the Teukolsky-based Green function method for
computing gravitational radiation within the framework of black hole
perturbation theory. We refer our readers to Ref.~\cite{2003LRR.....6....6S}
for an overview of the Teukolsky formalism.  In addition, an
analytical scheme of calculating the generic time-like Kerr geodesics can be found
presented in Ref.~\cite{Fujita_2009}, and the method of generating waveform
snapshots for EMRIs on generic orbits is discussed in
Ref.~\cite{2006PhRvD..73b4027D,Fujita:2009us}.

The Teukolsky master equation provides a comprehensive framework for describing
perturbations to Kerr spacetime \cite{Teukolsky:1972my,1973ApJ...185..635T}. 
It is expressed as follows:
\begin{widetext}
    \begin{equation}\label{Eq.TeukolskyMasterEquation}
	\begin{aligned}
			\mathcal{D}_s^2\psi\equiv&\left[\frac{(r^2+a^2)^2}{\Delta}-a^2\sin^2\theta\right]\frac{\partial^2\psi}{\partial t^2}+\frac{4Mar}{\Delta}\frac{\partial^2\psi}{\partial t\partial \varphi}+\left[\frac{a^2}{\Delta}-\frac{1}{\sin^2\theta}\right]\frac{\partial^2\psi}{\partial\varphi^2}
			-\Delta^{-s}\frac{\partial}{\partial r}\left(\Delta^{s+1}\frac{\partial\psi}{\partial r}\right)\\
   &-\frac{1}{\sin\theta}\frac{\partial}{\partial\theta}\left(\sin\theta\frac{\partial\psi}{\partial\theta}\right)-2s\left[\frac{a(r-M)}{\Delta}+\frac{i\cos\theta}{\sin^2\theta}\right]\frac{\partial\psi}{\partial\varphi}-2s\left[\frac{M(r^2-a^2)}{\Delta}-r-ia\cos\theta\right]\frac{\partial\psi}{\partial t}\\
   &+s(s\cot^2\theta-1)\psi=4\pi\Sigma \hat{T}\equiv\mathcal{T},
	\end{aligned}
\end{equation}
\end{widetext}
where $s = -2$ is the spin weight corresponding to a gravitational field. The
master function $\psi$ is related to the Weyl scalar by $\psi = \rho^{-4}
\psi_4$, where $\rho=(r-ia\cos\theta)^{-1}$ and $\psi_4 = -C_{\alpha\beta\gamma\delta} n^\alpha \bar{m}^\beta
n^\gamma \bar{m}^\delta$ is the Weyl scalar given by the Newman-Penrose (NP)
formalism \cite{1962JMP.....3..566N}. The source term $\hat{T}=2\rho^{-4}T_4$ contains information about the two small BHs, which act as the perturbers. The expression for $T_4$ can be found in Ref.~\cite{1973ApJ...185..635T}.

Both the solution
$\psi$ and the source term $\mathcal{T}$ can be decomposed into multiple modes
by variable separation and Fourier transformation, and the result is
\begin{subequations}\label{Eq.VariableSeparation}
    \begin{align}
        &\psi=\sum_{l=2}^\infty\sum_{m=-l}^l\int_{-\infty}^{\infty}\mathrm{d}\omega\ e^{-i\omega t+im\varphi}R_{lm}(r,\omega)\frac{{_{-2}}S_{lm}^{a\omega}(\theta)}{\sqrt{2\pi}},\\
        &\mathcal{T}=\sum_{l=2}^\infty\sum_{m=-l}^l\int_{-\infty}^{\infty}\mathrm{d}\omega\ e^{-i\omega t+im\varphi}\mathcal{T}_{lm}(r,\omega)\frac{{_{-2}}S_{lm}^{a\omega}(\theta)}{\sqrt{2\pi}}.
    \end{align}
\end{subequations}
Here ${_{s}}S_{lm}^{a\omega}(\theta)$ is the spin-weighted spheroidal harmonics (SWSH) function. It satisfies the orthonormalization relation
\begin{equation}\label{Eq.NormalizationOrthogonalRelation}
    \langle lm|l'm'\rangle\equiv\int_{0}^{\pi}{_{s}}S_{lm}^{a\omega}(\theta){_{s}}\bar{S}^{a\omega}_{l'm'}(\theta)\sin\theta\ \mathrm{d}\theta=\frac{\delta_{ll'}\delta_{mm'}}{2\pi},
\end{equation}

The function $R_{lm}(r,\omega)$ is the solution to the
radial equation
\begin{equation}\label{Eq.TeukolskyRadialEquation}
    \left[\Delta^2\frac{\mathrm{d}}{\mathrm{d} r}\left(\frac{1}{\Delta}\frac{\mathrm{d}}{\mathrm{d} r}\right)-\mathcal{V}_{lm}(r,\omega)\right]R_{lm}(r,\omega)=\mathcal{T}_{lm}(r,\omega),
\end{equation}
where 
\begin{equation}\label{Eq.TeukolskyEffectivePotential}
\begin{aligned}
    \mathcal{V}_{lm}(r,\omega)=&-\frac{K^2+4i(r-M)K}{\Delta}\\
    &+8i\omega r+2ma\omega+\mathcal{A}_{lm}
\end{aligned}
\end{equation}
is the effective potential, $K = (r^2 + a^2)\omega - ma$, and
$\mathcal{A}_{lm}$ is the eigenvalue of the SWSH.  The homogeneous radial
equation allows two independent solutions, and in the so-called ``in-up basis''
the two solutions have the following asymptotic behavior,
\begin{subequations}\label{Eq.AsymptoticHomo}
    \begin{align}
        &R^{\mathrm{in}}_{lm}(r\to r_+,\omega)=B^{\mathrm{tran}}_{lm}\Delta^2e^{-ikr^*},\\
        &R^{\mathrm{in}}_{lm}(r\to\infty,\omega)=B^{\mathrm{ref}}_{lm}r^3e^{i\omega r^*}+\frac{B^{\mathrm{inc}}_{lm}}{r}e^{-i\omega r^*},\\
        &R^{\mathrm{up}}_{lm}(r\to r_+,\omega)=C^{\mathrm{inc}}_{lm}e^{ikr^*}+C^{\mathrm{ref}}_{lm}\Delta^2e^{-ikr^*},\\
        &R^{\mathrm{up}}_{lm}(r\to\infty,\omega)=C^{\mathrm{tran}}_{lm}r^3 e^{i\omega r^*},
    \end{align}
\end{subequations}
where $k=\omega-ma/(2Mr_+)$, $r_\pm=M\pm\sqrt{M^2-a^2}$ are the roots of
$\Delta=0$, and $r^*$ is the tortoise coordinate defined by
\begin{equation}\label{Eq.Tortoise}
    \begin{aligned}
        r^*=&\int\frac{r^2+a^2}{\Delta}\mathrm{d} r,\\
        =&r+\frac{2Mr_+}{r_+-r_-}\ln\frac{r-r_+}{2M}-\frac{2Mr_-}{r_+-r_-}\ln\frac{r-r_-}{2M}.
    \end{aligned}
\end{equation}
The solution to the inhomogeneous radial equation can be constructed from the homogeneous solutions
using the Green's function method, which results in
\begin{equation}\label{Eq.InhomogeneousTeukolskyRadial}
    R_{lm}(r,\omega)=Z^{\mathrm{in}}_{lm}(r,\omega)R^{\mathrm{up}}_{lm}(r,\omega)+Z^{\mathrm{up}}_{lm}(r,\omega)R^{\mathrm{in}}_{lm}(r,\omega).
\end{equation}
The most challenging and crucial step in our work is calculating the values of $Z^{\mathrm{in}}_{lm}(r, \omega)$ and $Z^{\mathrm{up}}_{lm}(r, \omega)$  defined as follows
\begin{subequations}\label{Eq.AmplitudesDefinition}
    \begin{align}
        &Z^{\mathrm{in}}_{lm}(r,\omega)=\frac{1}{\mathcal{W}_{lm}(\omega)}\int_{r_+}^r\mathrm{d} r^\prime\frac{R^{\mathrm{in}}_{lm}(r^\prime,\omega)}{\Delta^{\prime2}}\mathcal{T}_{lm}(r^\prime,\omega),\\
        &Z^{\mathrm{up}}_{lm}(r,\omega)=\frac{1}{\mathcal{W}_{lm}(\omega)}\int_{r}^\infty\mathrm{d}r^\prime\frac{R^{\mathrm{up}}_{lm}(r^\prime,\omega)}{\Delta^{\prime2}}\mathcal{T}_{lm}(r^\prime,\omega),
    \end{align}
\end{subequations}
where
\begin{equation}\label{Eq.Wronskian}
    \mathcal{W}_{lm}=2i\omega C^{\mathrm{tran}}_{lm}B^{\mathrm{inc}}_{lm}
\end{equation}
is the scaled Wronskian and 
\begin{equation}\label{Eq.FourierSourceTerm}
    \begin{aligned}
        \mathcal{T}_{lm}(r,\omega)=&\mu\int_{-\infty}^{\infty}\mathrm{d} t\ e^{i\omega t-im\varphi(t)}\Delta^2\\
        &\times\Big\{(A_{nn0}+A_{\bar{m}n0}+A_{\bar{m}\bar{m}0})\delta[r-r(t)]\\
        &+\left[(A_{\bar{m}n1}+A_{\bar{m}\bar{m}1})\delta[r-r(t)]\right]_{,r}\\
        &+\left[A_{\bar{m}\bar{m}2}\delta[r-r(t)]\right]_{,rr}\Big\}.
    \end{aligned}
\end{equation}
The expressions for $A_{nn0}$, $A_{\bar{m}n0}$, $A_{\bar{m}\bar{m}0}$,
$A_{\bar{m}n1}$, $A_{\bar{m}\bar{m}1}$, and $A_{\bar{m}\bar{m}2}$ can be found
in Appendix \ref{appB}. 

Here we emphasize that we have derived the source term in Eq.~\eqref{Eq.FourierSourceTerm} by modeling the IB's stress-energy as that of two free monopole particles, similar to the approach used in Ref.~\cite{PhysRevD.103.L081501}, for example. To this end, the model in this work neglects the stress-energy due to the interaction between the two particles in the IB, which would contribute additional corrections to Eq.~\eqref{Eq.FourierSourceTerm}. We further elucidate this point in the discussion in Sec.~\ref{VC} and construct the relevant interactive stress-energy in Appendix \ref{appD}. We remind readers that our new approach to modeling b-EMRI systems is sufficient to draw qualitative conclusions, but still needs improvement for high-precision parameter estimation.

Now we analyze the asymptotic behavior of Eq.~\eqref{Eq.AsymptoticHomo}.  When
$r \to r_+$, the lower and upper limits in the integration of $Z^{\mathrm{in}}_{lm}(r,
\omega)$ coincide, leading to $Z^{\mathrm{in}}_{lm}(r, \omega) = 0$.  For the same reason, 
when $r\to \infty$, $Z^{\mathrm{up}}_{lm}(r, \omega) = 0$. 
For an observer in the solar system, the distance to the source $r$ is much larger than the size of the 
b-EMRI system, $p$. Therefore, we have $r/p\to \infty$, and in this case
what is detectable by us is
\begin{equation}\label{Eq.AsymptoticInhomogeneousRadial}
    R_{lm}(r\to \infty,\omega)=Z^{\mathrm{in}}_{lm}(r\to\infty,\omega)R^{\mathrm{up}}_{lm}(r\to\infty,\omega).
\end{equation}
In the following, 
we will use $Z^{in}_{lm}(\omega) \equiv Z^{in}_{lm}(r \to \infty, \omega)$. 
At infinity, the leading order curvature and metric perturbations are related by
\begin{equation}\label{Eq.AsymptoticPsi}
    \psi_4(r\to\infty)=\frac{1}{2}\frac{\partial^2}{\partial t^2}(h_+-ih_\times).
\end{equation}
Now inserting Eqs.~\eqref{Eq.AsymptoticHomo},
\eqref{Eq.AsymptoticInhomogeneousRadial}, and \eqref{Eq.AsymptoticPsi} into
Eq.~\eqref{Eq.VariableSeparation}, we obtain the Fourier waveform
\begin{equation}\label{Eq.FourierWaveform}
    h_+-ih_\times=-\frac{2}{r}\sum_{lm}\int\mathrm{d}\omega\frac{Z^{\mathrm{in}}_{lm}}{\omega^2}\frac{{_{-2}}S_{lm}^{a\omega}(\theta)}{\sqrt{2\pi}}e^{-i\omega(t-r)+im\varphi}.
\end{equation}

\subsection{Expansion of the source term}

Eq.~\eqref{Eq.BLCompactCoordinate} inspires us to expand the Teukolsky source
term into a series in $d$.  A direct consequence is that the amplitude
$Z^{\mathrm{in}}_{lm}$ defined in Eq.~\eqref{Eq.AmplitudesDefinition} also becomes a
series in $d$. To facilitate the expansion of $Z^{\mathrm{in}}_{lm}$, we first notice
that in the most general case it can be rewritten as 
\begin{equation}\label{Eq.ZinReWrite}
\begin{aligned}
    Z^{\mathrm{in}}_{lm}(\omega)=&\frac{\mu}{\mathcal{W}_{lm}(\omega)}\int_{-\infty}^{\infty}\mathrm{d}\tau\frac{\mathrm{d}t}{\mathrm{d}\tau}\\
    &\times e^{i\omega t(\tau,d)-im\varphi(\tau,d)} \mathcal{Z}_{lm}\left[r(\tau,d),\theta(\tau,d)\right],
\end{aligned}
\end{equation}
where the function $\mathcal{Z}_{lm}$ inside the integrand is determined by the
property of the source, and 
its exact form is derived in Appendix~\ref{appB}. 
Using Eq.~\eqref{Eq.BLCompactCoordinate}, the exponential term is expanded as
\begin{equation}\label{Eq.Exp}
\begin{aligned}
    \exp[i\omega t-im\varphi]=&\mathcal{E}_{lm}^{(0)}(\tau,\omega)+\mathcal{E}_{lm}^{(1)}(\tau,\omega)d\\
    &+\mathcal{E}_{lm}^{(2)}(\tau,\omega)d^2+\mathcal{O}(d^3),
\end{aligned}
\end{equation}
where
\begin{subequations}\label{Eq.Exp012}
\begin{align}
    \mathcal{E}^{(0)}_{lm}(\tau,\omega)=&\exp\left[i\left(\omega u^t-mu^\varphi\right)\tau\right],\\
    \mathcal{E}^{(1)}_{lm}(\tau,\omega)=&\exp\left[i\left(\omega u^t-mu^\varphi\right)\tau\right]\times i(\omega c_{01}-mc_{31}),\\
    \mathcal{E}^{(2)}_{lm}(\tau,\omega)=&\exp\left[i\left(\omega u^t-mu^\varphi\right)\tau\right]\notag\\
    &\times\left[i(\omega c_{02}-mc_{32})-\frac{1}{2}(\omega c_{01}-mc_{31})^2\right].
\end{align}
\end{subequations}
We also expand $\mathcal{Z}_{lm}$ as
\begin{equation}\label{Eq.Z}
\begin{aligned}
    \mathcal{Z}_{lm}[r,\theta]=&\mathcal{Z}_{lm}^{(0)}(r_0,\theta_0)+\mathcal{Z}_{lm}^{(1)}(r_0,\theta_0)d\\
    &+\mathcal{Z}_{lm}^{(2)}(r_0,\theta_0)d^2+\mathcal{O}(d^3),
\end{aligned}
\end{equation}
where
\begin{subequations}\label{Eq.Z012}
    \begin{align}
        \mathcal{Z}^{(0)}_{lm}(r_0,\theta_0)=&\mathcal{Z}_{lm}(r,\theta,d)\Big|_{d=0},\\
        \mathcal{Z}^{(1)}_{lm}(r_0,\theta_0)=&\frac{\mathrm{d} \mathcal{Z}_{lm}(r,\theta,d)}{\mathrm{d} d}\Big|_{d=0}\notag\\    =&\left(\partial_r\mathcal{Z}_{lm}\right)_{d=0}c_{11}+\left(\partial_\theta\mathcal{Z}_{lm}\right)_{d=0}c_{21}\notag\\
        &+\left(\partial_d\mathcal{Z}_{lm}\right)_{d=0},\\
    \mathcal{Z}^{(2)}_{lm}(r_0,\theta_0)=&\frac{1}{2}\frac{\mathrm{d}^2\mathcal{Z}_{lm}(r,\theta,d)}{\mathrm{d} d^2}\Big|_{d=0}\notag\\
    =&\frac{1}{2}\left(\partial_r^2\mathcal{Z}_{lm}\right)_{d=0}c_{11}^2+\frac{1}{2}\left(\partial_\theta\partial_r\mathcal{Z}_{lm}\right)_{d=0}c_{11}c_{21}\notag\\
    &+\frac{1}{2}\left(\partial_d\partial_r\mathcal{Z}_{lm}\right)_{d=0}c_{11}+\left(\partial_r\mathcal{Z}_{lm}\right)_{d=0}c_{12}\notag\\
    &+\frac{1}{2}\left(\partial_{\theta}^2\mathcal{Z}_{lm}\right)_{d=0}c_{21}^2+\frac{1}{2}\left(\partial_r\partial_\theta\mathcal{Z}_{lm}\right)_{d=0}c_{21}c_{11}\notag\\
    &+\frac{1}{2}\left(\partial_d\partial_\theta\mathcal{Z}_{lm}\right)_{d=0}c_{21}+\left(\partial_\theta\mathcal{Z}_{lm}\right)_{d=0}c_{22}\notag\\
    &+\frac{1}{2}\left(\partial_r\partial_d\mathcal{Z}_{lm}\right)_{d=0}c_{11}+\frac{1}{2}\left(\partial_\theta\partial_{d}\mathcal{Z}_{lm}\right)_{d=0}c_{21}\notag\\
    &+\frac{1}{2}\left(\partial_d^2\mathcal{Z}_{lm}\right)_{d=0}.
    \end{align}
\end{subequations}
By inserting Eqs.~\eqref{Eq.Exp} and \eqref{Eq.Z} into Eq.~\eqref{Eq.ZinReWrite} and expanding the result into a series of $d$, we obtain
\begin{equation}\label{Eq.ZExpansion}
\begin{aligned}
    Z^{\mathrm{in}}_{lm}(\omega)=&\frac{\mu}{\mathcal{W}_{lm}(\omega)}\int_{-\infty}^{\infty} \Big[Z_{lm}^{(0)}(\tau,\omega)+Z_{lm}^{(1)}(\tau,\omega)d\\
    &+Z_{lm}^{(2)}(\tau,\omega)d^2+\mathcal{O}(d^3)\Big]\mathrm{d}\tau
\end{aligned}
\end{equation}
where
\begin{subequations}\label{Eq.ExpZ012}
    \begin{align}
        Z^{(0)}_{lm}(\tau,\omega)=&\mathcal{E}^{(0)}_{lm}(\tau,\omega)\mathcal{Z}^{(0)}_{lm}(r_0,\theta_0),\\
        Z^{(1)}_{lm}(\tau,\omega)=&\mathcal{E}^{(0)}_{lm}(\tau,\omega)\mathcal{Z}^{(1)}_{lm}(r_0,\theta_0)\notag\\
        &+\mathcal{E}^{(1)}_{lm}(\tau,\omega)\mathcal{Z}^{(0)}_{lm}(r_0,\theta_0),\\
        Z_{lm}^{(2)}(\tau,\omega)=&\mathcal{E}^{(0)}_{lm}(\tau,\omega)\mathcal{Z}^{(2)}_{lm}(r_0,\theta_0)\notag\\
        &+\mathcal{E}^{(1)}_{lm}(\tau,\omega)\mathcal{Z}^{(1)}_{lm}(r_0,\theta_0)\notag\\
        &+\mathcal{E}^{(2)}_{lm}(\tau,\omega)\mathcal{Z}^{(0)}_{lm}(r_0,\theta_0).
    \end{align}
\end{subequations}
Here, we evaluate $r$ and $\theta$ at the CM ($d=0$) of the IB which, as mentioned 
before, follows approximately a
geodesic circular motion. Hence, we have $r_0 = p$ and $\theta_0 = \pi/2$. 

For the zeroth order expansion, the integration is straightforward
\begin{equation}\label{Eq.ZZ0}
    \begin{aligned}
        Z^{\mathrm{in}(0)}_{lm}(\omega)=&\frac{\mu}{\mathcal{W}_{lm}(\omega)}\int_{-\infty}^{\infty}Z_{lm}^{(0)}(\tau,\omega)\ \mathrm{d}\tau\\
        =&\frac{\mu}{\mathcal{W}_{lm}(\omega)}\int_{-\infty}^{\infty}e^{i(\omega u^t-mu^\varphi)\tau}\mathcal{Z}^{(0)}_{lm}(r_0,\theta_0,\omega)\ \mathrm{d}\tau\\
    =&\frac{2\pi\mu}{u^t\mathcal{W}_{lm}(\omega)}\delta\left(\omega-m\frac{u^\varphi}{u^t}\right)\mathcal{Z}_{lm}^{(0)}(r_0,\theta_0,\omega).
    \end{aligned}
\end{equation}
It reveals that the frequencies corresponding to the zeroth order 
gravitational radiation are
\begin{equation}\label{Eq.ZeroFrequency}
    \omega_m^{(0)}=m\frac{u^\varphi}{u^t},\qquad m=-l,\dots,l.
\end{equation}

Now we turn to the first and second order terms defined in
Eqs.~\eqref{Eq.ZExpansion} and \eqref{Eq.ExpZ012}. 
In fact, the presence of the expansion
coefficients $c_{01}$, $c_{02}$, $c_{11}$, $c_{12}$, $c_{21}$, $c_{22}$,
$c_{31}$, $c_{32}$ introduces extra trigonometric terms with respect to $\tau$.
We use
\begin{equation}\label{II33}
    \sin A=\frac{e^{iA}-e^{-iA}}{2i},\qquad\cos A=\frac{e^{iA}+e^{-iA}}{2},
\end{equation}
to turn these trigonometric terms into exponential forms. Then by transforming
the integration into delta functions as is exemplified in Eq.~\eqref{Eq.ZZ0},
we obtain the frequencies and amplitudes of the first and second order
waveforms. 

For the first order waveform, we find that the frequencies satisfy the condition
\begin{equation}\label{Eq.FirstFrequency}
    \omega_{mwv}^{(1)}=m\frac{u^\varphi}{u^t}+w\frac{\Omega}{u^t}+v\frac{\omega_{\mathrm{IB}}}{u^t},
\end{equation}
where $m = -l, \dots, l$, $w = 0, \pm 1$, and $v = \pm 1$. Therefore, we have six split 
first order modes for each $m$. The corresponding amplitudes are
\begin{equation}\label{Eq.Z1}
    Z^{\mathrm{in}(1)}_{lm}=\frac{2\pi\mu}{u^t}\sum_{wv}\frac{Z^{(1)}_{lmwv}}{\mathcal{W}_{lmwv}}.
\end{equation}
where $Z^{(1)}_{lmwv} \equiv Z^{(1)}_{lm}(r_0, \theta_0, \omega_{mwv}^{(1)})$ is the split first order amplitude and $\mathcal{W}_{lmwv} \equiv \mathcal{W}_{lm}(\omega_{mwv})$ is the Wronskian.

For the second-order waveform, there are fifteen modes for each $m$, which are given by
\begin{equation}\label{Eq.SecondFrequency}
    \omega_{mwv}^{(2)}=m\frac{u^\varphi}{u^t}+w\frac{\Omega}{u^t}+v\frac{\omega_{\mathrm{IB}}}{u^t},
\end{equation}
where $m = -l, \dots, l$, $w = 0, \pm 1, \pm 2$, and $v = 0, \pm 2$. The amplitudes are
\begin{equation}\label{Eq.Z2}
    Z^{\mathrm{in}(2)}_{lm}=\frac{2\pi\mu}{u^t}\sum_{wv}\frac{Z^{(2)}_{lmwv}}{\mathcal{W}_{lmwv}}.
\end{equation}
where $Z^{(2)}_{lmwv} \equiv Z^{(2)}_{lm}(r_0, \theta_0, \omega_{mwv}^{(2)})$
is the split second order amplitude. The analytical expressions for the six
$Z^{(1)}_{lmwv}$ and fifteen $Z^{(2)}_{lmwv}$ are derived in
Appendix~\ref{appC}.

\subsection{Waveform snapshot}

Given the above frequency spectra and split amplitudes,
Eq.~\eqref{Eq.FourierWaveform} can be expressed as a summation of discrete modes,
and the resulting waveform for one small black hole becomes 
\begin{equation}\label{Eq.HalfSnapshot}
    \begin{aligned}
        h_+-ih_\times=&-\frac{\mu}{r}\sum_{lmwv}\frac{S_{lmwv}(\theta)}{\omega_{mwv}^2}\\
        &\times \left(Z_{lm}^{(0)}+Z_{lmwv}^{(1)}d+Z_{lmwv}^{(2)}d^2\right)\\
        &\times e^{-i\omega_{mwv}(t-r)+im\varphi},
    \end{aligned}
\end{equation}
where $S_{lmwv}(\theta) \equiv {_{-2}}S_{lm}^{a\omega_{mwv}}(\theta)$. For the
other small black hole, the waveform can be obtained by changing the sign of
$d$, since we assume equal mass  for the two small BHs and use
Eq.~\eqref{Eq.FFFCircular} to describe their inner orbital motion.

Summing up the two waveforms of the two small BHs, we
derive the total waveform as
\begin{equation}\label{Eq.FullSnapshot}
\begin{aligned}
    h_+-ih_\times=&-\frac{2\mu}{r}\sum_{lmwv}\frac{S_{lmwv}(\theta)}{\omega_{mwv}^2}\\
        &\times\left(Z^{(0)}_{lm}+Z^{(2)}_{lmwv}d^2+\mathcal{O}(d^4)\right)\\
        &\times e^{-i\omega_{mwv}(t-r)+im\varphi},
\end{aligned}
\end{equation}
where all odd-order terms cancel out due to our assumption of a circular,
equal-mass IB. However, what we have derived so far is an ideal waveform
snapshot, which means that the orbital parameters of the b-EMRI are not evolving.
For instance, the outer radius $p$ remains constant. If we consider GW
radiation, as we will discuss in the next subsection, the orbital parameters will
evolve, resulting in a deviation of the waveform phase from the ideal one on a
timescale of $M/\sqrt{\epsilon}$ \cite{PhysRevD.78.064028}. 

\subsection{\label{IIG} Adiabatic evolution of the system}

Since GWs carry away energy and angular momentum, the b-EMRI
parameters evolve. At leading order in the multi-scale expansion, the evolution can be treated as `adiabatic', in which the local rate of change in the leading order orbital binding energy/angular momentum balances with the leading total asymptotic gravitational wave energy/angular momentum fluxes. While we have highlighted that the b-EMRI correction scales as a 1PA term, its contribution to the inspiral evolution may be incorporated as a sub-leading correction to the adiabatic evolution formula. This is analogous to the inclusion of the secondary's spin in standard EMRI waveforms.

To calculate the energy and angular momentum carried away by GW, one can
start from the effective stress-energy tensor of GW. In the TT gauge it
can be written as
\begin{equation}\label{Eq.EnergyMomentumTensorDefinition}
T^{GW}_{\alpha\beta}=\frac{1}{32\pi}\left\langle\frac{\partial
h_{TT}^{ij}}{\partial x^\alpha}\frac{\partial h_{ij}^{TT}}{\partial
x^\beta}\right\rangle.  \end{equation}
Integrating the $01$ component over the solid angle yields the average energy
fluxes at infinity and at the event horizon $r_+$. They can be expressed as
\begin{subequations}\label{Eq.EnergyFluxes} \begin{align}
&\left\langle\frac{\mathrm{d} \mathcal{E}}{\mathrm{d}
t}\right\rangle^{\infty}=\sum_{lmwv}\frac{1}{4\pi\omega_{mwv}^2}\left|Z^{\mathrm{in}}_{lmwv}\right|^2,\\
&\left\langle\frac{\mathrm{d} \mathcal{E}}{\mathrm{d}
t}\right\rangle^{H}=\sum_{lmwv}\frac{1}{4\pi\omega_{mwv}^2}\alpha_{lmwv}\left|Z^{\mathrm{up}}_{lmwv}\right|^2.
\end{align} \end{subequations}
The angular momentum fluxes are given by
\begin{subequations}\label{Eq.AngularMomentumFluxes} \begin{align}
&\left\langle\frac{\mathrm{d} \mathcal{L}_z}{\mathrm{d}
t}\right\rangle^\infty=\sum_{lmwv}\frac{m}{4\pi\omega_{mwv}^3}\left|Z_{lmwv}^{\mathrm{in}}\right|^2,\\
&\left\langle\frac{\mathrm{d} \mathcal{L}_z}{\mathrm{d}
t}\right\rangle^{H}=\sum_{lmwv}\frac{m}{4\pi\omega_{mwv}^3}\alpha_{lmwv}\left|Z^{\mathrm{up}}_{lmwv}\right|^2,
\end{align} \end{subequations}
where the coefficient $\alpha_{lmwv}$ is
\begin{equation}\label{Eq.Alpha}
\alpha_{lmwv}=\frac{256(2Mr_+)^5k(k^2+4\kappa^2)(k^2+16\kappa^2)\omega_{mwv}^3}{|C_{lmwv}|^2},
\end{equation}
with $k=\omega_{mwv}-ma/(2Mr_+)$, $\kappa=\sqrt{M^2-a^2}/4Mr_+$, and
\begin{equation}\label{Eq.C} \begin{aligned}
|C_{lmwv}|^2&=[(\lambda_{lmwv}+2)^2+4a\omega_{mwv}-4a^2\omega_{mwv}^2]\\
&\times(\lambda_{lmwv}^2+36ma\omega_{mwv}-36a^2\omega_{mwv}^2)\\
&+(2\lambda_{lmwv}+3)(96a^2\omega_{mwv}^2-48ma\omega_{mwv})\\
&+144\omega_{mwv}^2(M^2-a^2), \end{aligned} \end{equation}
where $\lambda_{lmwv}=\mathcal{A}_{lmwv}-2am\omega_{mwv}+a^2\omega_{mwv}^2-2$
with $\mathcal{A}_{lmwv}$ denoting the eigenvalue of SWSH
\cite{2014PhRvD..90l4039O}.

The conservation of energy and angular momentum requires that
\begin{subequations}\label{Eq.TotalFluxes}
    \begin{align}
        &\dot{\mathcal{E}}\equiv\left\langle\frac{\mathrm{d}\mathcal{E}}{\mathrm{d}t}\right\rangle=-\left\langle\frac{\mathrm{d}\mathcal{E}}{\mathrm{d}t}\right\rangle^H-\left\langle\frac{\mathrm{d}\mathcal{E}}{\mathrm{d}t}\right\rangle^\infty,\\
        &\dot{\mathcal{L}}_z\equiv\left\langle\frac{\mathrm{d}\mathcal{L}_z}{\mathrm{d}t}\right\rangle=-\left\langle\frac{\mathrm{d}\mathcal{L}_z}{\mathrm{d}t}\right\rangle^H-\left\langle\frac{\mathrm{d}\mathcal{L}_z}{\mathrm{d}t}\right\rangle^\infty.
    \end{align}
\end{subequations}
If we treat the inspiral process as an ``osculating" slide through a series of
geodesics, the above $\dot{\mathcal{E}}$ and $\dot{\mathcal{L}}_z$ 
determine how the system evolves from one geodesic to the next. In the
circular-equatorial case, the geodesic parameters are $M$, $a$, $p$. The evolution of mass and spin of the SMBH formally appears at 1PA order in the multi-scale expansion though it is known to be numerically subdominant (see e.g. Ref.~\cite{Wardell:2021fyy}). The only parameter we need to evolve is
the outer radius $p$. Now that the energy $\mathcal{E}(t)$ and
angular momentum $\mathcal{L}_z(t)$ at each moment are known, we can derive
\begin{equation}\label{Eq.MapToPdot}
    \dot{p}\equiv\frac{\mathrm{d} p}{\mathrm{d} t}=q_1\dot{\mathcal{E}}+q_2\dot{\mathcal{L}}_z,
\end{equation}
where $q_1$ and $q_2$ are functions of $\mathcal{E}$, $\mathcal{L}_z$ and $p$, and
these two functions are derived as follows. Since circular orbits remain
circular during radiation reaction \cite{PhysRevD.47.5376,Kennefick:1995za,Mino:1997bx,Ryan:1995zm},
we have $\dot{R}=0$ and $\dot{R}'=0$ according to the radial geodesic equation
defined in Eq.~\eqref{Eq.r-Kerr}.  The former equation relates the energy and
angular-momentum fluxes to the properties of the circular-equatorial orbits,
which is 
\begin{equation}\label{Eq.DotEToDotLz}
    \frac{\dot{\mathcal{E}}}{\dot{\mathcal{L}}_z}=\frac{M^{1/2}}{aM^{1/2}+p^{3/2}}=\frac{u^\varphi}{u^t}.
\end{equation}
The latter gives the expressions for $q_1$ and $q_2$ as 
\begin{subequations}\label{Eq.Q1Q2}
    \begin{align}
        &q_1=-\frac{2\left[2\mathcal{E}p^3-aM\mathcal{L}_z+a^2\mathcal{E}(M+p)\right]}{a^2\left(\mathcal{E}^2-1\right)-\mathcal{L}_z^2+6p\left[M+\left(\mathcal{E}^2-1\right)p\right]},\\
        &q_2=-\frac{2\left[\mathcal{L}_z(M-p)-aM\mathcal{E}\right]}{a^2\left(\mathcal{E}^2-1\right)-\mathcal{L}_z^2+6p\left[M+\left(\mathcal{E}^2-1\right)p\right]}.
    \end{align}
\end{subequations}

Now we have the evolution of $p=p(t)$, we can compute the adiabatic waveform as
\begin{equation}\label{Eq.AdiabaticWaveform}
    h_+-ih_\times=\frac{\mu}{r}\sum_{lm}H_{lm}(t,\theta,\varphi)e^{-i\Phi_m(t)},
\end{equation}
where
\begin{subequations}\label{Eq.AdiabaticComponents}
    \begin{align}
        &H_{lm}(t,\theta,\varphi)=-\frac{2S_{lm}(t,\theta)}{\left(\omega_m(t)\right)^2}Z^{(0)}_{lm}(t)e^{im\varphi},\\
        &\Phi_m(t)=\int_0^t\omega_m(t')\mathrm{d}t'.
    \end{align}
\end{subequations}
Here all terms containing $p$ become a function of $t$. The phase term is accumulative so it is an integral. Note that above we only include zeroth order modes in adiabatic waveforms to illustrate the main features. 

\section{\label{III}Numerical Implementation}

\subsection{Frequencies and amplitudes of the modes}

Our waveform model reveals three key frequencies:
\begin{equation}\label{Eq.MainFrequency}
    \Omega_{G}\equiv\frac{u^\varphi}{u^t},\quad\Omega_{P}\equiv\frac{\Omega}{u^t},\quad\Omega_{I}\equiv\frac{\omega_{\mathrm{IB}}}{u^t}.
\end{equation}
In particular,
the geodesic frequency $\Omega_G$ and precession frequency $\Omega_P$ have
similar magnitudes, whereas the rescaled inner frequency $\Omega_I$ is
significantly higher than the previous two. For instance, when
the parameters are $a=0.9M$, $p=10M$, $\epsilon=10^{-5}$, and $d=4000\mu$, we
have $\Omega_G=0.0307/M$, $\Omega_P=-0.0268/M$, and $\Omega_I=0.3344/M$. 

The
frequency spectrum  given by Eqs.~\eqref{Eq.ZeroFrequency},
\eqref{Eq.FirstFrequency} and \eqref{Eq.SecondFrequency} takes the general form
\begin{equation}\label{Eq.FrequencySpectra}
    \omega_{mwv}=m\Omega_{G}+w\Omega_p+v\Omega_I.
\end{equation}
Notice that for the second order modes with $v=\pm2$, the frequency can reach
values as high as $\omega\sim\mathcal{O}(1/M)$. In this high-frequency regime,
it becomes challenging to compute the homogeneous Teukolsky radial solutions $R^{\mathrm{in}/\mathrm{up}}_{lm}$ using the conventional MST
method~\cite{10.1143/PTP.95.1079,10.1143/PTP.112.415,10.1143/PTP.113.1165,2003LRR.....6....6S}.

To address the problems posed by the high frequencies, we use the
Sasaki-Nakamura (SN) method instead~\cite{10.1143/PTP.67.1788}, which offers
smoother and faster performance in this high-frequency regime. A detailed
presentation of the SN method, along with the \texttt{Julia} package
\texttt{GeneralizedSasakiNakamura.jl} used for computing $R^{\mathrm{in}/\mathrm{up}}_{lm}$ can
be found in Ref.~\cite{PhysRevD.110.124070}. Additionally, we employ the
\texttt{SpinWeightedSpheroidalHarmonics.jl} package to compute the SWSH and
their eigenvalues. With these ingredients in place, the amplitudes
$Z^{(0)}_{lm}$, $Z^{(1)}_{lmwv}$, and $Z^{(2)}_{lmwv}$ can be calculated
according to the algorithm outlined in Appendix \ref{appC}.

\subsection{Waveform and truncation}

When calculating the waveform according to Eq.~\eqref{Eq.FullSnapshot}, in
principle an infinite number of modes are needed. In practice, the $l=2$ modes
already capture the main features of the waveform. Including higher $l$ modes
does improve precision, but the fractional error falls below $10^{-6}$ when we
truncate the modes at $l_{\mathrm{max}}=10$, and below $10^{-11}$ when $l_{\mathrm{max}}=15$.  To
save computational time, in this work we truncate the modes at $l_{\mathrm{max}}=10$
when computing the flux (also see \cite{2021PhRvD.103j4014H}) and at
$l_{\mathrm{max}}=5$ when plotting the waveforms.

To evolve the outer orbit, we select a set of grids starting from $p_{\mathrm{init}}=10.0M$ 
to a value close to the innermost stable circular orbit (ISCO) \cite{1972ApJ...178..347B}. 
When $a=0.9M$, ISCO is located at about $p=2.32M$. We stop at $p_{\mathrm{final}}=2.35M$, 
as the system transits rapidly from inspiral to plunge near the ISCO. 
The grid spacing is chosen according to
\begin{equation}\label{Eq.interval}
    \Delta p(/M)=0.05\tanh(p-p_{\mathrm{final}})+0.001.
\end{equation}
This results in 214 grid points between $p_{\mathrm{init}}=10.0M$ and $p_{\mathrm{final}}=2.35M$. 

At each grid point, we compute the energy and angular momentum fluxes
$\dot{\mathcal{E}}(p)$ and $\dot{\mathcal{L}}_z(p)$. Using
Eq.~\eqref{Eq.MapToPdot} we obtain $\dot{p}$ at these grids. Cubic spline
interpolation \cite{1986nras.book.....P} is used off the grid points to compute
$\dot{p}(p)$ in the inspiraling regime
$p\in[p_{\mathrm{final}},p_{\mathrm{init}}]$. Then we numerically integrate
Eq.~\eqref{Eq.MapToPdot} to recover the evolution of $p$ with respect to $t$.
By inserting $p(t)$ into Eqs.~\eqref{Eq.AdiabaticWaveform} and
\eqref{Eq.AdiabaticComponents} we obtain the adiabatic waveform. Note that we
exclude modes with $\omega_{mwv}<\Omega_G+\Omega_P$ to avoid the so-called
transient resonance, as discussed in
Ref.~\cite{2014PhRvD..89h4036R,Lynch_2024}.

\section{\label{IV}Results}

\subsection{Amplitude, Flux, and resonant excitation of QNMs\label{IVA}}

From the analysis in the previous section, the first order amplitudes cancel, leaving us with
\begin{equation}\label{Eq.ZeroPlusSecondOrderAmplitudes}
    \begin{aligned}
        Z_{lmwv}^{\mathrm{in}}=&Z_{lm}^{(0)}+Z_{lmwv}^{(2)}d^2\\
        =&\begin{cases}
        Z_{lm}^{(0)}+Z_{lm00}^{(2)}d^2\qquad &w=v=0,\\
		Z_{lmwv}^{(2)}d^2\qquad &{\rm others}.
    \end{cases}
    \end{aligned}
\end{equation}

For $w=v=0$, the fluxes are proportional to
\begin{equation}\label{eq41}
    \begin{aligned}
        \left|Z_{lm00}^{\mathrm{in}}\right|^2=&\left|Z_{lm}^{(0)}\right|^2+\left(Z_{lm}^{(0)}\bar{Z}_{lm00}^{(2)}+\bar{Z}_{lm}^{(0)}Z_{lm00}^{(2)}\right)d^2\\
        &+\left|Z_{lm00}^{(2)}\right|^2d^4,
    \end{aligned}
\end{equation}
while for $w\neq 0$ or $v\neq 0$
\begin{equation}
    \left|Z_{lmwv}^{\mathrm{in}}\right|^2=\left|Z_{lmwv}^{(2)}\right|^2d^4.
\end{equation}

We first consider the modes with $v=0$.
In this case, the leading-order correction to the amplitude is of the order of
$\sim\mathcal{O}(d^2)$. We define a factor
\begin{equation}\label{Eq.MagnificationFactor}
    A_{lmwv}(\omega)=\frac{|Z^{(2)}_{lmwv}|d^2}{|Z^{(0)}_{lm}|}
\end{equation}
to quantify the relative importance of this correction.
Table~\ref{tab:AmplitudesOfModes} shows the amplitudes of the main modes
and the corresponding leading-order corrections when $v=0$. 
We can draw the conclusion that
$A_{lm00}\sim\epsilon$.  
To understand this relation, we note that although the corrections are
proportional to $d^2$, the inner diameter $d$ is of the order of thousands of
$\mu$. Therefore,  we have $d^2=(\mbox{thousands of }M\epsilon)^2\approx
M^2\epsilon$.

\begin{table}[htb]
    \centering
    \begin{ruledtabular}
    \begin{tabular}{ccccc}
    l&m&$\left|Z^{(0)}_{lm}\right|$&$\left|Z^{(2)}_{lm00}\right|\cdot d^2$&$A_{lm00}$\\
    \hline
        2 & 2 & $1.03\times10^{-3}$& $3.74\times 10^{-8}$&$3.63\times 10^{-5}$\\
        3 & 3 & $5.19\times 10^{-4}$&$2.32\times 10^{-8}$&$4.47\times 10^{-5}$\\
        4 & 4 & $2.59\times 10^{-4}$ & $1.40\times10^{-8}$ & $5.40\times 10^{-5}$\\
        5 & 5 & $1.26\times10^{-4}$ & $8.05\times 10^{-8}$ & $6.40\times 10^{-5}$\\
        6 & 6 & $5.95\times 10^{-5}$ & $4.44\times 10^{-9}$ & $7.45\times 10^{-5}$\\
        7 & 7 & $2.76\times 10^{-5}$ & $2.36 \times 10^{-9}$ & $8.53\times 10^{-5}$\\
        8 & 8 & $1.26\times 10^{-5}$ & $1.22\times 10^{-9}$ & $9.64 \times 10^{-5}$\\
    \end{tabular}
    \end{ruledtabular}
    \caption{The zeroth and second order amplitudes and their fractions of main modes, with parameters $a=0.9M$, $p=10M$, $\epsilon=10^{-5}$, $d=4000\mu$, $\tilde{\varphi}=\pi/4$ and $\tilde{\theta}=\pi/3$.}
    \label{tab:AmplitudesOfModes}
\end{table}

\begin{figure*}[htb]
    \centering
    \includegraphics[width=1.0\linewidth]{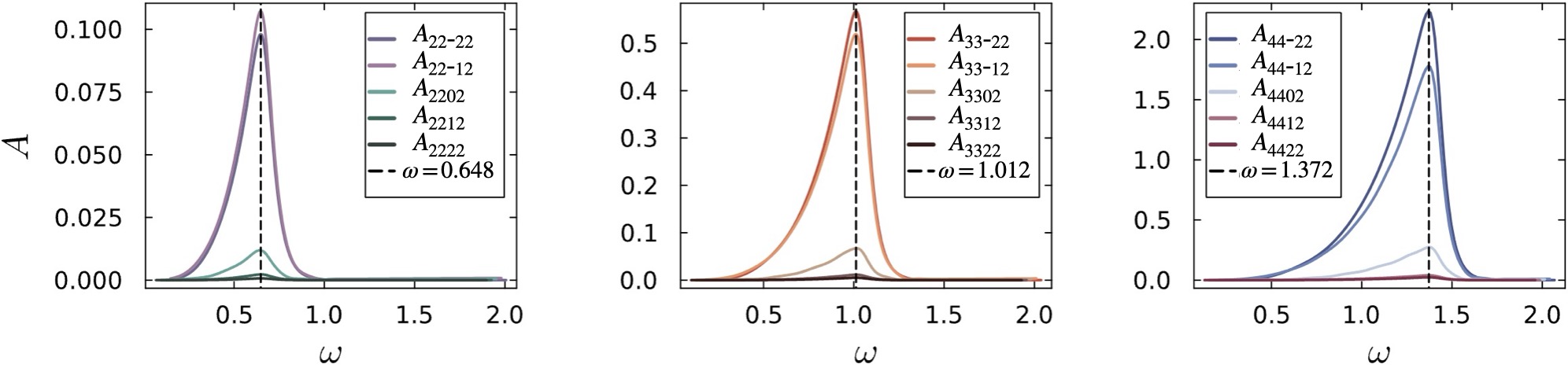}
    \caption{The factors $A_{22w2}$, $A_{33w2}$, and $A_{44w2}$ with $w=0,\pm1,\pm2$ as functions of the frequency $\omega$. Peaks appear at $\omega=0.648/M$ for $l=m=2$ modes, at $\omega=1.012/M$ for $l=m=3$ modes, and at $\omega=1.372/M$ for $l=m=4$ modes.}
    \label{fig:QNM_Excitation}
\end{figure*}

When $v=\pm2$ (high-frequency modes), the factors are much larger,
especially when the frequency approaches the QNM frequencies of the SMBH.
Fig.~\ref{fig:QNM_Excitation} illustrates the dependence of $A_{lmwv}$ on the
frequency $\omega$. The parameters are the same as in
Table.~\ref{tab:AmplitudesOfModes}. Peaks are observed near the 
corresponding QNM modes. The corresponding frequencies of the peaks
are given in Table~\ref{tab:QNMVsPeak}. We also give the
frequencies of the QNMs for comparison, which are calculated
using the \texttt{Python} package \texttt{qnm.py}
\cite{2019JOSS....4.1683S}. Notably, the peaks occur slightly below
the real part of the leading QNM frequencies. These features agree well with
Fig.~2 of Ref.~\cite{PhysRevD.103.L081501}, where this behavior is interpreted
as a resonant excitation of QNMs.

\begin{table}[htb]
    \centering
    \begin{ruledtabular}
    \begin{tabular}{cccc}
        l  & m & $\ \ \omega_{lm0}^{\mathrm{QNM}}(M)$& $\omega_{lm}^{\mathrm{peak}}(M)$\\
       \hline
       2  & 2 & 0.672-0.0649$i$& 0.648\\
       3 & 3 &1.045-0.0655$i$&1.012\\
       4 & 4 &1.410-0.0662$i$&1.372\\
    \end{tabular}
    \end{ruledtabular}
    \caption{QNM frequencies of the leading tones and peak frequencies of the $l=m=2,3,4$ modes.}
    \label{tab:QNMVsPeak}
\end{table}

It is remarkable that the ratios $A_{lmwv}$ for the high-frequency modes with
$v=\pm2$ can approach unity in some cases. These second-order relative amplitudes are about
$\epsilon^{-1}$ times larger than those of the low-frequency modes.  Moreover,
for each $m$, there are ten high-frequency modes corresponding to
$w=0,\pm1,\pm2,\ v=\pm2$. Therefore, the fluxes are amplified by another order
of magnitude.  Such a large flux due to the high-frequency modes should cause
the b-EMRI system to evolve faster than a standard EMRI. We notice that an
earlier study using the MPD equation to solve the evolution of b-EMRI also
revealed the appearance of two high-frequency modes (see Eq.~(30) in
Ref.~\cite{2024EPJC...84..478J}). However, the authors disregarded these two
terms in the calculation of the flux, which potentially could result in a
slower evolution of their b-EMRI.

To better see the importance of the high-frequency modes, we 
calculate the energy fluxes due to different modes,
\begin{subequations}\label{Eq.Configurations}
    \begin{align}
        &\dot{\mathcal{E}}^{\infty(0)}=\sum_{l=2}^{10}\sum_{m=-l}^l\frac{1}{4\pi\omega_m^2}\left|Z^{(0)}_{lm}\right|^2,\label{Eq.ConfigurationsA}\\
        &\dot{\mathcal{E}}^{\infty(1)}=\sum_{l=2}^{10}\sum_{m=-l}^l\frac{1}{4\pi\omega_m^2}\left|Z^{(0)}_{lm}+Z^{(2)}_{lm00}d^2\right|^2,\label{Eq.ConfigurationsB}\\
        &\dot{\mathcal{E}}^{\infty(2)}=\sum_{l=2}^{10}\sum_{mwv}\frac{1}{4\pi\omega_{mwv}^2}\left|Z^{(0)}_{lm}+Z^{(2)}_{lmwv}d^2\right|^2,\label{Eq.ConfigurationsC}
    \end{align}
\end{subequations}
where we have truncated the calculation at $l_{\mathrm{max}}=10$.  Here,
$\dot{\mathcal{E}}^{\infty(0)}$ represents the energy flux of a standard EMRI,
$\dot{\mathcal{E}}^{\infty(1)}$ accounts for low-frequency corrections similar
to those discussed in Ref.~\cite{2024EPJC...84..478J}, and
$\dot{\mathcal{E}}^{\infty(2)}$ includes the contributions from high-frequency
modes. The energy fluxes down to the horizon $\dot{\mathcal{E}}^{H(0)}$, $\dot{\mathcal{E}}^{H(1)}$, $\dot{\mathcal{E}}^{H(2)}$ are also computed, and all the results
are given in Table~\ref{tab:CatalogOfEnergyFluxes}.

From the table we find that the contributions from the b-EMRI
high-frequency modes are substantial, particularly during the early inspiral
stage. This leads to a faster evolution of the b-EMRI system compared to a
standard EMRI. In contrast, the low-frequency modes contribute only at the
$\mathcal{O}(\epsilon)$ order.  The behavior of the angular momentum fluxes
mirrors that of the energy fluxes and hence are not shown here.

\begin{figure}[htpb]
    \centering
    
    \subfloat[Low-frequency inner motion with $d=8000\mu$]{\label{fig:4a}\includegraphics[width=0.95\linewidth]{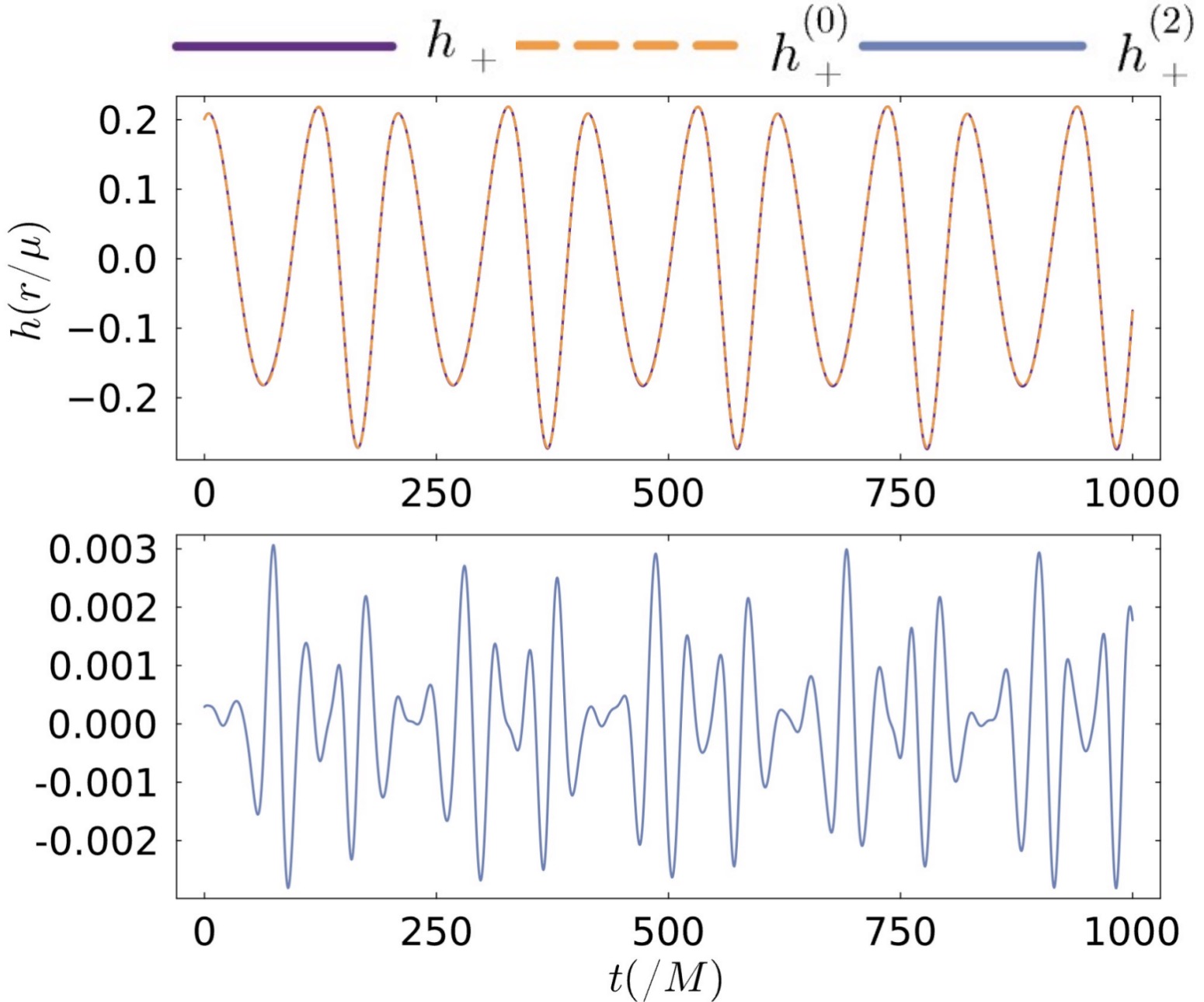}}

    \subfloat[Inband-frequency inner motion with $d=4000\mu$]{\label{fig:4b}\includegraphics[width=0.95\linewidth]{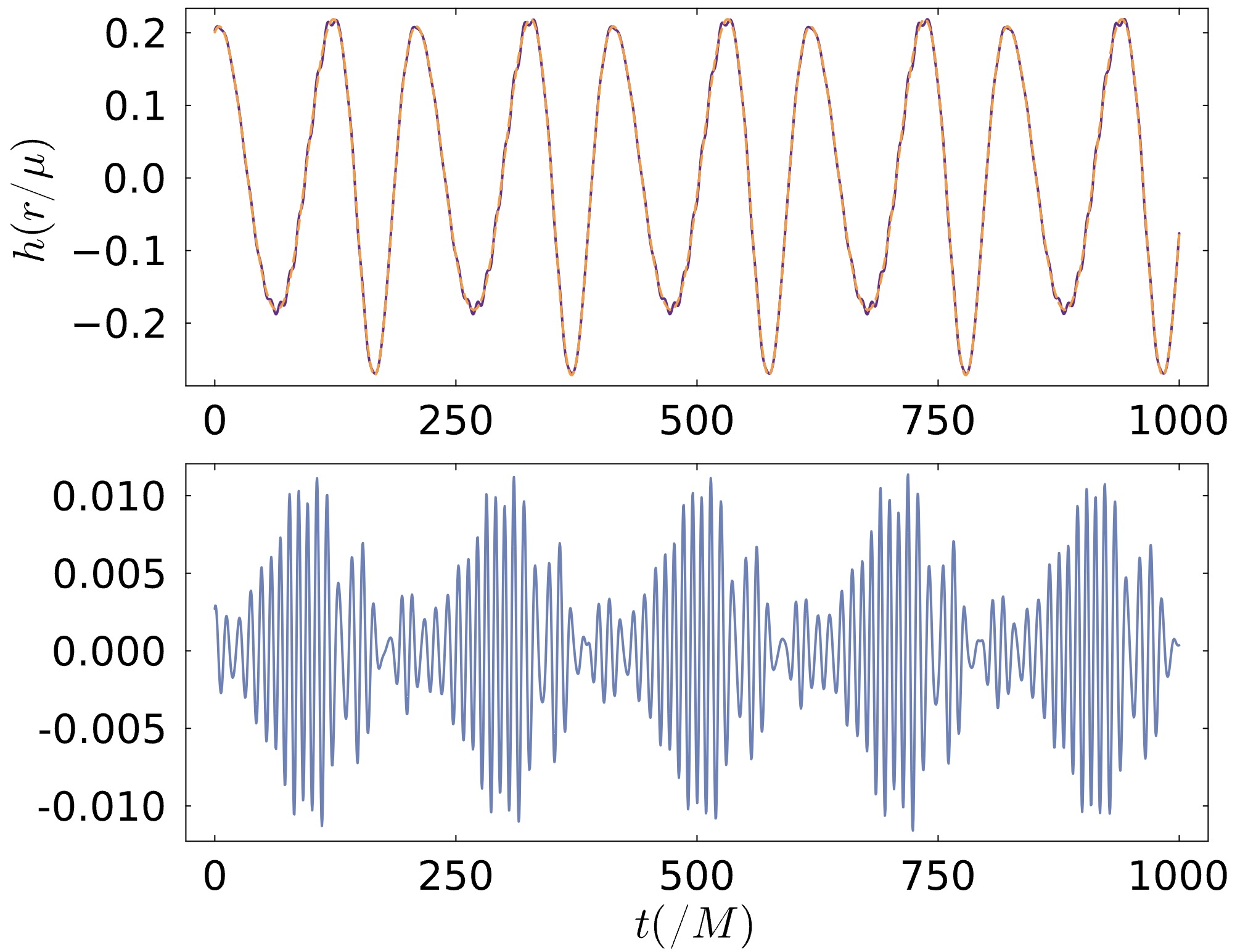}}

    \subfloat[High-frequency inner motion with $d=2000\mu$]{\label{fig:4c}\includegraphics[width=0.95\linewidth]{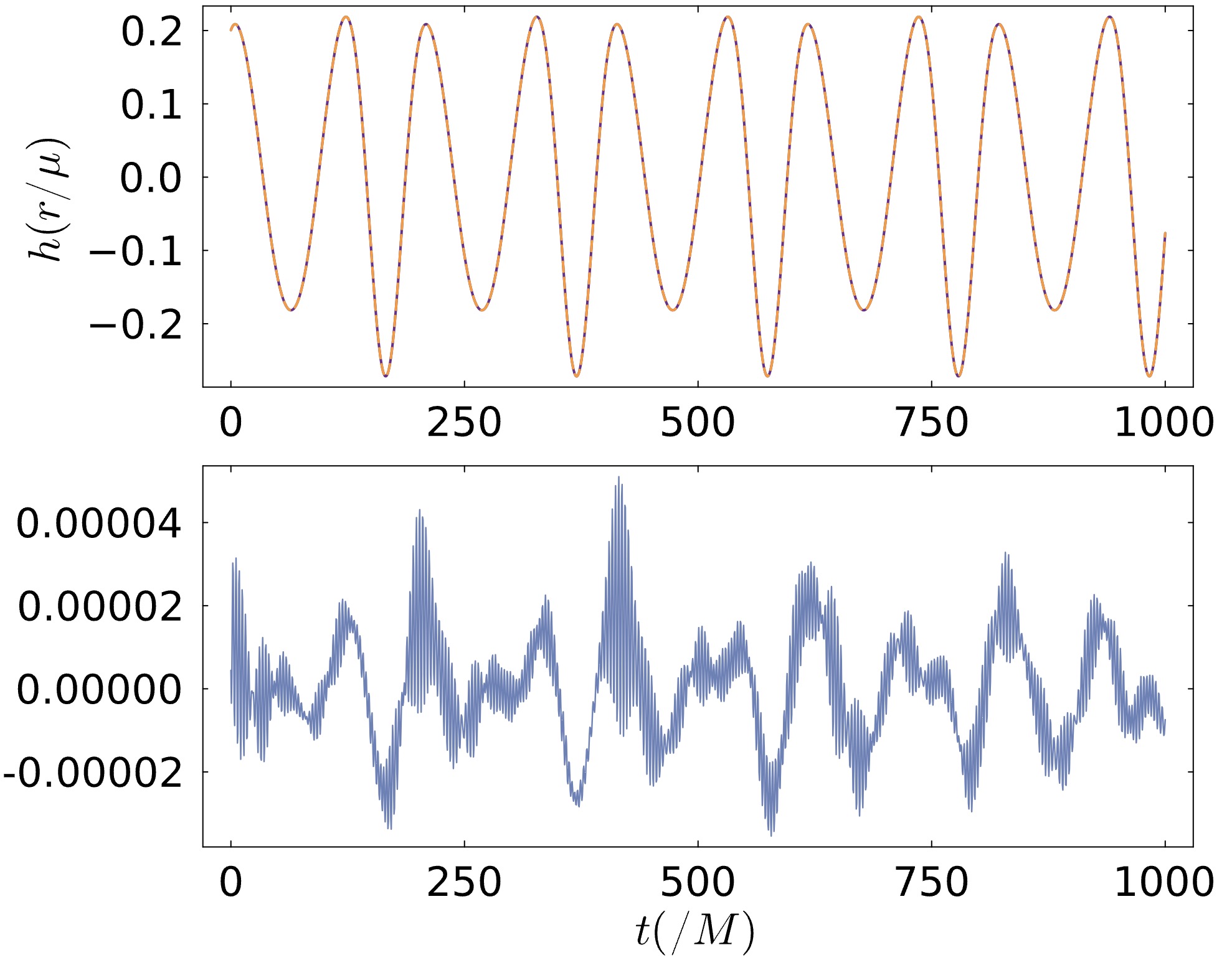}}

    \caption{Waveform snapshots with different setting of $d$. Other parameters are fixed at $a=0.9M$, $p=10M$, $\epsilon=10^{-5}$, $\tilde{\varphi}=\pi/4$, $\tilde{\theta}=\pi/3$ with viewing angles $\theta=\pi/3$ and $\varphi=\pi/6$.}
    \label{fig:WaveformSnapshots}
\end{figure}

Fig.~\ref{fig:WaveformSnapshots} shows more clearly the resonant excitation of
the QNMs by the high-frequency modes of the b-EMRI.  Here we choose
$\epsilon=10^{-5}$ and use different values of $d$ to change the total
frequency $\omega$.  Therefore, $d=8000\mu$ corresponds to a low frequency of
$2\Omega_I=0.236/M$, $d=4000\mu$ corresponds to an in-band frequency of
$2\Omega_I=0.669/M$, and $d=2000\mu$ corresponds to a high frequency of
$2\Omega_I=1.892/M$. The term ``in-band" means that the frequency lies within
the band that resonantly excites the QNMs. From Fig.~\ref{fig:QNM_Excitation}
and Table.~\ref{tab:QNMVsPeak}, it is clear that the frequencies associated
with $d=4000\mu$ center around the excitation band of the $l=m=2$ QNM.

When the frequency is within the
excitation band, we find that the second-order amplitude $h^{(2)}_+$ is significantly
excited, reaching about $10\%$ of that of $h^{(0)}_+$, which results in a visible fluctuation
in the composite waveform $h_+$ in panel (b) of Fig.~\ref{fig:WaveformSnapshots}.
Conversely, when the frequency is outside the excitation band (either lower or
higher), the QNMs are not excited, and the amplitude of $h^{(2)}_+$ is
relatively small. It is also worth noting that $h^{(2)}_+$ in panel (a) of
Fig.~\ref{fig:WaveformSnapshots} is larger than that in panel (c). This is because,
as $d$ increases, the $\mathcal{O}(d^2)$ factor also increases.

\subsection{Adiabatic trajectory and waveform}

In this subsection, we study the adiabatic evolution of the b-EMRI system with
different configurations and compare it with a standard EMRI. We continue to
use the parameters $a=0.9M$, $p=10M$, $\epsilon=10^{-5}$, $d=4000\mu$,
$\tilde{\varphi}=\pi/4$, and $\tilde{\theta}=\pi/3$.  We also adopt the
configurations defined in Eq.~\eqref{Eq.Configurations}. 
Eq.~\eqref{Eq.ConfigurationsA} corresponds to the evolution of a standard EMRI.
Eq.~\eqref{Eq.ConfigurationsB} represents the evolution of a b-EMRI driven only by the
low-frequency modes, and we refer to this case as ``Configuration 1''. Eq.~\eqref{Eq.ConfigurationsC} includes all modes, which we refer to as ``Configuration 2''.

\begin{figure}[htpb]
    \centering
    \includegraphics[width=1.0\linewidth]{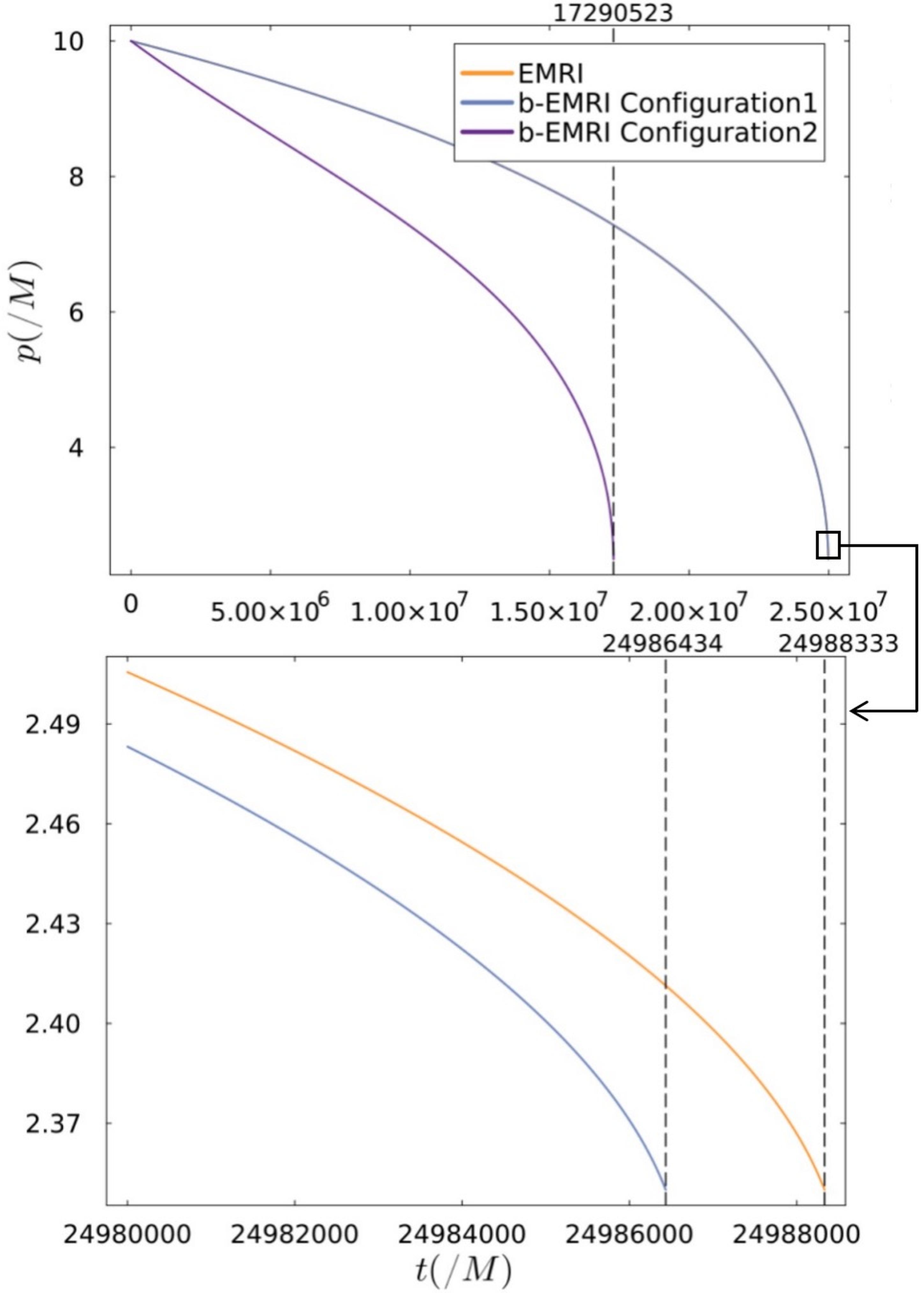}
    \caption{Evolution of the outer radius $p$ with three different configurations. The upper panel
	shows the long-term evolution and the lower one zooms in at the final stage of the system.}
    \label{fig:EvolutionalTrajectoryOfP}
\end{figure}

\begin{figure*}[htpb]
    \centering
    
    \subfloat[Adiabatic waveforms of standard EMRI and b-EMRI in Configuration 1]{\label{fig:6a}\includegraphics[width=1.0\linewidth]{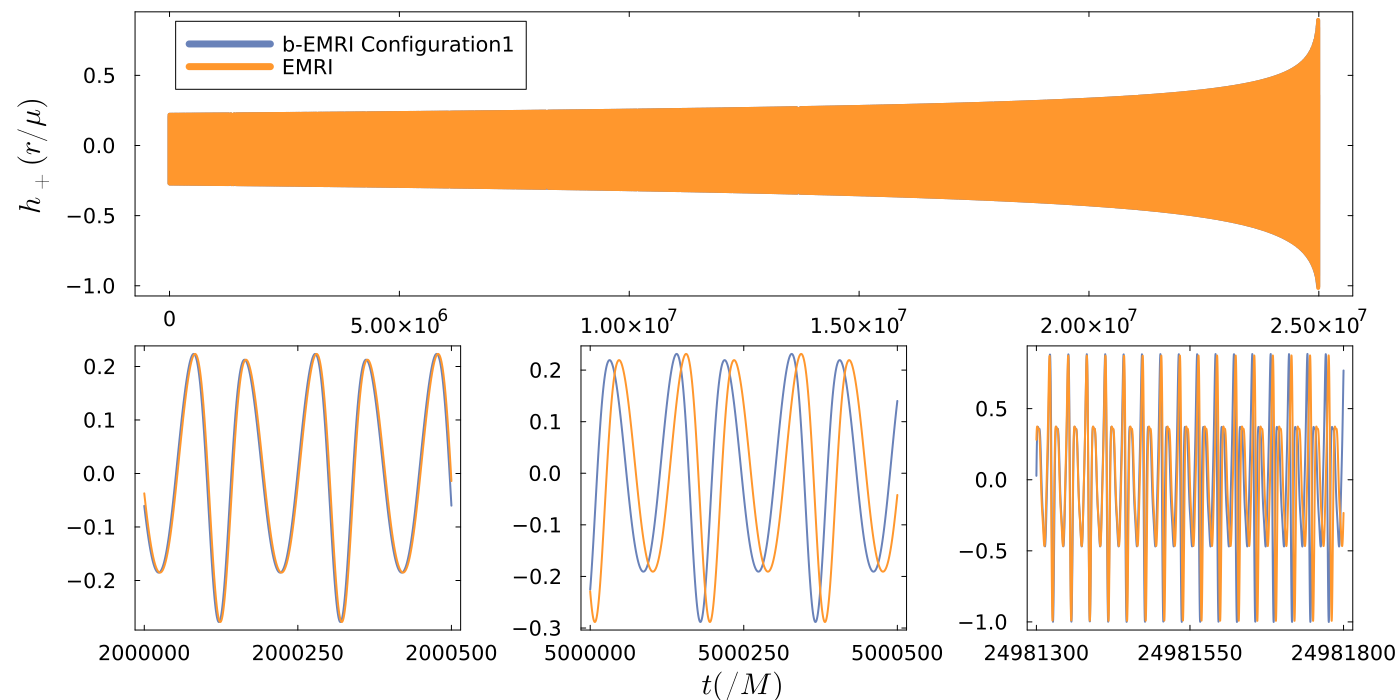}}

    \subfloat[Adiabatic waveforms of standard EMRI and b-EMRI in Configuration 2]{\label{fig:6b}\includegraphics[width=1.0\linewidth]{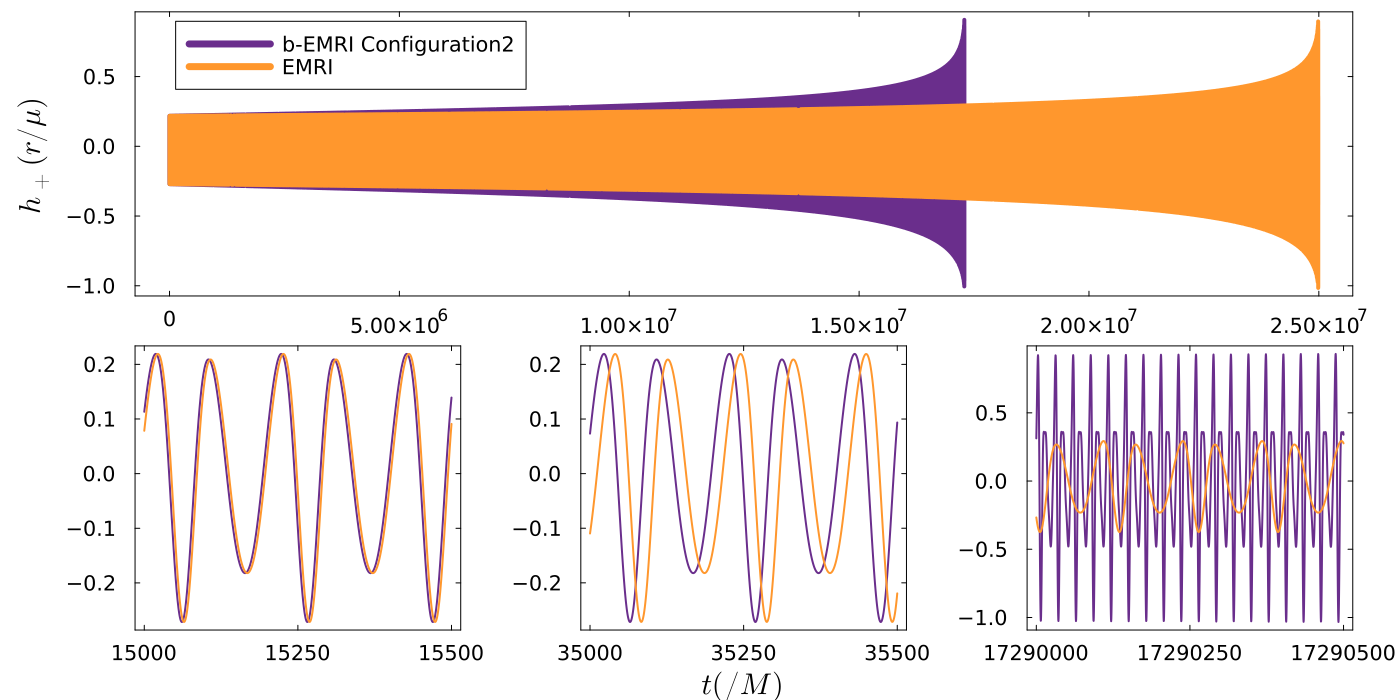}}
    
\caption{The upper panels of (a) and (b) show the full adiabatic waveforms of the EMRI 
(orange curve) and the b-EMRI with different configurations (blue curve and purple curve). 
The time span covers the evolution from $p_{\mathrm{init}}=10M$ to $p_{\mathrm{final}}=2.35M$. 
The lower three panels of (a) and (b) are chosen to show the how the deviation in phase (dephasing) progresses at the beginning, in the middle and at the end.
}
    \label{fig:AdiabaticWaveforms}
\end{figure*}

Fig.~\ref{fig:EvolutionalTrajectoryOfP} shows the evolution of the parameter
$p$. With our choice of parameters, it takes $69\%$ of the evolution time of
the EMRI for the b-EMRI system to evolve from $p_{\mathrm{init}}=10M$ to
$p_{\mathrm{final}}=2.35M$ if we account for high-frequency modes. During the early inspiral stage, the
b-EMRI system under ``Configuration 2'' evolves significantly faster than both
the standard EMRI and the b-EMRI with ``Configuration 1''. For example, if we take $M=10^6M_\odot$, the inspiral timescale for an EMRI in our fiducial model is $3.9$ years and is $2.7$ years for a b-EMRI under ``Configuration 2''. The difference in the
evolution time between the EMRI and the b-EMRI with ``Configuration 1'' is small,
roughly $1899M\approx 2.6\mathrm{h}$ in this case.

With the trajectory $p(t)$ in hand, we further compute the adiabatic waveforms
for both the EMRI and the b-EMRI systems. Fig.~\ref{fig:AdiabaticWaveforms}
shows the results. When including only the low-frequency modes into the fluxes,
the waveforms of the standard EMRI and b-EMRI systems start to deviate in phase
after about $5\times10^4$ outer orbital periods. In contrast, when including
the high-frequency modes, the deviation in phase already appears only after
$40$ outer orbital periods.

These results suggest that the high-frequency modes contribute significantly
more to the evolution of the b-EMRI system than the low-frequency modes.  As a
result, the b-EMRI system evolves much faster than a standard EMRI in the early
inspiral phase. This observation extends beyond the current understanding of
b-EMRI system evolution.  For example, in Fig.~2 of
Ref.~\cite{2024EPJC...84..478J} where the contributions of the high-frequency
modes are not included, we see a much slower evolution of the b-EMRI system
which is consistent with what we have derived in
Fig.~\ref{fig:AdiabaticWaveforms}(a).

We also find that the influence of the high-frequency modes diminishes as the
parameter $p$ decreases due to radiation reaction effects. This is caused
probably by the fact that the averaged outer circular orbit starts to produce
modes with higher frequencies as $p$ shrinks. These modes could be as strong as
the modes produced by the $d^2$ terms of the b-EMRI.  As a result, a key
feature of the b-EMRI waveform is its faster evolution during the early
inspiral stage, and the converging to the standard EMRI rate as $p$ becomes
smaller. This characteristic might be used to distinguish b-EMRIs from standard
EMRI systems.

\begin{table*}[htb]
    \centering
    \begin{ruledtabular}
    \begin{tabular}{cc|cccccc}
         $p/M$& $d/\mu$& $\dot{\mathcal{E}}^{\infty(0)}(/\epsilon^2)$& $\dot{\mathcal{E}}^{\infty(1)}(/\epsilon^2)$&$\dot{\mathcal{E}}^{\infty(2)}(/\epsilon^2)$ &  $\dot{\mathcal{E}}^{H(0)}(/\epsilon^2)$& $\dot{\mathcal{E}}^{H(1)}(/\epsilon^2)$&$\dot{\mathcal{E}}^{H(2)}(/\epsilon^2)$ \\
         \hline
         10&2000& $5.04909\times10^{-5}$ & $5.04969\times 10^{-5}$ & $1.40386\times 10^{-4}$ & $-1.43895\times10^{-6}$ & $-1.43910\times 10^{-6}$ & $1.10771\times 10^{-3}$ \\
         10&4000& $5.04909\times10^{-5}$ & $5.04945\times10^{-5}$ & $7.01751\times10^{-5}$ & $-1.43895\times10^{-6}$ & $-1.43903\times10^{-6}$ & $2.12752\times10^{-5}$ \\
         10&6000& $5.04909\times10^{-5}$ & $5.04937\times10^{-5}$ & $5.31631\times10^{-5}$ & $-1.43895\times10^{-6}$ & $-1.43901\times10^{-6}$ & $-1.44617\times10^{-6}$ \\
         10&8000& $5.04909\times10^{-5}$ & $5.04934\times10^{-5}$ & $5.08327\times10^{-5}$ & $-1.43895\times10^{-6}$ & $-1.43902\times10^{-6}$ & $-1.44713\times10^{-6}$ \\
         8&2000& $1.46335\times10^{-4}$ & $1.46352\times10^{-4}$ & $1.98487\times10^{-4}$ & $-7.51153\times10^{-6}$ & $-7.51229\times10^{-6}$ & $5.96528\times10^{-4}$ \\
         8&4000& $1.46335\times10^{-4}$ & $1.46346\times10^{-4}$ & $1.58943\times10^{-4}$ & $-7.51153\times10^{-6}$ & $-7.51195\times10^{-6}$ & $2.16421\times10^{-6}$ \\
         8&6000& $1.46335\times10^{-4}$ & $1.46344\times10^{-4}$ & $1.47468\times10^{-4}$ & $-7.51153\times10^{-6}$ & $-7.51192\times10^{-6}$ & $-7.46815\times10^{-6}$ \\
         8&8000& $1.46335\times10^{-4}$ & $1.46343\times10^{-4}$ & $1.46488\times10^{-4}$ & $-7.51153\times10^{-6}$ & $-7.51200\times10^{-6}$ & $-7.51337\times10^{-6}$ \\
         6&2000& $5.65865\times10^{-4}$ & $5.65934\times10^{-4}$ & $5.90544\times10^{-4}$ & $-6.04253\times10^{-5}$ & $-6.04311\times10^{-5}$ & $3.84389\times10^{-4}$ \\
         6&4000& $5.65865\times10^{-4}$ & $5.65910\times10^{-4}$ & $5.71402\times10^{-4}$ & $-6.04253\times10^{-5}$ & $-6.04289\times10^{-5}$ & $-5.81383\times10^{-5}$ \\
         6&6000& $5.65865\times10^{-4}$ & $5.65903\times10^{-4}$ & $5.66266\times10^{-4}$ & $-6.04253\times10^{-5}$ & $-6.04294\times10^{-5}$ & $-6.04366\times10^{-5}$ \\
         6&8000& $5.65865\times10^{-4}$ & $5.65900\times10^{-4}$ & $5.65955\times10^{-4}$ & $-6.04253\times10^{-5}$ & $-6.04311\times10^{-5}$ & $-6.04320\times10^{-5}$ \\
         4&2000& $3.59721\times10^{-3}$ & $3.59767\times10^{-3}$ & $3.60627\times10^{-3}$ & $-7.76063\times10^{-4}$ & $-7.76124\times10^{-4}$ & $-6.75433\times10^{-4}$ \\
         4&4000& $3.59721\times10^{-3}$ & $3.59753\times10^{-3}$ & $3.59869\times10^{-3}$ & $-7.76063\times10^{-4}$ & $-7.76120\times10^{-4}$ & $-7.76084\times10^{-4}$ \\
         4&6000& $3.59721\times10^{-3}$ & $3.59746\times10^{-3}$ & $3.59756\times10^{-3}$ & $-7.76063\times10^{-4}$ & $-7.76157\times10^{-4}$ & $-7.76162\times10^{-4}$ \\
         4&8000& $3.59721\times10^{-3}$ & $3.59738\times10^{-3}$ & $3.59742\times10^{-3}$ & $-7.76063\times10^{-4}$ & $-7.76221\times10^{-4}$ & $-7.76234\times10^{-4}$ \\
    \end{tabular}
    \end{ruledtabular}
    \caption{Energy fluxes due to different modes when $a=0.9M$, $\epsilon=10^{-5}$, $\tilde{\varphi}=\pi/4$, and $\tilde{\theta}=\pi/3$}
    \label{tab:CatalogOfEnergyFluxes}
\end{table*}

\section{\label{V}Discussion}

\subsection{Comparisons with previous works}

Now we compare our results with those presented in three previous works
\cite{PhysRevD.103.L081501,2024EPJC...84..478J,2024arXiv240507113M}.  The first BH perturbation theory based model of b-EMRI waveforms appears in
Ref.~\cite{PhysRevD.103.L081501}. Although the source term of the perturbation 
is simplified and has only one frequency in that study, it already
shows that the perturbation can resonantly
excite the QNMs of the central SMBH. Moreover, in that study the excitation ratio is defined as
\begin{equation}\label{Eq.CardosoRatio}
    {_s}\mathcal{R}_{lm}={_s}\dot{E}_{lm}/{_s}\dot{E}_{Nlm},
\end{equation}
so that ${_s}\mathcal{R}_{lm}$ tracks the flux excited by the high-frequency
motion of the IB. The behavior of this ratio as a function of frequency
$\omega$ is similar to our $A_{lmwv}$ defined in Eq.~\eqref{Eq.EnergyFluxes}.
For example, Fig.~\ref{fig:QNM_Excitation} in our work agrees with Fig.~2 of
Ref.~\cite{PhysRevD.103.L081501}. Both figures show that the positions of the
peaks $\omega^{\mathrm{peak}}_{lm}$ are slightly smaller than the real parts of
the QNMs frequencies $\omega^{\mathrm{Q\!N\!M}}_{lm0}$, and the curves are
flatter on the left of the peaks while steeper on the right.

To better interpret the above results, we need to understand the nature of
QNMs. They are solutions to the homogeneous Teukolsky radial equation
Eq.~\eqref{Eq.TeukolskyRadialEquation} satisfying purely outgoing boundary
condition at infinity and purely ingoing boundary condition at the horizon.
According to these solutions, the asymptotic amplitudes $B^{\mathrm{inc}}_{lm}$
and $C^{\mathrm{inc}}_{lm}$ in Eq.~\eqref{Eq.AsymptoticHomo} go zero when
$\omega=\omega^{\mathrm{Q\!N\!M}}_{lm0}$. This will make the Wronskian
$\mathcal{W}_{lm}(\omega)$ vanish and the amplitude $Z^{(2)}_{lmwv}$ diverge,
as one can see in Eqs.~\eqref{Eq.AmplitudesDefinition} and
\eqref{Eq.Wronskian}.  However, $\omega^{\mathrm{Q\!N\!M}}_{lm0}$ is a complex
frequency but $\omega_{mwv}$, the frequency associated with the b-EMRI, is a
real one. So the system cannot achieve
$B^{\mathrm{inc}}_{lm}=C^{\mathrm{inc}}_{lm}=0$ even when the real parts of
these two frequencies completely match.  Such a mismatch results in a
finite peak. 

Later, Ref.~\cite{2024EPJC...84..478J}  treats the inner binary as a mass
distribution and expands it into multipole moments to compute the source term. 
To evolve the outer orbit, the authors treat the BBH as a single spinning body
and use the MPD equation to compute its orbit around the SMBH. 
This approach ensures conservation of the
stress-energy tensor and inspires us to discuss the validity of our model in Sec.~\ref{VC}. However, 
in their study the authors omitted the high-frequency source terms
because the frequencies
are outside the LISA band. Our results, nevertheless, shows that
these dropped high-frequency modes, if resonate with the central SMBH's QNMs, will enhance energy dissipation.  
Notably, the IB is allowed to evolve and merger in
Ref.~\cite{2024EPJC...84..478J}, so the b-EMRI will be an important multi-band
GW source which can produce both high-frequency GW in the LIGO/Virgo band and
low-frequency GW in the LISA band \cite{2018CmPhy...1...53C}. The merger of the
IB also leads to a sudden mass loss making the low-frequency waveform more
distinguishable. These potential multi-band observational aspects are not discussed in our
work. 

More recently, Ref.~\cite{2024arXiv240507113M} uses the ``Numerical Kludge''
template to generate the waveform of b-EMRI. The OB in that work takes a
generic orbit and the IB is eccentric, making the b-EMRI more realistic.
However, the trajectories of the small BHs around the SMBH are still derived
from an algebraic combination of the inner and outer motion. Therefore, the
method cannot fully deal with the relativistic effects on the (three-body) GW
source in the strong-gravity regime.  Interestingly, the waveform snapshots
shown in Ref.~\cite{2024arXiv240507113M} also exhibit strong modulation by the
existence of the IB. However, this modulation is not associated with the
excitation of QNMs, but due to the large inner orbital motion of the two
stellar-mass BHs, since the semi-latus rectum of their IB (which is analogous
to $d$ in this work) is a significant fraction of the size of the OB.  Such a
large IB raises the issue of stability for the three-body system. We will
discuss this issue in the next subsection.

\subsection{\label{VB}Parameter space}

Throughout this work, we assume no evolution of the IB, which means that the IB
is neither tidally disrupted nor excited to high eccentricities. The stability
of a hierarchical three-body system is a classical problem 
\cite{1988Natur.331..687H}.  The maximum distance for the two bodies in
the IB to remain stable is determined by the Hill radius,
$R_H:=(\epsilon/3)^{1/3}p$, where Newtonian gravity is assumed.  Since here we
are working in the strong-gravity regime, we require $d<\eta\cdot R_H$, where $0<\eta<1$. The determination of $\eta$ involves more delicate considerations of relativistic three-body dynamics. We refer interested readers to our previous work \cite{2022PhRvD.106j3040C} and a more recent work \cite{10.1093/mnras/stae1093}. Here, we simply set $\eta=1/3$ as an example. Given our fiducial model in which $p=10M$ and $\epsilon=10^{-5}$,
we find $d<0.05M$. In other words, an IB with $d=4000\mu=0.04M$ is stable in
this case. Additionally, since we assume Newtonian motion for the inner orbital
motion, $d$ should be sufficiently large to avoid strong relativistic
correction. For this latter reason, we set $d>1000\mu=1000M\epsilon$. The leading post-Newtonian effect to the inner orbit is of the order $\mathcal{O}(\mu\sqrt{\mu/d^3})\sim 10^{-6}$, which is negligible. Moreover, the evolution timescale of the IB due to GW radiation is much longer than that of the OB, so we do not the IB here.

Another important assumption in this work is that the inner binary follows a circular orbit.
However, dynamical effects in triple systems, 
such as the Kozai-Lidov mechanism \cite{1962AJ.....67..591K,1962P&SS....9..719L}, could
excite the eccentricity of the IB if the inclination angle $\tilde{\theta}$ between
the inner and outer orbits exceeds
a critical value \cite{10.1093/mnras/stae1093}. The timescale of the eccentric
Kozai-Lidov mechanism (EKM) was derived in
Ref.~\cite{PhysRevLett.107.181101,2015MNRAS.452.3610A}, which gives
\begin{equation}\label{Eq.EKMOriginal}
\tau_{\mathrm{E\!K\!M}}=\frac{256\sqrt{10}}{15\pi\sqrt{\epsilon_{\mathrm{oct}}}}t_{\mathrm{sec}},
\end{equation} 
where
\begin{subequations}\label{Eq.EKMFunction}
    \begin{align}
        &t_{\mathrm{sec}}=\frac{\sqrt{M_{C}d}}{\Phi_0},\\
        &\epsilon_{\mathrm{oct}}=\frac{d}{p}\frac{e}{1-e^2},\\
        &\Phi_0=\frac{Md^2}{p^3(1-e^2)^{3/2}},
    \end{align}
\end{subequations}
$M_C=\mu/4$ is the reduced mass of the IB, and $e$ is an eccentricity parameter of the
OB. If we assume a circular outer orbit but allow a small
eccentricity $e\ll1$ to study the EKM timescale, inserting
Eq.~\eqref{Eq.EKMFunction} into Eq.~\eqref{Eq.EKMOriginal}, we find
\begin{equation}\label{Eq.EKMTimescale}
    \tau_{\mathrm{E\!K\!M}}=\frac{128\sqrt{10}}{15\pi}\frac{(1-e^2)^2}{\sqrt{e}}\tilde{p}^{7/2}\tilde{d}^{-2}\epsilon^{-3/2}M,
\end{equation}
where $\tilde{p}\equiv p/M$ and $\tilde{d}\equiv d/\mu$. To ensure that the
inner orbit remains circular for a sufficiently long time, we require that
$\tau_{\mathrm{E\! K\!  M}}>2\pi\times10^4/\Omega_{g}$, i.e., the inner orbit
stays circular for at least $10^4$ outer orbital periods. This timescale
suffice to allow significant dephasing between b-EMRI and EMRI, as shown in
Fig.~\ref{fig:AdiabaticWaveforms}. In the following analysis, we will set
$e=0.1$ as an example.

\begin{figure}[]
    \centering
    \includegraphics[width=1\linewidth]{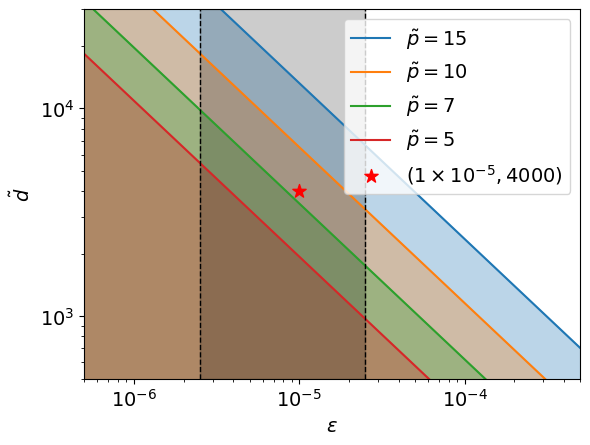}
    \caption{Parameter space of a b-EMRI. The regions below the solid (dashed) lines satisfy the 
EKM (Hill's) condition when $\tilde{p}=20, 10, 5$. The grey shaded region represents 
a IB in the mass range of $10M_\odot<\mu<100M_\odot$, where we have assumed 
$M=4\times 10^6M_\odot$. The red star denotes our fiducial model 
	in which $\epsilon=10^{-5},\ \tilde{d}=4000$.} 
    \label{fig:Parameterspace}
\end{figure}

In Fig.~\ref{fig:Parameterspace} we show the parameter space where the Hill's condition $d<R_H/3$ and the EKM condition
$\tau_{\mathrm{E\!K\!M}}>2\pi\times 10^4/\Omega_g$ can be met. We find that
for larger $p$ (e.g., $\tilde{p}=20$), the Hill's condition plays a more
important role and sets the upper limit for $d$. The EKM condition is more
important for smaller $p$. The red star in the plot indicates the parameters
used in our fiducial model. It lies in a region where both conditions can be
met as long as $p\ge10M$. 

It is worth noting that even in the case where the IB has some eccentricity,
the resonant excitation of QNMs is still valid as long as the inner orbital
frequencies match the QNM frequencies.  Technically speaking, we can perform a
Fourier expansion of the inner orbit, treating the Fourier modes as a series of
circular motions. Then our analysis shown above remains applicable to eccentric
in binaries.

\subsection{\label{VC}Validity of the model}

In Sec.~\ref{IIG} we demonstrated how to track the evolution of the system `quasi-adiabatically'. In particular, the evolution of the parameter $p$ is derived assuming osculating geodesic motion of the OB. In this work, however, we follow
Ref.~\cite{PhysRevD.103.L081501,2021PhLB..82236654F}
and write the stress-energy tensor of the IB as
\begin{equation}\label{Eq.TwoMonopoleEMT} \begin{aligned}
	T^{\mu\nu}(x)=&m_1\int\mathrm{d}\tau
	\frac{\delta^4\left(x-z_1(\tau)\right)}{\sqrt{-g(x)}}\frac{\mathrm{d}
	z^\mu_1(\tau)}{\mathrm{d}\tau}\frac{\mathrm{d}
	z^\nu_1(\tau)}{\mathrm{d}\tau}\\ &+m_2\int\mathrm{d}\tau
\frac{\delta^4\left(x-z_2(\tau)\right)}{\sqrt{-g(x)}}\frac{\mathrm{d}
z^\mu_2(\tau)}{\mathrm{d}\tau}\frac{\mathrm{d} z^\nu_2(\tau)}{\mathrm{d}\tau},
\end{aligned} \end{equation}
where $m_1=m_2=\mu/2$ and $z^\mu_{1,2}$ are the trajectories of the two small
BHs in the BL coordinates derived as Eq.~\eqref{Eq.BLCompactCoordinate}.  This
$T^{\mu\nu}$ is missing information about the internal energy and angular momentum of the
IB. Because we constructed the background motion of the IB independently of the stress-energy, we cannot guarantee its conservation without adding a term accounting for the interaction of the two monopole particles. For the equal mass IB we consider here, we have $\nabla_\mu T^{\mu \nu}\propto \mu d^2$, meaning we are missing some terms at the same order which we keep in the waveform amplitudes and inspiral evolution. In Appendix~\ref{appD}, we show how to construct the correction to the stress-energy due to the interaction between the two monopole particles. When this interaction term is included, the stress-energy is conserved through to the order at which we truncate our waveform model and in that sense the model is fully relativistic to that order. 

While we have not yet included this interaction term in the stress-energy when solving the Teukolsky equation, we remark that it is straight forward to add in follow up work as a linear correction to the Teukolsky amplitudes (and therefore fluxes). Without accounting for that term, the waveform model should not be used in parameter estimation studies for example. However, we stress that its neglect does not change the qualitative features of our findings. 

\section{\label{VI}Conclusion and future work}

In this work we have modeled the trajectories of the two stellar-mass BHs in a
b-EMRI system using a relativistic coordinate transformation.  In turn, we derived the source term in the Teukolsky equation, allowing us to compute
the waveform of a b-EMRI using black hole perturbation theory. We find that the
transformation of the orbits from the  $FFF$ into $BL$ coordinates
introduces both low- and high-frequency perturbations in the source term of the Teukolsky
equation. It induces six additional modes proportional to the IB's separation, $d$, and fifteen additional modes proportional to $d^2$ in the waveform. Our assumption of equal-mass IBs results in
the cancellation of the linear $d$ waveform contributions. Nevertheless, more
generic choices of mass for the two small BHs does not introduce significant
difficulties in the application of our method.

Our results confirm that the high-frequency perturbation
resonantly excites
the QNMs of the central SMBH, which was first discovered in Ref.~\cite{PhysRevD.103.L081501}. 
We further analyzed the condition for the mode excitation and attributed the finite
amplitudes of the QNMs to a mismatch between the real perturbation frequencies and the
complex QNM frequencies. The most important finding in this work is that 
when the QNMs are excited,
b-EMRIs evolve faster than standard EMRIs with the same outer parameters ($M$, $a$,
$p$, $\epsilon$), because the excited QNMs enhance the energy and angular-momentum fluxes. 

Two potential extensions of this work can be carried out in the future. First,
the model could be extended to account for the interaction stress-energy of the IB and to a wider range of astrophysical scenarios.  For
example, one can incorporate eccentricity in the outer binary orbit, as is
expected by the formation model of b-EMRIs
\cite{2015arXiv150107856A,2018CmPhy...1...53C}, or evolve the inner orbit
dynamically, as has been modeled in Ref.~\cite{2022PhRvD.106j3040C}.  Second,
one can study the distinguishability of b-EMRI systems from normal EMRIs.
While this has been examined in Ref.~\cite{2024EPJC...84..478J} and
Ref.~\cite{2024arXiv240507113M}, our model has revealed new signatures of
b-EMRIs, particularly related to $p$, such that the b-EMRI system evolves
faster during the early inspiral phase before converging to the standard EMRI
evolution.

\begin{acknowledgments}
This work is supported by the National Key Research and Development Program of
China (Grant No. 2021YFC2203002) and the National Natural Science Foundation of
China (Grant No. 12473037). We would like to acknowledge our referee for thoroughly reading the manuscript and providing constructive suggestions. Y.Y. acknowledges NUS Physics Department for providing office environment and care. Y.Y. acknowledges Haolan Zheng and Han Yan for several inspirational discussions. The authors thank Soichiro Isoyama, Wenbiao Han, and Xilong Fan for insightful feedback on a draft of this paper. J.M. acknowledges support by the NUS Faculty of Science, under the research grant 22-5478-A0001.
\end{acknowledgments}

\appendix

\section{\label{appA}Transformation from the LIF to BL coordinates}

The transformation equations between different coordinates have been derived in Ref.~\cite{2005CQGra..22.4729B}. A more general
set of transformation equations from Fermi coordinates to other coordinate
systems can be found in Ref.~\cite{2008CQGra..25n5019K}. In this work, we
use the results from Ref.~\cite{2005CQGra..22.4729B} for comoving
observers on circular orbits. For any circular orbit, the coordinates measured
by such an observer are $(T,X^1,X^2,X^3)$, where $T$ is the proper time along the
observer's worldline, and $X^1,X^2,X^3$ represent the three spatial coordinates
in the vicinity of the observer. The spatial coordinates $X^1,X^2,X^3$ are
assumed to be small compared to the local curvature radius.
The transformation equations between  $(T,X^1,X^2,X^3)$ and the BL coordinates
$(t,r,\theta,\varphi)$ are
\begin{subequations}\label{Eq.BiniOriginal1}
    \begin{align}
        DT=&T-\Omega(\nu)X^1X^3+\mathcal{O}(3),\\
        DR=&X^1\left(1+\frac{1}{2}{^{(0)}}\kappa(r,m)^{\hat{r}}X^1\right)\notag\\
        &-\frac{1}{2}{^{(0)}}\kappa(\theta,m)^{\hat{r}}(X^2)^2\notag\\
        &-\frac{1}{2}{^{(0)}}\kappa(\varphi,U)^{\hat{r}}(X^3)^2+\mathcal{O}(3),\\
        D\Theta=&X^2\left(1+{^{(0)}}\kappa(\theta,m)^{\hat{r}}X^1\right)+\mathcal{O}(3),\\
        D\Phi=&X^3\left(1+{^{(0)}}\kappa(\varphi,U)^{\hat{r}}X^1\right)+\mathcal{O}(3),
    \end{align}
\end{subequations}
where
\begin{subequations}\label{Eq.BiniOriginal2}
    \begin{align}
        &DT=\gamma\Big[{^{(0)}}M(t-t_0)\notag\\
        &\qquad\qquad-\left({^{(0)}}M{^{(0)}}M_\varphi+\nu{^{(0)}}\gamma_{\varphi\varphi}^{1/2}\right)(\varphi-\varphi_0)\Big],\\
        &DR={^{(0)}}\gamma_{rr}^{1/2}(r-r_0),\\
        &D\Theta={^{(0)}}\gamma_{\theta\theta}^{1/2}(\theta-\theta_0),\\
        &D\Phi=\gamma\Big[-{^{(0)}}M\nu(t-t_0)\notag\\
        &\qquad\qquad+\left({^{(0)}}M{^{(0)}}M_\varphi\nu+{^{(0)}}\gamma_{\varphi\varphi}^{1/2}\right)(\varphi-\varphi_0)\Big].
    \end{align}
\end{subequations}
If the circular orbit is geodesic, some of the coefficients are in simpler forms, we denote them as
\begin{equation}\label{Eq.TransformationCoefficients}
    \begin{aligned}
        &\kappa_r\equiv{^{(0)}}\kappa(r,m)^{\hat{r}},\quad \kappa_\theta\equiv{^{(0)}}\kappa(\theta,m)^{\hat{r}},\\
        & \kappa_\varphi\equiv{^{(0)}}\kappa(\varphi,U)^{\hat{r}},\quad \mathcal{M}\equiv{^{(0)}}M,\quad\mathcal{M}_\varphi\equiv{^{(0)}}M_\varphi,\\
        &\gamma^{1/2}_{rr}\equiv{^{(0)}}\gamma_{rr}^{1/2},\quad\gamma^{1/2}_{\theta\theta}\equiv{^{(0)}}\gamma_{\theta\theta}^{1/2},\quad\gamma^{1/2}_{\varphi\varphi}\equiv{^{(0)}}\gamma_{\varphi\varphi}^{1/2}.
    \end{aligned}
\end{equation}
Sort $t,r,\theta,\varphi$ out and write them on the left hand side of the equations, we obtain
\begin{subequations}\label{Eq.EquationOfMotionBoyerLindquist}
    \begin{align}
        t=&t_0+\frac{\gamma_{\varphi\varphi}^{1/2}+\mathcal{M}\mathcal{M}_\varphi \nu}{\gamma_{\varphi\varphi}^{1/2}\mathcal{M}/\gamma}T+\frac{\mathcal{M}\mathcal{M}_{\varphi}+\gamma_{\varphi\varphi}^{1/2}\nu}{\gamma_{\varphi\varphi}^{1/2}\mathcal{M}/\gamma}X^3\notag\\
        &+\frac{(\mathcal{M}\mathcal{M}_\varphi+\gamma_{\varphi\varphi}^{1/2}\nu)\kappa_\varphi-(\gamma_{\varphi\varphi}^{1/2}+\mathcal{M}\mathcal{M}_\varphi\nu)\Omega}{\gamma_{\varphi\varphi}^{1/2}\mathcal{M}/\gamma}X^1X^3,\\
        r=&r_0+\frac{1}{\gamma_{rr}^{1/2}}X^1\notag+\frac{\kappa_r}{2\gamma_{rr}^{1/2}}\left(X^1\right)^2\\
        &-\frac{\kappa_\theta}{2\gamma_{rr}^{1/2}}\left(X^2\right)^2-\frac{\kappa_\varphi}{2\gamma_{rr}^{1/2}}\left(X^3\right)^2,\\
        \theta=&\theta_0+\frac{1}{\gamma_{\theta\theta}^{1/2}}X^2+\frac{\kappa_\theta}{\gamma_{\theta\theta}^{1/2}}X^1X^2,\\
        \varphi=&\varphi_0+\frac{\nu\gamma}{\gamma_{\varphi\varphi}^{1/2}}T+\frac{\gamma}{\gamma_{\varphi\varphi}^{1/2}}X^3+\frac{\kappa_{\varphi}-\nu\Omega}{\gamma_{\varphi\varphi}^{1/2}/\gamma}X^1X^3.
    \end{align}
\end{subequations}
where
\begin{equation}\label{A5}
    \begin{aligned}
        &\mathcal{M}=\sqrt{1-\frac{2M}{p}},\quad \mathcal{M}_\varphi=-\frac{2aM}{p-2M},\\
        &\Delta_0=p^2-2Mp+a^2,\quad \Omega=-\sqrt{\frac{M}{p^3}}\\
        &\gamma_{rr}^{1/2}=\frac{p}{\sqrt{\Delta_0}},\quad\gamma_{\theta\theta}^{1/2}=\sqrt{\Delta_0},\quad\gamma_{\varphi\varphi}^{1/2}=\sqrt{\frac{p\Delta_0}{p-2M}},\\
        &\kappa_r=\frac{Mp-a^2}{p^2\sqrt{\Delta_0}},\quad\kappa_\theta=-\frac{\sqrt{\Delta_0}}{p^2},\\
        &\kappa_{\varphi}=\frac{a^2Mp-4M^2p^2+4Mp^3-p^4-M\Delta_0}{p^2(p-2M)\sqrt{\Delta_0}},\\
        &\nu=\frac{\sqrt{\Delta_0}}{a+(p-2M)\sqrt{p/M}},\quad\gamma=\frac{1}{\sqrt{1-\nu^2}}.
    \end{aligned}
\end{equation}
In our case, we have
\begin{equation}\label{A6}
    t_0=0,\qquad r_0=p,\qquad \theta_0=\frac{\pi}{2},\qquad \varphi_0=0.
\end{equation}
It is straightforward to verify that the coefficients of $T$ in the $t$ and $\varphi$ components of Eq.~ \eqref{Eq.EquationOfMotionBoyerLindquist} correspond to the 4-velocity components $u^t$ and $u^\varphi$.

Inserting Eqs.~\eqref{Eq.EulerRotation}-\eqref{Eq.FFFtoLIF} into Eq.~\eqref{Eq.EquationOfMotionBoyerLindquist}, we obtain the coefficients for the first- and the second-order expansions
\begin{subequations}\label{A9}
    \begin{align}
    c_{01}=&\frac{\mathcal{M}\mathcal{M}_{\varphi}+\gamma_{\varphi\varphi}^{1/2}\nu}{4\gamma_{\varphi\varphi}^{1/2}\mathcal{M}/\gamma}\Big\{(1+\cos\tilde{\theta})\sin\left[\left(\Omega+\omega_{\mathrm{IB}}\right)\tau-\tilde{\varphi}\right]\notag\\
        &-(1-\cos\tilde{\theta})\sin\left[\left(\Omega-\omega_{\mathrm{IB}}\right)\tau-\tilde{\varphi}\right]\Big\},\\
    c_{11}=&\frac{1}{4\gamma_{rr}^{1/2}}\Big\{(1+\cos\tilde{\theta})\cos\left[\left(\Omega+\omega_{\mathrm{IB}}\right)\tau-\tilde{\varphi}\right]\notag\\
        &-(1-\cos\tilde{\theta})\cos\left[\left(\Omega-\omega_{\mathrm{IB}}\right)\tau-\tilde{\varphi}\right]\Big\},\\
        c_{21}=&\frac{1}{2\gamma_{\theta\theta}^{1/2}}\sin\tilde{\theta}\cos\left(\omega_{\mathrm{IB}}\tau\right),\\
        c_{31}=&\frac{\gamma}{4\gamma_{\varphi\varphi}^{1/2}}\Big\{(1+\cos\tilde{\theta})\sin\left[\left(\Omega+\omega_{\mathrm{IB}}\right)\tau-\tilde{\varphi}\right]\notag\\
        &-(1-\cos\tilde{\theta})\sin\left[\left(\Omega-\omega_{\mathrm{IB}}\right)\tau-\tilde{\varphi}\right]\Big\},\\
        c_{02}=&\frac{(\mathcal{M}\mathcal{M}_\varphi+\gamma_{\varphi\varphi}^{1/2}\nu)\kappa_\varphi-(\gamma_{\varphi\varphi}^{1/2}+\mathcal{M}\mathcal{M}_\varphi\nu)\Omega}{16\gamma_{\varphi\varphi}^{1/2}\mathcal{M}/\gamma}\notag\\
        &\times\Big\{2\cos^4\frac{\tilde{\theta}}{2}\sin\left[2\left(\Omega+\omega_{\mathrm{IB}}\right)\tau-2\tilde{\varphi}\right]\notag\\
        &-\sin^2\tilde{\theta}\sin\left(2\Omega\tau-2\tilde{\varphi}\right)\notag\\
        &+2\sin^4\frac{\tilde{\theta}}{2}\sin\left[2\left(\Omega-\omega_{\mathrm{IB}}\right)\tau-2\tilde{\varphi}\right]\Big\},\\
        c_{12}=&\frac{\kappa_r+\kappa_\varphi}{16\gamma_{rr}^{1/2}}\cos^4\frac{\tilde{\theta}}{4}\cos\left[2\left(\Omega+\omega_{\mathrm{IB}}\right)\tau-2\tilde{\varphi}\right]\notag\\
        &+\frac{\kappa_r+\kappa_\varphi}{16\gamma_{rr}^{1/2}}\sin^4\frac{\tilde{\theta}}{2}\cos\left[2\left(\Omega-\omega_{\mathrm{IB}}\right)\tau-2\tilde{\varphi}\right]\notag\\
        &-\frac{\kappa_r+\kappa_{\varphi}}{32\gamma_{rr}^{1/2}}\sin^2\tilde{\theta}\cos\left(2\Omega\tau-2\tilde{\varphi}\right)\notag\\
        &-\frac{\kappa_r+2\kappa_{\theta}-\kappa_\varphi}{32\gamma_{rr}^{1/2}}\sin^2\tilde{\theta}\cos\left(2\omega_{\mathrm{IB}}\tau\right)\notag\\
        &+\frac{\kappa_r-\kappa_{\varphi}}{32\gamma_{rr}^{1/2}}\left(1+\cos^2\tilde{\theta}\right)-\frac{\kappa_\theta}{16\gamma_{rr}^{1/2}}\sin^2\tilde{\theta},\\
        c_{22}=&\frac{\kappa_{\theta}\sin\tilde{\theta}}{16\gamma_{\theta\theta}^{1/2}}\Big\{\left(1+\cos\tilde{\theta}\right)\cos\left[\left(\Omega+2\omega_{\mathrm{IB}}\right)\tau-\tilde{\varphi}\right]\notag\\
        &-\left(1-\cos\tilde{\theta}\right)\cos\left[\left(\Omega-2\omega_{\mathrm{IB}}\right)\tau-\tilde{\varphi}\right]\notag\\
        &+2\cos\tilde{\theta}\cos\left(\Omega\tau-\tilde{\varphi}\right)\Big\},\\
        c_{32}=&\frac{\kappa_{\varphi}-\nu\Omega}{16\gamma_{\varphi\varphi}^{1/2}/\gamma}\times\Big\{2\cos^4\frac{\tilde{\theta}}{2}\sin\left[2\left(\Omega+\omega_{\mathrm{IB}}\right)\tau-2\tilde{\varphi}\right]\notag\\
        &-\sin^2\tilde{\theta}\sin\left(2\Omega\tau-2\tilde{\varphi}\right)\notag\\
        &+2\sin^4\frac{\tilde{\theta}}{2}\sin\left[2\left(\Omega-\omega_{\mathrm{IB}}\right)\tau-2\tilde{\varphi}\right]\Big\}.
    \end{align}
\end{subequations}
As we can see, 
these coefficients are all functions of $\tau$. 

\section{\label{appB}Teukolsky source term}
The source term in Eq.~\eqref{Eq.TeukolskyMasterEquation} is given by $\hat{T} = 2 \rho^{-4} T_4$, where $T_4$ is a combination of the projected stress-energy tensor defined by
\begin{equation}\label{Eq.TeukolskySourceTerm}
        \begin{aligned}
            T_4=&\frac{1}{2}\rho^8\bar{\rho}\hat{L}_{-1}[\rho^{-4}\hat{L}_0(\rho^{-2}\bar{\rho}^{-1}T_{nn})]\\
            &+\frac{1}{2\sqrt{2}}\rho^8\bar{\rho}\Delta^2\hat{L}_{-1}[\rho^{-4}\bar{\rho}^{2}\hat{J}_+(\rho^{-2}\bar{\rho}^{-2}\Delta^{-1}T_{\bar{m}n})]\\
            &+\frac{1}{4}\rho^8\bar{\rho}\Delta^2\hat{J}_+[\rho^{-4}\hat{J}_+(\rho^{-2}\bar{\rho}T_{\bar{m}\bar{m}})]\\
            &+\frac{1}{2\sqrt{2}}\rho^8\bar{\rho}\Delta^2\hat{J}_+[\rho^{-4}\bar{\rho}^{2}\Delta^{-1}\hat{L}_{-1}(\rho^{-2}\bar{\rho}^{-2}T_{\bar{m}n})],
        \end{aligned}
\end{equation}
where the differential operators $\hat{L}_s$ and $\hat{J}_+$ are defined as
\begin{subequations}\label{Eq.DifferentialOperators}
\begin{align}
    & \hat{L}_s=\partial_{\theta}-\frac{i}{\sin\theta}\partial_{\varphi}-ia\sin\theta\partial_t+s\cot\theta,\\
    &\hat{J}_+=\partial_r-\frac{1}{\Delta}\left((r^2+a^2)\partial_t+a\partial_{\varphi}\right).
\end{align}
\end{subequations}
Under the NP tetrad
\begin{subequations}\label{Eq.NPTetrad}
    \begin{align}
    		&l^{\mu}=\frac{1}{\Delta}\left(a^2+r^2,\Delta,0,a\right),\\
    		&n^{\mu}=\frac{1}{2\Sigma}\left(a^2+r^2,-\Delta,0,a\right),\\
    		&m^{\mu}=\frac{1}{\sqrt{2}(r+ia\cos\theta)}\left(ia\sin\theta,0,1,\frac{i}{\sin\theta}\right),\\
    		&\bar{m}^{\mu}=\frac{1}{\sqrt{2}(r-ia\cos\theta)}\left(-ia\sin\theta,0,1,-\frac{i}{\sin\theta}\right),
    \end{align}
\end{subequations}
the projected components of the stress-energy tensor read
\begin{subequations}\label{Eq.ProjectedT}
    \begin{align}
        T_{nn}&=\mu\frac{C_{nn}}{\sin\theta}\delta\left[r-r(t)\right]\delta\left[\theta-\theta(t)\right]\delta\left[\varphi-\varphi(t)\right],\\
        T_{\bar{m}n}&=\mu\frac{C_{\bar{m}n}}{\sin\theta}\delta\left[r-r(t)\right]\delta\left[\theta-\theta(t)\right]\delta\left[\varphi-\varphi(t)\right],\\
        T_{\bar{m}\bar{m}}&=\mu\frac{C_{\bar{m}\bar{m}}}{\sin\theta}\delta\left[r-r(t)\right]\delta\left[\theta-\theta(t)\right]\delta\left[\varphi-\varphi(t)\right],
    \end{align}
\end{subequations}
where
\begin{subequations}\label{Eq.ProjectedC}
    \begin{align}
        C_{nn}=&\frac{1}{4\Sigma^3u^t}\left[\mathcal{E}(r^2+a^2)-a\mathcal{L}_z+\Sigma\frac{\mathrm{d} r}{\mathrm{d}\tau}\right]^2,\\
        C_{\bar{m}n}=&-\frac{\rho}{2\sqrt{2}\Sigma u^t}\left[\mathcal{E}(r^2+a^2)-a\mathcal{L}_z+\Sigma\frac{\mathrm{d} r}{\mathrm{d}\tau}\right]\notag\\
        &\times\left[i\sin\theta\left(a\mathcal{E}-\frac{\mathcal{L}_z}{\sin^2\theta}\right)+\Sigma\frac{\mathrm{d}\theta}{\mathrm{d}\tau}\right],\\
        C_{\bar{m}\bar{m}}=&\frac{\rho^2}{2\Sigma u^t}\left[i\sin\theta\left(a\mathcal{E}-\frac{\mathcal{L}_z}{\sin^2\theta}\right)+\Sigma\frac{\mathrm{d}\theta}{\mathrm{d}\tau}\right]^2.
    \end{align}
\end{subequations}

Start from Eqs.~\eqref{Eq.AmplitudesDefinition} and \eqref{Eq.FourierSourceTerm}, 
doing integration by part, we obtain
\begin{equation}\label{Eq.AmplitudeZ}
    \begin{aligned}
        Z^{\mathrm{in}}_{lm}(\omega)=&\frac{1}{\mathcal{W}_{lm}(\omega)}\int_{r_+}^{\infty}\mathrm{d} r^\prime\frac{R^{\mathrm{in}}_{lm}(r^\prime,\omega)}{\Delta^{\prime2}}\mathcal{T}_{lm}(r^\prime,\omega),\\
        =&\frac{\mu}{\mathcal{W}_{lm}(\omega)}\int_{-\infty}^{\infty}\mathrm{d} t\ e^{i\omega t-im\varphi(t)}\\
        &\times\Big\{R^{\mathrm{in}}_{lm}(r,\omega)[A_{nn0}(r,\theta(t))\\
        &\qquad\qquad\qquad\ +A_{\bar{m}n0}(r,\theta(t))+A_{\bar{m}\bar{m}0}(r,\theta(t))]\\
        &\quad-\frac{\mathrm{d} R^{\mathrm{in}}_{lm}(r,\omega)}{\mathrm{d} r}[A_{\bar{m}n1}(r,\theta(t))+A_{\bar{m}\bar{m}1}(r,\theta(t))]\\
        &\quad+\frac{\mathrm{d}^2 R^{\mathrm{in}}_{lm}(r,\omega)}{\mathrm{d} r^2}A_{\bar{m}\bar{m}2}(r,\theta(t))\Big\}_{r=r(t)},
    \end{aligned}
\end{equation}
where
\begin{subequations}\label{Eq.ProjectedA}
    \begin{align}
        A_{nn0}(r,\theta(t))=&\frac{2\sqrt{2\pi}}{\Delta^2}\rho^{-2}\bar{\rho}^{-1}C_{nn}L_1^\dagger\left[\rho^{-4}L_2^\dagger\left(\rho^3S\right)\right],\\
        A_{\bar{m}n0}(r,\theta(t))=&-\frac{4\sqrt{\pi}}{\Delta}\rho^{-3}C_{\bar{m}n}\notag\\
        &\times\Big[\left(L_2^\dagger S\right)\left(\frac{iK}{\Delta}+\rho+\bar{\rho}\right)\notag\\
        &\quad-a\sin\theta S\frac{K}{\Delta}(\bar{\rho}-\rho)\Big],\\
        A_{\bar{m}\bar{m}0}(r,\theta(t))=&\sqrt{2\pi}\rho^{-3}\bar{\rho}C_{\bar{m}\bar{m}}S\notag\\
        &\times\left[-i\left(\frac{K}{\Delta}\right)_{,r}-\frac{K^2}{\Delta^2}+2i\rho\frac{K}{\Delta}\right],\\
        A_{\bar{m}n1}(r,\theta(t))=&-\frac{4\sqrt{\pi}}{\Delta}\rho^{-3}C_{\bar{m}n}\notag\\
        &\times\left[L_2^\dagger S+ia\sin\theta\left(\bar{\rho}-\rho\right)S\right],\\
        A_{\bar{m}\bar{m}1}(r,\theta(t))=&2\sqrt{2\pi}\rho^{-3}\bar{\rho}C_{\bar{m}\bar{m}}S\left(i\frac{K}{\Delta}+\rho\right),\\
        A_{\bar{m}\bar{m}2}(r,\theta(t))=&\sqrt{2\pi}\rho^{-3}\bar{\rho}C_{\bar{m}\bar{m}}S.
    \end{align}
\end{subequations}
Here the operator in Eq.~\eqref{Eq.DifferentialOperators} takes a new form due to variable separation
\begin{subequations}\label{Eq.OperatorNeq}
    \begin{align}
        &L_s^\dagger=\partial_\theta+\frac{m}{\sin\theta}-a\omega\sin\theta+s\cot\theta,\\
        & J_+=\partial_r+iK/\Delta.
    \end{align}
\end{subequations}
Then by replacing the variable of integration from $t$ to $\tau$, we obtain
Eq.~\eqref{Eq.ZinReWrite}. Note that $\mathrm{d}t/\mathrm{d}\tau$ cancels 
the $u^t$ in the denominator of Eq.~\eqref{Eq.ProjectedC}.

\section{\label{appC}Analytical expressions of split amplitudes}
Inserting  Eqs.~\eqref{Eq.Exp012} and \eqref{Eq.Z012} into Eq.~\eqref{Eq.ExpZ012}, we derive the expression of  $Z_{lm}^{(1)}$ and $Z_{lm}^{(2)}$. The first order amplitudes take the form of
\begin{equation}\label{C1}
    \begin{aligned}
        Z^{(1)}_{lm}(\tau,\omega)=&\big[\alpha_1 c_{01}+\alpha_2 c_{11}+\alpha_3 c_{21}\\
        &+\alpha_4 c_{31}+\alpha_5\partial_\tau c_{11}+\alpha_6\partial_\tau c_{21}\big]\\
        &\times\exp\left[i\left(\omega u^t-mu^\varphi\right)\tau\right],
    \end{aligned}
\end{equation}
where $\alpha_1$, $\alpha_2$, $\alpha_3$, $\alpha_4$, $\alpha_5$, $\alpha_6$ are some constants when we fix $r_0=p$ and $\theta_0=\pi/2$.  Terms like $\partial_\tau c_{11}$ and $\partial_\tau c_{21}$ appear because there are $\mathrm{d} r/\mathrm{d}\tau,\ \mathrm{d}\theta/\mathrm{d}\tau$ in $C_{nn},C_{\bar{m}n},C_{\bar{m}\bar{m}}$ as defined in Eq.~\eqref{Eq.ProjectedC}. By matching the analytical expressions of $c_{01}$, $c_{11}$, $c_{21}$, $c_{31}$, $\partial_\tau c_{11}$, $\partial_\tau c_{21}$ using \texttt{Mathematica}, it's not hard to obtain their expressions. However, due to their length, we will not show them here. Finally, by applying Eq.~\eqref{A9}, we can further expand our results as combinations of 
$\sin$ and $\cos$ terms, 
\begin{equation}\label{C2}
\begin{aligned}
    Z^{(1)}_{lm}(\tau,\omega)=&\big\{\tilde{\alpha}_1\sin\left[\left(\Omega+\omega_{\mathrm{IB}}\right)\tau-\tilde{\varphi}\right]\\
    &+\tilde{\alpha}_2\cos\left[\left(\Omega+\omega_{\mathrm{IB}}\right)\tau-\tilde{\varphi}\right]\\
    &+\tilde{\alpha}_3\sin\left[\left(\Omega-\omega_{\mathrm{IB}}\right)\tau-\tilde{\varphi}\right]\\
    &+\tilde{\alpha}_4\cos\left[\left(\Omega-\omega_{\mathrm{IB}}\right)\tau-\tilde{\varphi}\right]\\
    &+\tilde{\alpha}_5\sin\left(\omega_{\mathrm{IB}}\tau\right)+\tilde{\alpha}_6\cos\left(\omega_{\mathrm{IB}}\tau\right)\big\}\\
    &\times\exp\left[i\left(\omega u^t-mu^\varphi\right)\tau\right].
\end{aligned}
\end{equation}
Then using Eq.~\eqref{II33} we obtain
\begin{equation}
    \begin{aligned}
        Z^{(1)}_{lm}(\tau,\omega)=&Z^{(1)}_{lm-1-1}\exp\left[i\left(\omega u^t-m u^\varphi+\Omega+\omega_{\mathrm{IB}}\right)\tau\right]\\
        &+Z^{(1)}_{lm11}\exp\left[i\left(\omega u^t-m u^\varphi-\Omega-\omega_{\mathrm{IB}}\right)\tau\right]\\
        &+Z^{(1)}_{lm-11}\exp\left[i\left(\omega u^t-m u^\varphi+\Omega-\omega_{\mathrm{IB}}\right)\tau\right]\\
        &+Z^{(1)}_{lm1-1}\exp\left[i\left(\omega u^t-m u^\varphi-\Omega+\omega_{\mathrm{IB}}\right)\tau\right]\\
        &+Z^{(1)}_{lm0-1}\exp\left[i\left(\omega u^t-mu^\varphi+\omega_{\mathrm{IB}}\right)\tau\right]\\
        &+Z^{(1)}_{lm01}\exp\left[i\left(\omega u^t-mu^\varphi-\omega_{\mathrm{IB}}\right)\tau\right].
    \end{aligned}
\end{equation}
This leads to the frequency spectrum when we do the integration in Eq.~\eqref{Eq.ZExpansion}. Meanwhile, the coefficients of the exponential terms are the split amplitudes
\begin{subequations}\label{C4}
    \begin{align}
        &Z^{(1)}_{lm-1-1}=\frac{-i\tilde{\alpha}_1+\tilde{\alpha}_2}{2}e^{-i\tilde{\varphi}},\\
        &Z^{(1)}_{lm-11}=\frac{-i\tilde{\alpha}_3+\tilde{\alpha}_4}{2}e^{-i\tilde{\varphi}},\\
        &Z^{(1)}_{lm0-1}=\frac{-i\tilde{\alpha}_5+\tilde{\alpha}_6}{2},\\
        &Z^{(1)}_{lm01}=\frac{i\tilde{\alpha}_5+\tilde{\alpha}_6}{2},\\
        &Z^{(1)}_{lm1-1}=\frac{i\tilde{\alpha}_3+\tilde{\alpha}_4}{2}e^{+i\tilde{\varphi}},\\
        &Z^{(1)}_{lm11}=\frac{i\tilde{\alpha}_1+\tilde{\alpha}_2}{2}e^{+i\tilde{\varphi}},
    \end{align}
\end{subequations}
where
\begin{subequations}\label{C5}
    \begin{align}
        \tilde{\alpha}_1=&\frac{\gamma\left(1+\cos\tilde{\theta}\right)}{4\gamma_{\varphi\varphi}^{1/2}}\left[\alpha_1\frac{\mathcal{M}\mathcal{M}_\varphi+\gamma_{\varphi\varphi}^{1/2}\nu}{\mathcal{M}}+\alpha_4\right]\notag\\
        &-\frac{(\Omega+\omega_{\mathrm{IB}})(1+\cos\tilde{\theta})}{4\gamma_{rr}^{1/2}}\alpha_5,\\
        \tilde{\alpha}_2=&\frac{1+\cos\tilde{\theta}}{4\gamma_{rr}^{1/2}}\alpha_2,\\
        \tilde{\alpha}_3=&\frac{\gamma\left(1-\cos\tilde{\theta}\right)}{4\gamma_{\varphi\varphi}^{1/2}}\left[\alpha_1\frac{\mathcal{M}\mathcal{M}_\varphi+\gamma_{\varphi\varphi}^{1/2}\nu}{\mathcal{M}}+\alpha_4\right]\notag\\
        &-\frac{(\Omega-\omega_{\mathrm{IB}})(1-\cos\tilde{\theta})}{4\gamma_{rr}^{1/2}}\alpha_5,\\
        \tilde{\alpha}_4=&\frac{1-\cos\tilde{\theta}}{4\gamma_{rr}^{1/2}}\alpha_2,\\
        \tilde{\alpha}_5=&\frac{\sin\tilde{\theta}}{2\gamma_{\theta\theta}^{1/2}}\alpha_3,\\
        \tilde{\alpha}_6=&\frac{\omega_{\mathrm{IB}}\sin\tilde{\theta}}{2\gamma_{\theta\theta}^{1/2}}\alpha_6.
    \end{align}
\end{subequations}

For second order amplitudes, we use the same method. Analogous to Eq.~\eqref{C1}, we have
\begin{widetext}
    \begin{equation}\label{C6}
\begin{aligned}
    Z^{(2)}_{lm}(\tau,\omega)=&\big\{\beta_{00}c_{01}c_{01}+\beta_{01}c_{11}c_{11}+\beta_{02}c_{21}c_{21}+\beta_{03}c_{31}c_{31}+\beta_{04}c_{01}c_{11}+\beta_{05}c_{01}c_{21}+\beta_{06}c_{01}c_{31}\\
    &+\beta_{07}c_{11}c_{21}+\beta_{08}c_{11}c_{31}+\beta_{09}c_{21}c_{31}+\beta_{0a}c_{02}+\beta_{0b}c_{12}+\beta_{0c}c_{22}+\beta_{0d}c_{32}\\
    &+\beta_{10}c_{01}c_{11}^\prime+\beta_{11}c_{01}c_{21}^\prime+\beta_{12}c_{11}c_{11}^\prime+\beta_{13}c_{11}c_{21}^\prime+\beta_{14}c_{21}c_{11}^\prime+\beta_{15}c_{21}c_{21}^\prime+\beta_{16}c_{31}c_{11}^\prime+\beta_{17}c_{31}c_{21}^\prime\\
    &+\beta_{20}c_{11}^\prime c_{11}^\prime+\beta_{21}c_{21}^\prime c_{21}^\prime+\beta_{22}c_{11}^\prime c_{21}^\prime+\beta_{23}c_{12}^\prime+\beta_{24}c_{22}^\prime\big\}\exp\left[i\left(\omega u^t-mu^\varphi\right)\tau\right].
\end{aligned}
\end{equation}
\end{widetext}
Here prime denotes derivative with respect to $\tau$. By matching coefficients,
we obtain the analytical expressions of these $\beta$. Now we need to
incorporate the $\sin$ and $\cos$ terms into the exponential terms to gain
the frequency spectrum. To shorten the expressions, we will update the notations
used in Eq.~\eqref{A9}, as the original one are too lengthy to present
in full. We rewrite them as
\begin{subequations}\label{C7}
    \begin{align}
        c_{01}=&\vartheta_{11}\sin\left[(\Omega+\omega_{\mathrm{IB}})\tau-\tilde{\varphi}\right]\notag\\
        &+\vartheta_{12}\sin\left[(\Omega-\omega_{\mathrm{IB}})\tau-\tilde{\varphi}\right],\\
        c_{11}=&\vartheta_{21}\cos\left[(\Omega+\omega_{\mathrm{IB}})\tau-\tilde{\varphi}\right]\notag\\
        &+\vartheta_{22}\cos\left[(\Omega-\omega_{\mathrm{IB}})\tau-\tilde{\varphi}\right],\\
        c_{11}^\prime=&\vartheta_{21}^\prime\sin\left[(\Omega+\omega_{\mathrm{IB}})\tau-\tilde{\varphi}\right]\notag\\
        &+\vartheta_{22}^\prime\sin\left[(\Omega-\omega_{\mathrm{IB}})\tau-\tilde{\varphi}\right],\\
        c_{21}=&\vartheta_{31}\sin(\omega_{\mathrm{IB}}\tau),\\
        c_{21}^\prime=&\vartheta_{31}^\prime\cos(\omega_{\mathrm{IB}}\tau),\\
        c_{31}=&\vartheta_{41}\sin\left[(\Omega+\omega_{\mathrm{IB}})\tau-\tilde{\varphi}\right]\notag\\
        &+\vartheta_{42}\sin\left[(\Omega-\omega_{\mathrm{IB}})\tau-\tilde{\varphi}\right],\\
    c_{02}=&\vartheta_{51}\sin\left[2(\Omega+\omega_{\mathrm{IB}})\tau-2\tilde{\varphi}\right]\notag\\
    &+\vartheta_{52}\sin\left(2\Omega\tau-2\tilde{\varphi}\right)\notag\\
    &+\vartheta_{53}\sin\left[2(\Omega-\omega_{\mathrm{IB}})\tau-2\tilde{\varphi}\right],\\
        c_{12}=&\vartheta_{61}\cos\left[2(\Omega+\omega_{\mathrm{IB}})\tau-2\tilde{\varphi}\right]\notag\\
        &+\vartheta_{62}\cos\left(2\Omega\tau-2\tilde{\varphi}\right)\notag\\
        &+\vartheta_{63}\cos\left[2(\Omega-\omega_{\mathrm{IB}})\tau-2\tilde{\varphi}\right]\notag\\
        &+\vartheta_{64}\cos(2\omega_{\mathrm{IB}}\tau)+\vartheta_{65},\\
        c_{12}^\prime=&\vartheta_{61}^\prime\sin\left[2(\Omega+\omega_{\mathrm{IB}})\tau-2\tilde{\varphi}\right]\notag\\
        &+\vartheta_{62}^\prime\sin\left(2\Omega\tau-2\tilde{\varphi}\right)\notag\\
        &+\vartheta_{63}^\prime\sin\left[2(\Omega-\omega_{\mathrm{IB}})\tau-2\tilde{\varphi}\right]\notag\\
        &+\vartheta_{64}^\prime\sin(2\omega_{\mathrm{IB}}\tau),\\
        c_{22}=&\vartheta_{71}\cos\left[(\Omega+2\omega_{\mathrm{IB}})\tau-\tilde{\varphi}\right]\notag\\
        &+\vartheta_{72}\cos\left[(\Omega-2\omega_{\mathrm{IB}})\tau-\tilde{\varphi}\right]\notag\\
        &+\vartheta_{73}\cos\left(\Omega\tau-\tilde{\varphi}\right),\\
        c_{22}^\prime=&\vartheta_{71}^\prime\sin\left[(\Omega+2\omega_{\mathrm{IB}})\tau-\tilde{\varphi}\right]\notag\\
        &+\vartheta_{72}^\prime\sin\left[(\Omega-2\omega_{\mathrm{IB}})\tau-\tilde{\varphi}\right]\notag\\
        &+\vartheta_{73}^\prime\sin\left(\Omega\tau-\tilde{\varphi}\right),\\
        c_{32}=&\vartheta_{81}\sin\left[2(\Omega+\omega_{\mathrm{IB}})\tau-2\tilde{\varphi}\right]\notag\\
        &+\vartheta_{82}\sin\left(2\Omega\tau-2\tilde{\varphi}\right)\notag\\
        &+\vartheta_{83}\sin\left[2(\Omega-\omega_{\mathrm{IB}})\tau-2\tilde{\varphi}\right].
    \end{align}
\end{subequations}
where
\begin{subequations}\label{C8}
    \begin{align}
        &\vartheta_{11}=\frac{\mathcal{M}\mathcal{M}_\varphi+\gamma_{\varphi\varphi}^{1/2}\nu}{4\gamma_{\varphi\varphi}^{1/2}\mathcal{M}/\gamma}\left(1+\cos\tilde{\theta}\right),\\
        &\vartheta_{12}=-\frac{\mathcal{M}\mathcal{M}_\varphi+\gamma_{\varphi\varphi}^{1/2}\nu}{4\gamma_{\varphi\varphi}^{1/2}\mathcal{M}/\gamma}\left(1-\cos\tilde{\theta}\right),\\
        &\vartheta_{21}=\frac{1}{4\gamma_{rr}^{1/2}}\left(1+\cos\tilde{\theta}\right),\\
        &\vartheta_{22}=-\frac{1}{4\gamma_{rr}^{1/2}}\left(1-\cos\tilde{\theta}\right),\\
        &\vartheta_{21}^\prime=-(\Omega+\omega_{\mathrm{IB}})\vartheta_{21},\\
        &\vartheta_{22}^\prime=-(\Omega-\omega_{\mathrm{IB}})\vartheta_{22},\\
        &\vartheta_{31}=\frac{1}{2\gamma_{\theta\theta}^{1/2}}\sin\tilde{\theta},\\
        &\vartheta_{31}^\prime=-\omega_{\mathrm{IB}}\vartheta_{31},\\
        &\vartheta_{41}=\frac{\gamma}{4\gamma_{\varphi\varphi}^{1/2}}\left(1+\cos\tilde{\theta}\right),\\
        &\vartheta_{42}=-\frac{\gamma}{4\gamma_{\varphi\varphi}^{1/2}}\left(1-\cos\tilde{\theta}\right),\\
        &\vartheta_{51}=2\Theta_0\cos^4\frac{\tilde{\theta}}{2},\\
        &\vartheta_{52}=-\Theta_0\sin^2\tilde{\theta},\\
        &\vartheta_{53}=2\Theta_0\sin^4\frac{\tilde{\theta}}{2},\\
        &\Theta_0=\frac{\left(\mathcal{M}\mathcal{M}_\varphi+\gamma_{\varphi\varphi}^{1/2}\nu\right)\kappa_\varphi-\left(\gamma_{\varphi\varphi}^{1/2}+\mathcal{M}\mathcal{M}_\varphi\nu\right)\Omega}{16\gamma_{\varphi\varphi}^{1/2}\mathcal{M}/\gamma},\\
        &\vartheta_{61}=\frac{\kappa_r+\kappa_{\varphi}}{16\gamma_{rr}^{1/2}}\cos^4\frac{\tilde{\theta}}{2},\\
        &\vartheta_{62}=\frac{\kappa_r+\kappa_\varphi}{16\gamma_{rr}^{1/2}}\sin^4\frac{\tilde{\theta}}{2},\\
        &\vartheta_{63}=-\frac{\kappa_r+\kappa_\varphi}{32\gamma_{rr}^{1/2}}\sin^2\tilde{\theta},\\
        &\vartheta_{64}=-\frac{\kappa_r+2\kappa_{\theta}-\kappa_\varphi}{32\gamma_{rr}^{1/2}}\sin^2\tilde{\theta},\\
        &\vartheta_{65}=\frac{\kappa_r-\kappa_{\varphi}}{32\gamma_{rr}^{1/2}}\left(1+\cos^2\tilde{\theta}\right)-\frac{\kappa_\theta}{16\gamma_{rr}^{1/2}}\sin^2\tilde{\theta},\\
        &\vartheta_{61}^\prime=-2(\Omega+\omega_{\mathrm{IB}})\vartheta_{61},\\
        &\vartheta_{62}^\prime=-2\Omega\vartheta_{62},\\
        &\vartheta_{63}^\prime=-2(\Omega-\omega_{\mathrm{IB}})\vartheta_{63},\\
        &\vartheta_{64}^\prime=-2\omega_{\mathrm{IB}}\vartheta_{64},\\
        &\vartheta_{71}=\frac{\kappa_{\theta}\cos\tilde{\theta}}{16\gamma_{\theta\theta}^{1/2}}\left(1+\cos\tilde{\theta}\right),\\
        &\vartheta_{72}=-\frac{\kappa_{\theta}\cos\tilde{\theta}}{16\gamma_{\theta\theta}^{1/2}}\left(1-\cos\tilde{\theta}\right),\\
        &\vartheta_{73}=\frac{\kappa_{\theta}}{8\gamma_{\theta\theta}^{1/2}}\cos\left(2\tilde{\theta}\right),\\
        &\vartheta_{71}^\prime=-\left(\Omega+2\omega_{\mathrm{IB}}\right)\vartheta_{71},\\
        &\vartheta_{72}^\prime=-\left(\Omega-2\omega_{\mathrm{IB}}\right)\vartheta_{72},\\
        &\vartheta_{73}^\prime=-\Omega\vartheta_{73},\\
        &\vartheta_{81}=\frac{\kappa_\varphi-\nu\Omega}{8\gamma_{\varphi\varphi}^{1/2}/\gamma}\cos^4\frac{\tilde{\theta}}{2},\\
        &\vartheta_{82}=-\frac{\kappa_\varphi-\nu\Omega}{16\gamma_{\varphi\varphi}^{1/2}/\gamma}\sin^2\tilde{\theta},\\
        &\vartheta_{83}=\frac{\kappa_\varphi-\nu\Omega}{6\gamma_{\varphi\varphi}^{1/2}/\gamma}\sin^4\frac{\tilde{\theta}}{2}.
    \end{align}
\end{subequations}
Analogous to Eq.~\eqref{C4}, the split amplitudes for the second order expansion are given by
\begin{widetext}
    \begin{subequations}\label{C9}
        \begin{align}
            Z_{lm-2-2}^{(2)}=&-\frac{1}{4}\left[\beta_{00}\vartheta_{11}^2-\beta_{01}\vartheta_{21}^2+\beta_{20}\vartheta_{21}'^2+\beta_{03}\vartheta_{41}^2+\beta_{10}\vartheta_{11}\vartheta_{21}'+\beta_{06}\vartheta_{11}\vartheta_{41}+\beta_{16}\vartheta_{21}'\vartheta_{41}\right]\notag\\
        &-\frac{1}{4}i\left[\beta_{04}\vartheta_{11}\vartheta_{21}+\beta_{12}\vartheta_{21}\vartheta_{21}'+\beta_{08}\vartheta_{21}\vartheta_{41}\right]+\frac{1}{2}\left(\beta_{0a}+\beta_{0b}\right)\vartheta_{61}-\frac{1}{2}i\left(\beta_{21}\vartheta_{61}'+\beta_{0d}\vartheta_{81}\right),\\
        Z_{lm-20}^{(2)}=&-\frac{1}{4}\left[\beta_{10}\left(\vartheta_{12}\vartheta_{21}'+\vartheta_{11}\vartheta_{22}'\right)+\beta_{06}\left(\vartheta_{12}\vartheta_{41}+\vartheta_{11}\vartheta_{42}\right)+\beta_{16}\left(\vartheta_{22}'\vartheta_{41}+\vartheta_{21}'\vartheta_{42}\right)\right]\notag\\
      & -\frac{1}{4}i\left[\beta_{04}\left(\vartheta_{12}\vartheta_{21}+\vartheta_{11}\vartheta_{22}\right)+\beta_{12}\left(\vartheta_{21}'\vartheta_{22}+\vartheta_{21}\vartheta_{22}'\right)+\beta_{08}\left(\vartheta_{22}\vartheta_{41}+\vartheta_{21}\vartheta_{42}\right)\right]\notag\\
      &+\frac{1}{2}\left[-\beta_{00}\vartheta_{11}\vartheta_{12}+\beta_{01}\vartheta_{21}\vartheta_{22}-\beta_{20}\vartheta_{21}'\vartheta_{22}'-\beta_{03}\vartheta_{41}\vartheta_{42}+\left(\beta_{0a}+\beta_{0b}\right)\vartheta_{62}\right]\notag\\
      &-\frac{1}{2}i\left(\beta_{23}\vartheta_{62}'+\beta_{0d}\vartheta_{82}\right),\\
      Z_{lm-22}^{(2)}=&-\frac{1}{4}\left[\beta_{00}\vartheta_{12}^2-\beta_{01}\vartheta_{22}^2+\beta_{20}\vartheta_{22}'^2+\beta_{03}\vartheta_{42}^2+\beta_{10}\vartheta_{12}\vartheta_{22}'+\beta_{06}\vartheta_{12}\vartheta_{42}+\beta_{16}\vartheta_{22}'\vartheta_{42}\right]\notag\\
      &-\frac{1}{4}i\left[\beta_{04}\vartheta_{12}\vartheta_{22}+\beta_{12}\vartheta_{22}\vartheta_{22}'+\beta_{08}\vartheta_{22}\vartheta_{42}\right]+\frac{1}{2}\left(\beta_{0a}+\beta_{0b}\right)\vartheta_{63}-\frac{1}{2}i\left(\beta_{23}\vartheta_{63}'+\beta_{0d}\vartheta_{83}\right),\\
      Z_{lm-1-2}^{(2)}=& -\frac{1}{4} \left[ 
\beta_{05} \vartheta_{11} \vartheta_{31} + 
\beta_{09} \vartheta_{31} \vartheta_{41} + 
\beta_{14} \vartheta_{21}' \vartheta_{31} - 
\beta_{13} \vartheta_{21} \vartheta_{31}'
\right] + \frac{1}{2} \beta_{24} \vartheta_{71}' \notag\\
& - \frac{1}{4}i \left[ 
\beta_{07} \vartheta_{21} \vartheta_{31} + 
\beta_{11} \vartheta_{11} \vartheta_{31}' + 
\beta_{17} \vartheta_{31}' \vartheta_{41} + 
\beta_{22} \vartheta_{21}' \vartheta_{31}'
\right]- \frac{1}{2}i \beta_{0c} \vartheta_{71},\\
Z_{lm-10}^{(2)}=&\frac{1}{4}\left[\beta_{05}\vartheta_{31}\left(\vartheta_{11}-\vartheta_{12}\right)+\beta_{14}\vartheta_{31}\left(\vartheta_{21}'-\vartheta_{22}'\right)+\beta_{13}\vartheta_{31}'\left(\vartheta_{21}+\vartheta_{22}\right)+\beta_{09}\vartheta_{31}\left(\vartheta_{41}-\vartheta_{42}\right)\right]\notag\\
&-\frac{1}{4}i\left[\beta_{07}\vartheta_{31}\left(\vartheta_{22}-\vartheta_{21}\right)+\beta_{11}\vartheta_{31}'\left(\vartheta_{11}+\vartheta_{12}\right)+\beta_{22}\vartheta_{31}'\left(\vartheta_{21}'+\vartheta_{22}'\right)+\beta_{17}\vartheta_{31}'\left(\vartheta_{41}+\vartheta_{42}\right)\right]\notag\\
&-\frac{1}{2}i\beta_{0c}\vartheta_{73}+\frac{1}{2}\beta_{24}\vartheta_{73}',\\
Z_{lm-12}^{(2)}=&\frac{1}{4}\left[\beta_{05}\vartheta_{12}\vartheta_{31}+\beta_{14}\vartheta_{22}'\vartheta_{31}+\beta_{13}\vartheta_{22}\vartheta_{31}'+\beta_{09}\vartheta_{31}\vartheta_{42}\right]+\frac{1}{2}\beta_{24}\vartheta_{72}'\notag\\
&+\frac{1}{4}i\left[\beta_{11}\vartheta_{12}\vartheta_{31}'-\beta_{07}\vartheta_{22}'\vartheta_{31}+\beta_{22}\vartheta_{22}'\vartheta_{31}'+\beta_{17}\vartheta_{31}'\vartheta_{42}\right]+\frac{1}{2}i\beta_{0c}\vartheta_{72},\\
Z_{lm0-2}^{(2)}=&\frac{1}{2}\left[ 
\beta_{00}\vartheta_{11}\vartheta_{12} + 
\beta_{01}\vartheta_{21}\vartheta_{22} + 
\beta_{20}\vartheta_{21}'\vartheta_{22}' + 
\beta_{03}\vartheta_{41}\vartheta_{42} + 
\left( \beta_{0a} + \beta_{0b} \right) \vartheta_{64}
\right]- \frac{1}{2} i\beta_{23} \vartheta_{64}'\notag \\
&+ \frac{1}{4}\left[ 
\beta_{10}\left( \vartheta_{12}\vartheta_{21}' + \vartheta_{11}\vartheta_{22}' \right) + 
\beta_{21}\vartheta_{31}'^2 + 
\beta_{16}\left( \vartheta_{22}'\vartheta_{41} + \vartheta_{21}'\vartheta_{42} \right) + 
\beta_{06}\left( \vartheta_{11}\vartheta_{42} + \vartheta_{12}\vartheta_{41} \right) - 
\beta_{02}\vartheta_{31}^2
\right] \notag\\
&+ \frac{1}{4} i\left[ 
\beta_{04}\left( \vartheta_{12}\vartheta_{21} - \vartheta_{11}\vartheta_{22} \right) + 
\beta_{12}\left( \vartheta_{21}\vartheta_{22}' - \vartheta_{21}'\vartheta_{22} \right) + 
\beta_{08}\left( - \vartheta_{22}\vartheta_{41} + \vartheta_{21}\vartheta_{42} \right) - 
\beta_{15}\vartheta_{31}\vartheta_{31}'
\right],\\
Z^{(2)}_{lm00}=&\frac{1}{2}\big[\beta_{00}\left(\vartheta_{11}^2+\vartheta_{12}^2\right)+\beta_{01}\left(\vartheta_{21}^2+\vartheta_{22}^2\right)+\beta_{10}\left(\vartheta_{11}\vartheta_{21}^\prime+\vartheta_{12}\vartheta_{22}^\prime\right)+\beta_{20}\left(\vartheta_{21}^{\prime2}+\vartheta_{22}^{\prime2}\right)\notag\\
&+\beta_{02}\vartheta_{31}^2+\beta_{21}\vartheta_{31}^{\prime2}+\beta_{06}\left(\vartheta_{11}\vartheta_{41}+\vartheta_{12}\vartheta_{42}\right)+\beta_{16}\left(\vartheta_{21}^{\prime}\vartheta_{41}+\vartheta_{22}^\prime\vartheta_{42}\right)+\beta_{03}\left(\vartheta_{41}^2+\vartheta_{42}^2\right)\big]\notag\\
&+\left(\beta_{0a}+\beta_{0b}\right)\vartheta_{65},\\
Z_{lm02}^{(2)}=&\frac{1}{2} \left[ 
\beta_{00}\vartheta_{11}\vartheta_{12} + 
\beta_{01}\vartheta_{21}\vartheta_{22} + 
\beta_{20}\vartheta_{21}'\vartheta_{22}' + 
\beta_{03}\vartheta_{41}\vartheta_{42} + 
\left( \beta_{0a} + \beta_{0b} \right) \vartheta_{64}
\right]+ \frac{1}{2}i \beta_{23}\vartheta_{64}' \notag\\
&+ \frac{1}{4} \left[ 
\beta_{10}\left( \vartheta_{12}\vartheta_{21}' + \vartheta_{11}\vartheta_{22}' \right) + 
\beta_{21}\vartheta_{31}'^2 + 
\beta_{16}\left( \vartheta_{22}'\vartheta_{41} + \vartheta_{21}'\vartheta_{42} \right) + 
\beta_{06}\left( \vartheta_{11}\vartheta_{42} + \vartheta_{12}\vartheta_{41} \right) - 
\beta_{02}\vartheta_{31}^2
\right]\notag \\
&+ \frac{1}{4}i \left[ 
\beta_{04}\left( \vartheta_{11}\vartheta_{22} - \vartheta_{12}\vartheta_{21} \right) + 
\beta_{12}\left( \vartheta_{21}'\vartheta_{22} - \vartheta_{21}\vartheta_{22}' \right) + 
\beta_{08}\left( \vartheta_{22}\vartheta_{41} - \vartheta_{21}\vartheta_{42} \right) + 
\beta_{15}\vartheta_{31}\vartheta_{31}'\right],\\
Z_{lm1-2}^{(2)}=&\frac{1}{4}\left[\beta_{05}\vartheta_{12}\vartheta_{31}+\beta_{14}\vartheta_{22}'\vartheta_{31}+\beta_{13}\vartheta_{22}\vartheta_{31}'+\beta_{09}\vartheta_{31}\vartheta_{42}\right]+\frac{1}{2}\beta_{24}\vartheta_{72}'\notag\\
       &+\frac{1}{4}i\left[-\beta_{07}\vartheta_{22}\vartheta_{31}+\beta_{11}\vartheta_{12}\vartheta_{31}'+\beta_{22}\vartheta_{22}'\vartheta_{31}'+\beta_{17}\vartheta_{31}'\vartheta_{42}\right]+\frac{1}{2}i\beta_{0c}\vartheta_{72},\\
       Z_{lm10}^{(2)}=&\frac{1}{4}\left[\beta_{05}\vartheta_{31}\left(\vartheta_{11}-\vartheta_{12}\right)+\beta_{14}\vartheta_{31}\left(\vartheta_{21}'-\vartheta_{22}'\right)+\beta_{13}\vartheta_{31}'\left(\vartheta_{21}+\vartheta_{22}\right)+\beta_{09}\vartheta_{31}\left(\vartheta_{41}-\vartheta_{42}\right)\right]\notag\\
&+\frac{1}{4}i\left[\beta_{07}\vartheta_{31}\left(\vartheta_{22}-\vartheta_{21}\right)+\beta_{11}\vartheta_{31}'\left(\vartheta_{11}+\vartheta_{12}\right)+\beta_{22}\vartheta_{31}'\left(\vartheta_{21}'+\vartheta_{22}'\right)+\beta_{17}\vartheta_{31}'\left(\vartheta_{41}+\vartheta_{42}\right)\right]\notag\\
      &+\frac{1}{2}i\beta_{0c}\vartheta_{73}+\frac{1}{2}\beta_{24}\vartheta_{73}',\\
       Z_{lm12}^{(2)}=&-\frac{1}{4}\left[\beta_{05}\vartheta_{11}\vartheta_{31}+\beta_{14}\vartheta_{21}'\vartheta_{31}-\beta_{13}\vartheta_{21}\vartheta_{31}'+\beta_{09}\vartheta_{31}\vartheta_{41}\right]+\frac{1}{2}\beta_{24}\vartheta_{71}'\notag\\
       &+\frac{1}{4}i\left[\beta_{07}\vartheta_{21}\vartheta_{31}+\beta_{11}\vartheta_{11}\vartheta_{31}'+\beta_{22}\vartheta_{21}'\vartheta_{31}'+\beta_{17}\vartheta_{31}'\vartheta_{41}\right]+\frac{1}{2}i\beta_{0c}\vartheta_{71},\\
       Z_{lm2-2}^{(2)}=&-\frac{1}{4}\left[\beta_{00}\vartheta_{12}^2-\beta_{01}\vartheta_{22}^2+\beta_{20}\vartheta_{22}'^2+\beta_{03}\vartheta_{42}^2+\beta_{10}\vartheta_{12}\vartheta_{22}'+\beta_{06}\vartheta_{12}\vartheta_{42}+\beta_{16}\vartheta_{22}'\vartheta_{42}\right]\notag\\
        &+\frac{1}{4}i\left[\beta_{04}\vartheta_{12}\vartheta_{22}+\beta_{12}\vartheta_{22}\vartheta_{22}'+\beta_{08}\vartheta_{22}\vartheta_{42}\right]+\frac{1}{2}\left(\beta_{0a}+\beta_{0b}\right)\vartheta_{63}+\frac{1}{2}i\left(\beta_{23}\vartheta_{63}'+\beta_{0d}\vartheta_{83}\right),\\
        Z_{lm20}^{(2)}=&-\frac{1}{4}\left[\beta_{10}\left(\vartheta_{12}\vartheta_{21}'+\vartheta_{11}\vartheta_{22}'\right)+\beta_{06}\left(\vartheta_{12}\vartheta_{41}+\vartheta_{11}\vartheta_{42}\right)+\beta_{16}\left(\vartheta_{22}'\vartheta_{41}+\vartheta_{21}'\vartheta_{42}\right)\right]\notag\\
        &+\frac{1}{4}i\left[\beta_{04}\left(\vartheta_{12}\vartheta_{21}-\vartheta_{11}\vartheta_{22}\right)+\beta_{12}\left(\vartheta_{21}'\vartheta_{22}+\vartheta_{21}\vartheta_{22}'\right)+\beta_{08}\left(\vartheta_{22}\vartheta_{41}+\vartheta_{21}\vartheta_{42}\right)\right]\notag\\
        &+\frac{1}{2}\left[-\beta_{00}\vartheta_{11}\vartheta_{12}+\beta_{01}\vartheta_{21}\vartheta_{22}-\beta_{20}\vartheta_{21}'\vartheta_{22}'-\beta_{03}\vartheta_{41}\vartheta_{42}+\left(\beta_{0a}+\beta_{0b}\right)\vartheta_{62}\right]+\frac{1}{2}i\left(\beta_{23}\vartheta_{62}'+\beta_{0d}\vartheta_{82}\right),\\
        Z_{lm22}^{(2)}=&-\frac{1}{4}\left[\beta_{00}\vartheta_{11}^2-\beta_{01}\vartheta_{21}^2+\beta_{20}\vartheta_{21}'^2+\beta_{03}\vartheta_{41}^2\right]+\frac{1}{4}i\left[\beta_{04}\vartheta_{11}\vartheta_{21}+\beta_{12}\vartheta_{21}\vartheta_{21}'+\beta_{08}\vartheta_{21}'\vartheta_{41}\right]\notag\\
      &+\frac{1}{2}\left(\beta_{0a}+\beta_{0b}\right)\vartheta_{61}+\frac{1}{2}i\left(\beta_{23}\vartheta_{61}'+\beta_{0d}\vartheta_{81}\right).
        \end{align}
    \end{subequations}
\end{widetext}

\section{\label{appD} Conserved stress-energy constructed in the FFF}
Typically when considering the motion of extended compact bodies, we specify their multipole structure and CM. This specifies the structure of the stress-energy. The equations of motion that determine the worldline follow from the conservation of stress-energy --- which we then solve. This is the approach of the Mathisson-Papapetrou-Dixon equations. In this work, we reverse the logic. We construct worldlines resembling two point particles of equal mass in a circular Newtonian orbit whose CM is in free fall. We ask that the stress-energy resembles that of two free point particles up to an interaction term. We then write down the interaction term that conserves the stress-energy to the order that we require.\\

We can take the metric to be the Minkowski metric ($\eta_{\alpha\beta}$) in the FFF coordinates $\tilde x^{\alpha}=(\tilde{t},\tilde{x}, \tilde{y},\tilde{z})$ to $\mathcal{O}(\tilde x^\alpha{}\tilde x^\beta)$~\cite{Poisson:2009pwt};
\begin{equation}
    ds^2=\left(\eta_{\alpha \beta}+h_{\alpha\beta}\right)d \tilde x^\alpha d\tilde x^\beta.
\end{equation}

The worldline of each body in the IB is
\begin{equation}
\label{eq:FFFworldline}
    z_i^\mu=\left(\tau, (-1)^i\frac{d}{2}\cos \omega_{\mathrm{IB}}\tau,0,(-1)^i\frac{d}{2}\sin \omega_{\mathrm{IB}}\tau\right).
\end{equation}
We define $\dot z_i^\mu=\frac{d z_i^\mu}{d\tau}$ while $ \frac{d z_i^\mu}{d\tau_i}=\dot z_i^\mu\frac{d\tau}{d\tau_i}$. We impose the timelike normalisation of the tangent vector of each worldline, which implies
\begin{align}
    \frac{d \tau}{d \tau_i}=&\left(1-h_{\tilde t \tilde t}|_{z_i}-\left(\frac{d\omega_{\mathrm{IB}}}{2}\right)^2\right)^{-2}+\mathcal{O}(d^3),
\end{align}
having partially expanded in $d$ at fixed $\omega_{\mathrm{IB}}$. Now consider the monopole stress-energy term
\begin{equation}
    T^{\mu \nu}_i=m_i\int\left(\frac{d\tau}{d\tau_i}\right)^2 \dot z_i^\mu \dot z_i^\nu \frac{\delta^4(\tilde{x}^\alpha-z_i^\alpha)}{\sqrt{-g}}d\tau_i.
\end{equation}
By itself, it is not conserved given our constructed worldline:
\begin{align}
    \nabla_\nu T^{\mu \nu}_i\doteq&m_i\int \left(\frac{d\tau}{d\tau_i}\right)\dot z_i^\mu \dot z_i^\nu \partial_\nu\frac{\delta^4(\tilde{x}^\alpha- z_i^\alpha)}{\sqrt{-g}}d\tau+\mathcal{O}(m_i d^3),
\end{align}
where the dot on the equals sign is to signal that this statement is specialised to the FFF since we have used the fact that the Christoffel connection is $O(\tilde x^\alpha)$. By the chain rule
\begin{equation}
    \frac{d}{d\tau_i}\delta^4(x^\alpha- z_i^\alpha(\tau_i))=-\frac{d \tau}{d\tau_i}\dot z_i^\nu\partial_\nu \delta^4(x^\alpha- z_i^\alpha(\tau_i)), 
\end{equation}
thus
\begin{align}
    \nabla_\nu T^{\mu \nu}_i\doteq&-m_i\int \dot z_i^\mu\frac{d}{d\tau}\frac{\delta^4(\tilde{x}^\alpha- z_i^\alpha)}{\sqrt{-g}}d\tau+\mathcal{O}( m_i d^3).
\end{align}
Integrating by parts and discarding the (vanishing) boundary term, we have 
\begin{align}
    \nabla_\nu T^{\mu \nu}_i\doteq& m_i\int \ddot z_i^\mu \frac{\delta^4(\tilde{x}^\alpha- z_i^\alpha)}{\sqrt{-g}}d\tau+\mathcal{O}(d^3),\nonumber\\
    \doteq& m_i \int \ddot z_i^\mu \frac{\delta^3(\tilde{x}^a- z_i^a)\delta(\tilde t-\tau)}{\sqrt{-g}}d\tau+\mathcal{O}(m_i d^3),
\end{align}
for $\tilde x^a=(\tilde x, \tilde y, \tilde z)$. Clearly this equation becomes trivially satisfied if the trajectories are instead geodesics of the Minkowski metric.
We now expand the delta function in small $d$ at fixed $\omega_{\mathrm{IB}}$;
\begin{equation}
    \nabla_\nu T^{\mu \nu}_i\doteq m_i \int\ddot z_i^\mu\left(\delta^3(\tilde x^a)- z_i^a\partial_a\delta^3(\tilde x^a)\right)\frac{\delta(\tilde t-\tau)}{\sqrt{-g}}d\tau+\mathcal{O}(m_i d^3).
\end{equation}
In the sum of two monopole particles with equal masses $m_i=\mu/2$;
\begin{equation}
    \sum_i\nabla_\nu T^{\mu \nu}_i\doteq -\mu \int\ddot z_i^\mu  z_i^a\partial_a\delta^3(\tilde x^a)\frac{\delta(\tilde t-\tau)}{\sqrt{-g}}d\tau+\mathcal{O}(\mu d^3),
\end{equation}
since our constructed trajectories satisfy $\ddot z_i^{\tilde{t}}\doteq0$, $\ddot z_1^{a}\doteq~-\ddot z_2^{a}$ and $ z_1^{a}\doteq- z_2^{a}$.
We now look for a correction to the stress-energy tensor so that it is conserved to $O(\mu d^3)$. 
We use the relation 
\begin{equation}
P^{\mu}{}_ {\nu}\ddot z_i^\nu\doteq-\omega_{\mathrm{IB}}^2P^{\mu}{}_ {\nu} z_i^\nu,
\end{equation}
that follows from Eq.~\eqref{eq:FFFworldline}, having defined the projection operator
\begin{equation}
P^{\mu}{}_ {\nu}\equiv \delta^\mu{}_\nu-\tilde{t}^{\mu}\tilde{t}_{\nu}, 
\end{equation}
and the vector $\tilde{t}^{\nu}\doteq(1,0,0,0)$.
Then
\begin{align}
     T_{\text{Int}}^{\mu \nu}\doteq& -\mu \omega_{\mathrm{IB}}^2\int P^{\mu}{}_ {\alpha} z_i^\alpha P^{\nu}{}_ {\beta} z_i^\beta\frac{\delta^4(\tilde x^\alpha-z^\alpha_c)}{\sqrt{-g}}d\tau\nonumber+\mathcal{O}(\mu d^3),
\end{align}
is the required interaction term in the stress-energy to ensure its conservation though $\mathcal{O}(\mu d^3)$ at fixed $\omega_{\mathrm{IB}}$ having also introduced $z_c^\alpha\doteq\tau \tilde t^\alpha$ as the centre-of-mass worldline. Instead of using the projections, we could also write
\begin{equation}
     T_{\text{Int}}^{\mu \nu}\doteq -\mu \omega_{\mathrm{IB}}^2\int (z_i^\mu-z_c^\mu)(z_i^\nu-z_c^\nu)\frac{\delta^4(\tilde x^\alpha-z^\alpha_c)}{\sqrt{-g}}d\tau+\mathcal{O}(\mu d^3),
\end{equation}
and we notice the trace of this term could be neatly expressed in terms of Synge's world function at this order. The approximated total stress-energy including the interaction term is
\begin{equation}
 T^{\mu \nu}=\mu\int \dot z_i^\mu \dot z_i^\nu \frac{\delta^4(\tilde{x}^\alpha-z_c^\alpha)}{\sqrt{-g}}d\tau+T_{\text{Int}}^{\mu \nu}+\mathcal{O}(\mu d^3).
\end{equation}
We can now compute the stress-energy of the IB in an arbitrary coordinate system. In hindsight, the form of the interaction term is not so mysterious --- it is reminiscent of the Newtonian point particle rotational energy, $\mu \omega_{\mathrm{IB}}^2 r^2$, in which $r$ is the distance to the axis of rotation. As $d\rightarrow0$, we recover the stress-energy of a free monopole particle of mass $\mu$. As a final remark, constructing such a conserved stress-energy tensor may seem arduous in the general case (un-equal mass eccentric IBs). However the relevant stress-energy tensor can also be constructed from the variation of the Lagrangian defined in Ref.~\cite{10.1063/1.3382338,PhysRevD.108.123041} (see e.g Chapter 4 of Ref.~\cite{Poisson:2009pwt}).

\bibliography{apssamp}

\end{document}